\documentclass[aps,prd,amssymb,eqsecnum,nofootinbib,floatfix, a4paper,twocolumn]{revtex4}

\usepackage{mathrsfs}
\usepackage{latexsym,amsmath,amsfonts,amssymb}
\usepackage{bm}

\usepackage{graphicx}

\usepackage{color}

\allowdisplaybreaks

\newcommand{\be}{\begin{equation}}
\newcommand{\ee}{\end{equation}}
\newcommand{\bea}{\begin{eqnarray}}
\newcommand{\eea}{\end{eqnarray}}

\def\tal{{\widetilde{\alpha}}}
\def\bal{{\overline{\alpha}}}

\def\lam{{\lambda}}
\def\hlam{{\widehat \lambda}}
\def\bx{{\bar x}}

\def\d{\partial}
\def\l{\left(}
\def\r{\right)}

\def\t0{\tilde{0}}
\def\ta{\tilde{a}}
\def\tb{\tilde{b}}
\def\tc{\tilde{c}}

\def\cA{{\cal A}}
\def\cF{{\cal F}}
\def\cR{{\cal R}}
\def\cG{{\cal G}}
\def\cT{{\cal T}}

\def\hcf{{\widehat c_5}}
\def\hcs{{\widehat c_6}}
\def\hcthree{{\widehat c_3}}
\def\hcfour{{\widehat c_4}}

\def\del{{\delta}}
\def\bd{{\bar \delta}}

\newcommand{\bg}{\begin{gather}}
\newcommand{\eg}{\end{gather}}
\newcommand{\bseq}{\begin{subequations}}
\newcommand{\eseq}{\end{subequations}}

\begin{document}

\title{Infrared modified gravity with propagating torsion: \\instability of torsionfull de Sitter-like solutions}

\author{Vasilisa \surname{Nikiforova}$^1$}
\author{Thibault \surname{Damour}$^2$}
 
\affiliation{$^1$Institute for Nuclear Research of the Russian Academy of Sciences, 60th October Anniversary Prospect 7a, 117312, Moscow, Russia\\
$^2$Institut des Hautes Etudes Scientifiques, 91440 Bures-sur-Yvette, France}

\date{\today}

\begin{abstract}
We continue the exploration of the consistency of a modified-gravity theory that generalizes
General Relativity by including a dynamical torsion in addition to the dynamical metric. The six-parameter
theory we consider was found to be consistent  around arbitrary torsionless Einstein backgrounds, in spite of 
its containing a (notoriously delicate) massive spin-2 excitation. 
At zero bare cosmological constant, this theory was found to admit a self-accelerating
solution whose exponential expansion is sustained by a non-zero torsion background.
The scalar-type perturbations of the latter torsionfull self-accelerating solution were recently studied and were
found to preserve the number of propagating scalar  degrees of freedom, but to exhibit, for some values of
the torsion background some exponential instabilities (of a rather mild type). Here, we study the tensor-type and
vector-type perturbations of the torsionfull self-accelerating solution, and of its deformation by a non-zero bare
cosmological constant. We find strong, ``gradient" instabilities in the vector sector. No tuning of the parameters
of the theory can kill these instabilities without creating instabilities in the other sectors. Further work is
needed to see whether generic torsionfull backgrounds are prone to containing gradient instabilities, or if 
the instabilities we found are mainly due to the (generalized) self-accelerating nature of the special 
de Sitter backgrounds we considered.
\end{abstract}

\maketitle

\section{Introduction}
 
Since its discovery more than a century ago \cite{Einstein:1915ca,Einstein:1916vd}, General Relativity (GR) has been found 
to be in excellent agreement with all gravitational observations and experiments (for
reviews of tests of GR, see, e.g., \cite{Patrignani:2016xqp,Will:2014kxa}). However, the search for modified theories
of gravity (incorporating GR in some limit) has been an active research topic nearly
since the formulation of GR. There are several motivations for looking for modified, or extended, gravity theories,
notably: (i) the desire to unify gravity with other interactions (or with matter fields); (ii) the usefulness
of having foils to devise, or to interpret, experimental tests; (iii) the search for a natural explanation 
of several remarkable cosmological facts, such as the need to postulate both dark matter and dark energy,
well in excess of the visible matter content of the universe. 

In particular, many modified gravity theories
have been suggested to try to explain the observed late acceleration of the universe as being due to
a dynamical self-acceleration mechanism linked to some infra-red physics, instead of resulting from
the addition of an extremely tiny cosmological constant in Einstein's equations. Among such
self-accelerating universes, we can mention the ones coming from: 
higher-order (i.e. non-quadratic) scalar kinetic terms \cite{ArmendarizPicon:1999rj,ArmendarizPicon:2000dh}, 
gravity leaking to extra dimensions \cite{Deffayet:2000uy,Deffayet:2001pu}, 
bigravity \cite{Damour:2002wu,Volkov:2013roa},
massive gravity \cite{deRham:2010tw,Volkov:2013roa}, galileons \cite{Nicolis:2008in}, and generalized scalar-tensor
theories \cite{Lombriser:2015sxa}. For a recent reviews of modified-gravity models of dark energy see
Refs. \cite{Joyce:2016vqv,Nojiri:2017ncd}.

An endemic problem of modified-gravity self-accelerating cosmological models is the presence of
{\it instabilities} of various sorts: tachyons, ghosts, excitation of new degrees of freedom, gradient instabilities.
Instabilities seem to be a necessary consequence of many self-acceleration mechanisms. For examples of
instabilities in self-accelerated universes see, e.g., \cite{ArmendarizPicon:1999rj,Garriga:1999vw,
Hsu:2004vr,Gorbunov:2005zk,Comelli:2012db,DeFelice:2012mx,Konnig:2014xva,Lagos:2014lca,Cusin:2014psa,Konnig:2015lfa}. For reviews of these instabilities,
see, e.g., Refs. \cite{Rubakov:2014jja,deRham:2014zqa}. For more references and examples
of cures of these instabilities, see Refs. \cite{Gumrukcuoglu:2014xba,Akrami:2015qga,Heisenberg:2016spl,Gumrukcuoglu:2017ioy}.

Separately from the endemic stability problems of self-accelerating cosmological universes, the recent extremely
tight limit ($\sim 10^{-15} $ level) on the fractional difference between the speed of gravitational waves and
the speed of light derived from combining the observations of GW170817 and GRB 170817A
\cite{TheLIGOScientific:2017qsa,GBM:2017lvd,Monitor:2017mdv} has put stringent constraints on
many of the modified-gravity models featuring self-accelerating solutions 
\cite{Baker:2017hug,Creminelli:2017sry,Sakstein:2017xjx,Ezquiaga:2017ekz,Akrami:2018yjz}.
The latter constraints severely reduce the viable range of modified-gravity theories that have been proposed
as alternatives to GR.

In the present work, we shall consider a class of modified-gravity theories whose phenomenology
 has received relatively little attention
(compared to the models mentioned above), though it has many appealing theoretical features. The class of
theories that we shall study here is a subclass of the geometric theories that generalize GR by
including a dynamical torsion in addition to the dynamical metric (or vierbein) of GR. We shall refer to it as
{\it torsion gravity} (TG) in the following. These theories originated in the Einstein-Cartan theory
\cite{Cartan:1923zea,Cartan:1924yea}, defined by taking as Lagrangian density the curvature scalar 
considered  as a functional of both the metric and a metric-preserving,
but nonsymmetric, affine connection. In Einstein-Cartan theory, the torsion does not propagate so that,
in absence of sources for the torsion, the theory reduces to GR (for a detailed discussion, with historical references, of the Einstein-Cartan-Sciama-Kibble theory, viewed as a gauge theory of the Poincar\'e group, see Ref. \cite{Hehl:1976kj}). 
A simple class of TG theories
where the torsion propagates is obtained by adding to the curvature scalar terms quadratic either in the torsion
or in the curvature tensor \cite{Sezgin:1979zf,Sezgin:1981xs,Hayashi:1979wj,Hayashi:1980av,Hayashi:1980ir,Hayashi:1980qp}. Remarkably, when appropriately  restricting the arbitrary coefficients entering the action of these theories,
one obtains classes of ghost-free, and tachyon-free (around Minkowski spacetime) generalizations of GR including, besides the usual massless Einsteinian spin-2
field, several other possible massive fields \cite{Sezgin:1979zf,Sezgin:1981xs,Hayashi:1980ir,Hayashi:1980qp} . Here, following Refs. \cite{Nair:2008yh,Nikiforova:2009qr},
we shall consider a 5-parameter\footnote{or 6 parameters when adding a bare
cosmological constant,} class of TG theories which contains, as propagating massive fields (embodied
in the torsion) both  massive spin-2 and massive spin-0 excitations. This class of TG theories has so far proven to be
remarkably healthy and robust for an extension of GR containing a (notoriously delicate) massive spin-2 field.
Indeed, Ref.   \cite{Nair:2008yh} has shown that this model stayed consistent around de Sitter and anti-de Sitter
backgrounds, while Ref.  \cite{Nikiforova:2009qr} has shown not only that the number of propagating degrees of freedom
remains the same as in flat flat spacetime when considering the excitations around an arbitrary {\it torsionless}
Einstein background (in other words, the notoriously dangerous Boulware-Deser phenomenon \cite{Boulware:1973my} 
does not take place in such general backgrounds), but also that, at least for weakly curved backgrounds, there were
no ghosts in the spectrum. See also Ref. \cite{Deffayet:2011uk} which contrasts TG gravity with bigravity.

It has been recently found in Ref. \cite{Nikiforova:2016ngy} that this class of TG models admits, at zero bare cosmological
constant, a self-accelerating expanding de Sitter solution driven by a connection background involving {\it a non-zero
torsion}. In view of the endemic instability problems of self-accelerating universes, and of the fact that the
previous  stability results of  Refs. \cite{Nair:2008yh,Nikiforova:2009qr} were limited to torsionless
backgrounds of TG theory, the study of the stability of the self-accelerating torsionful universe  of Ref. \cite{Nikiforova:2016ngy}
is an open issue which deserves a detailed study. This study was started in Refs. \cite{Nikiforova:2017saf,Nikiforova:2017xww} by considering  the stability of scalar perturbations of the self-accelerating torsionful universe.
[Indeed, the scalar sector is usually considered as being the most
prone to exhibiting instabilities.]  On the one hand, the stability analysis of  
Refs. \cite{Nikiforova:2017saf,Nikiforova:2017xww} confirmed 
the good behavior of torsion gravity for what concerns the preservation of the number of degrees of freedom. 
Indeed, the number of scalar degrees of freedom around the torsion background of the self-accelerating universe
of Ref. \cite{Nikiforova:2016ngy} was found to be two, which is the same as in a Minkowski background.
On the other hand, while Ref. \cite{Nikiforova:2017saf} found that there existed exponentially growing modes
when the background torsion was large compared to the Hubble expansion rate, Ref. \cite{Nikiforova:2017xww}
concluded that the scalar perturbation modes were stable when the background torsion was comparable to
the Hubble expansion rate. 

The aim of the present work is to complete the stability analysis of Refs. \cite{Nikiforova:2017saf,Nikiforova:2017xww}
by studying the vector and tensor sectors of the perturbations of the  self-accelerating universe, and to extend it
to a study of the perturbations of a one-parameter family of torsionfull de Sitter-like solutions obtained 
by deforming the self-accelerating solution by a non-zero bare cosmological constant. We shall also, for
completeness, reexamine the scalar sector, thereby bringing some qualifications to the previous findings  \cite{Nikiforova:2017xww}. 

The organization of this paper is as follows. We review the formalism of torsion gravity (TG) in Sec. \ref{sec2},
and present its general field equations in Sec. \ref{sec3}. In \ref{sec4} we present two separate one-parameter
families of de Sitter-like solutions in TG parametrized by a bare vacuum energy $c_2$: one family (called ``first branch")
is torsionless (and was already introduced in Ref.   \cite{Nair:2008yh}), while the other family (called ``second branch")
is the $c_2$-deformation of the self-accelerating solution of Ref. \cite{Nikiforova:2016ngy}. The $SO(3)$-covariant
analysis of the perturbations of these  de Sitter-like solutions is presented in Sec. \ref{sec5}. We show in Sec. \ref{sec6}
how the large parameter $\tal/\lambda^2$ can be scaled out from the cosmological perturbation equations, and how
the perturbations can be expressed as functions of the dimensionless variable $z= k \eta $ measuring the ratio
between the physical wavenumber and the Hubble expansion rate. Our general method for studying the behavior
of cosmological perturbations in the large-$z$ limit (high-frequency, sub-horizon modes having wavelengths small compared to the Hubble horizon) is explained in Sec. \ref{sec6bis}, where we also summarize, in advance, our main results.
Then we successively present our analysis of the high-frequency dispersion laws   
for tensor, vector and scalar perturbations in Secs. \ref{sec7}, \ref{sec8}, and \ref{sec9}, respectively. 
Some concluding remarks are presented in Sec. \ref{sec10}. Finally, to relieve the tedium, the explicit forms of 
the (initial) perturbation equations (for tensor, vector and scalar perturbations) are presented in 
three corresponding Appendices (A, B and C).

\section{Formalism and action of Torsion Gravity (TG)} \label{sec2}

The 6-parameter class of Torsion Gravity (TG) theories considered here (following Refs. \cite{Nair:2008yh,Nikiforova:2009qr,Nikiforova:2016ngy}) is defined by an action whose basic fields (besides the ones describing matter, which
we shall not consider here) are a vierbein ${e_i}^\mu$ (with inverse ${e^i}_\mu$;  ${e_i}^\mu{e^j}_\mu =\delta^j_i$ ) and a Lorentz connection 
$A_{i j \mu}= - A_{j i \mu}$. Here, we follow the notation of Refs. \cite{Hayashi:1979wj,Hayashi:1980av,Hayashi:1980ir,Hayashi:1980qp}: the signature is mostly plus; indices $i, j, k, \ldots= 0,1,2,3 $ denote Lorentz-frame indices (moved
by the Minkowski metric $\eta_{ij}, \eta^{ij}$), while Greek indices $\mu, \nu, \ldots=0,1,2,3$ denote spacetime indices
linked to a coordinate system $x^\mu$, and moved by the coordinate-system metric
$g_{\mu \nu} \equiv \eta_{ij} {e^i}_\mu {e^j}_\nu$. The fact that the connection ${A^i}_{ j \mu}\equiv \eta^{i i'} A_{i' j \mu}$ 
is algebraically described by the 24 independent components of a local Lorentz ($\eta$-antisymmetric) connection 
automatically embodies the fact that $A$ preserves the metric $g_{\mu \nu}$. Indices are moved by their relevant metric, and we generally try, for clarity, to keep the Lorentz indices before the spacetime ones (``frame first"),
e.g. $e_{i \mu} = \eta_{ij} {e^j}_\mu = g_{\mu \nu}  {e_i}^\nu$. When there is a risk of confusion, we shall add a tilde
on the Lorentz indices: ${e^{\tilde{i}}}_\mu$. 

We work with the general action (where $ |e| \equiv \det {e^i}_\mu = \sqrt{ - \det g_{\mu \nu}}$)
\be
S[{e^i}_\mu, A_{i j \mu}] =\int d^4 x \, |e| \,  L[e, \d e, \d^2 e, A, \d A]\,,
\ee
with
\bea  \label{lagrangian}
L &=& \frac{3}{2}  \widetilde{\alpha}\, F[e, A, \d A]
+ \frac{3}{2} \overline\alpha \, R[e, \d e, \d^2 e]   + c_2\\ \nonumber
&+& c_3 F^{ij}F_{ij}
 + c_4 F^{ij}F_{ji} + c_5 F^2 + c_6 (\epsilon^{ijkl}  F_{ijkl})^2 \; ,  
 \eea
where $ \widetilde{\alpha}, \overline\alpha, c_2, c_3, c_4, c_5, c_6$ are coupling constants\footnote{
Here, $ \overline\alpha$ denotes the opposite ($ \overline\alpha \equiv - \alpha$) of the
parameter denoted $\alpha$ in Refs. \cite{Hayashi:1979wj,Hayashi:1980av,Hayashi:1980ir,Hayashi:1980qp,Nair:2008yh,Nikiforova:2009qr,Nikiforova:2016ngy}.
 We introduce this change of sign to have $ \overline\alpha >0$.},
 and where $\epsilon^{ijkl}$ denotes the Levi-Civita symbol (say with $\epsilon^{0123} = + 1$).
Apart from the term $R[e, \d e, \d^2 e]$ (which is the usual Einsteinian curvature scalar computed from the 
vierbein ${e^i}_\mu$), the various contributions to the Lagrangian $L$, Eq. \eqref{lagrangian}, are all constructed
from the frame components (with respect to ${e^i}_\mu$) of the curvature tensor $ F_{ijkl}[e, A, \d A]$ 
defined by the connection $A_{i j \mu}$ (see below for its definition). More precisely, 
$F_{ij}[e, A, \d A]\equiv \eta^{kl}F_{kilj}$ denotes the  Ricci tensor of $A_{i j \mu}$, while
$ F[e, A, \d A] \equiv \eta^{ij}F_{ij}$ denotes the curvature scalar of $A_{i j \mu}$. Note that the Lagrangian
is linear in the Einsteinian scalar curvature, while it contains terms linear and quadratic in the various
contractions of the curvature tensor $ F_{ijkl}[e, A, \d A]$ of $A_{i j \mu}$. As we shall explicitly see below,
the field equations obtained by independently varying ${e^i}_\mu$ and $A_{i j \mu}$ will contain at most
two derivatives of $e$ and $A$.  Let us note in passing that the term 
$\frac{3}{2} \overline\alpha \, R[e, \d e, \d^2 e]$ can be replaced
in $L$ (modulo a total divergence) by the sum of $ \frac{3}{2} \overline\alpha \,F[e, A, \d A]$ and of 
a Lagrangian contribution $L_T$ made of several terms quadratic in the 
torsion tensor ${T^i}_{jk}[e, \d e, A]$ (defined below), see, e.g., Eq. (7) in  Ref. \cite{Sezgin:1979zf},
and Sec. II of Ref. \cite{Nikiforova:2016ngy}. We shall, however, use here the form \eqref{lagrangian}
of the action.

Both the Riemannian curvature ${R^i}_{ j \mu \nu}$ associated with the vierbein ${e^i}_\mu$ and the 
curvature  ${F^i}_{ j \mu \nu}$ associated with
the connection ${A^i}_{j \mu}$ are most simply defined by Cartan's structure equations. On the one hand, we
have the Riemannian spin-connection one-form ${\omega^i}_j \equiv {\omega^i}_{j\mu}dx^\mu$
(which is just the Levi-Civita connection expressed in an orthonormal frame),
which defines the Riemannian curvature two-form 
${{\cR}^i}_{ j}= \frac12 {R^i}_{ j \mu \nu}dx^\mu \wedge dx^\nu$ via Cartan's  second structure formula
\be
{\cR^i}_{ j}=d {\omega^i}_{ j} + {\omega^i}_{ s}\wedge {\omega^s}_{ j}  .
\ee 
On the other hand, the non-Riemannian connection 
one-form ${{\cA}^i}_{ j}={A^i}_{ j \mu} dx^\mu$ defines its corresponding curvature two-form 
${{\cF}^i}_{ j}= \frac12 {F^i}_{ j \mu \nu}dx^\mu \wedge dx^\nu$ via exactly the same Cartan formula:
\be
{\cF^i}_{ j}=d {\cA^i}_{ j} + {\cA^i}_{ s}\wedge {\cA^s}_{ j}  .
\ee 
The corresponding frame components of these two curvature tensors,
namely ${R^i}_{jkl} \equiv  {R^i}_{ j \mu \nu} {e_k}^\mu {e_l}^\nu$
and ${F^i}_{jkl} \equiv  {F^i}_{ j \mu \nu} {e_k}^\mu {e_l}^\nu$,
can then be explicitly written 
(in their ``all indices down" forms: $R_{ijkl} \equiv \eta_{i i'}{R^{i'}}_{jkl}$
and $F_{ijkl} \equiv \eta_{i i'}{F^{i'}}_{jkl}$) as
\bea
R_{ijkl} &=& {e_k}^\mu {e_l}^\nu \l \partial_\mu \omega_{ij\nu} - \partial_\nu \omega_{ij\mu} \right.\\ \nonumber
&+& \left. \eta^{mn} \omega_{im\mu}\omega_{n j\nu} - \eta^{mn} \omega_{im\nu}\omega_{n j\mu} \r \; ,
\eea
\bea
F_{ijkl} &=& {e_k}^\mu {e_l}^\nu \l \partial_\mu A_{ij\nu} - \partial_\nu A_{ij\mu} \right.\\ \nonumber
&+& \left. \eta^{mn} A_{im\mu}A_{n j\nu} - \eta^{mn} A_{im\nu}A_{n j\mu} \r \; .
\eea
Then the other objects (Ricci tensors and scalars of, respectively, $\omega$ and   $\cA$: $R_{ij}$, $R$,
and  $F_{ij}$, $F$) entering the TG action above are defined as 
\be 
R_{ij}=\eta^{kl}R_{kilj} =\eta^{kl}R_{ikjl} \;,\;\; R=\eta^{ij}R_{ij} \; , 
\ee
\be 
F_{ij}=\eta^{kl}F_{kilj} =\eta^{kl}F_{ikjl} \;,\;\; F=\eta^{ij}F_{ij} \; .
\ee
Note that both $R_{ijkl}$ and  $F_{ijkl}$ are antisymmetric under $ i\leftrightarrow j$ and $k \leftrightarrow l$.
However, contrary to $R_{ijkl}$, $F_{ijkl}$ is not symmetric under the exchange $ij \leftrightarrow kl$, so that $F_{ij}$ a priori differs from $F_{ji}$.
For brevity, we will sometimes contract Lorentz indices without explicitly indicating
the use of the Minkowski metric: i.e. a term like $\eta^{mn} A_{im\mu}A_{n j\nu}$ will be simply
written as  $A_{im\mu}A_{m j\nu}$.

For completeness, let us also mention the form of Cartans's first structure equations. In the
Riemannian case, the absence of torsion yields $ d e^i + {\omega^i}_j e^j=0 $,
where $e^i\equiv {e^i}_\mu dx^\mu$. This equation allows one to express the spin-connection
in terms of the structure constants of the frame field, defined by $ d e^i \equiv \frac12 {C^i}_{[jk]} e^j \wedge e^k$
(with ${C^i}_{[jk]}= - {C^i}_{[kj]}$), or, equivalently
\be
C_ {i [jk]} \equiv (\d_\mu e_{i\nu} - \d_\nu e_{i\mu})  {e_j}^\mu {e_k}^\nu\; . 
\ee
Namely,
\be \label{omega=C}
 \omega_{ij\mu} = \omega_{ijk}e^k_\mu =\frac{1}{2}( C_{i [jk]}+ C_{j [ki]} - C_{k[ij]} ) e^k_\mu \; .  
\ee
We added some  brackets around the last two indices of $C_ {i [jk]}$  as a reminder of  its antisymmetry with
respect to these indices. We will use the same reminder for the torsion tensor $T_{i [jk]}$. By contrast, we
do not put an antisymmetry symbol on the first two indices of the several other three-index objects 
($A_{ijk}, \omega_{ijk}, K_{ijk},\ldots$) 
which are antisymmetric over their first two indices.

By contrast to the case of $\omega$, the application of the first Cartan structure equation to the torsionfull connection $A$
yields
\be
 d e^i + {\cA^i}_j e^j =  -\frac12 {T^i}_{[jk]} e^j \wedge e^k  \,,
\ee
where ${T^i}_{[jk]}= - {T^i}_{[kj]}$ are the frame components of the torsion tensor. 
This equation yields
\be
T_{i [jk]} = A_{ijk}- A_{ikj} - C_{i [jk]}.
\ee
One can solve these equations to express $A_{ijk}$ in terms of $T_{i [jk]} + C_{i [jk]}$,
with a result often written as
\be
A_{ijk}= \omega_{ijk} + K_{ijk}\,,
\ee
where $\omega_{ijk}$ was expressed in terms of $C_{i [jk]}$ in Eq. \eqref{omega=C},
and where $ K_{ijk} = -  K_{jik}$ are the frame components of the {\it contorsion} tensor.
This tensor is defined as
\be
K_{ijk}= \frac12 (  T_{i [jk]}+ T_{j [ki]} - T_{k[ij]}) \,,
\ee
whose inverse is
\be
T_{i [jk]} =  K_{ijk}- K_{ikj}\,.
\ee
Note also the expression of $K_{ij\mu} \equiv K_{ijk} {e^k}_\mu$ in terms of $\omega$ and $A$:
\be
K_{ij\mu}= A_{ij\mu} -\omega_{ij\mu}\,.
\ee
As discussed in Refs. \cite{Sezgin:1979zf,Sezgin:1981xs,Hayashi:1979wj,Hayashi:1980av,Hayashi:1980ir,Hayashi:1980qp,Nair:2008yh,Nikiforova:2009qr,Nikiforova:2016ngy}, the action above defines
a healthy theory (without ghosts or tachyons) about a Minkowski background involving, besides
a massless graviton, a massive spin-2 field and a massive pseudoscalar field if the parameters
entering the action satisfy the following inequalities
\be \label{ineqs}
\tal > 0, \;\;\;\; \bal > 0, \;\;\;\; c_5 < 0, \;\;\;\; c_6 > 0 \,,
\ee
and the equality
\be
c_3+c_4=-3 \, c_5 \; . 
\ee
 Then the squared mass of the massive spin-2 field is
\be \label{m2}
m_2^2=\frac{\tal (\tal  + \bal)}{2  \bal \, (- c_5)}  >0 \,,
\ee
 while that of the pseudoscalar field is
\be \label{m0}
m^2_0=\frac{\widetilde{\alpha}}{16c_6} >0\;.  
\ee
In addition, the strength of the matter coupling of the massless spin-2 field is 
\cite{Hayashi:1980ir,Nikiforova:2009qr}
\be
G_0=\frac1{24 \pi (\tal+\bal)}\,,
\ee
while that of the massive spin-2 field is
\be
G_2= \frac43 \frac{\tal}{\bal} G_0=\frac{\tal}{18 \pi  \bal (\tal+\bal)} \, .
\ee
In this work, we shall assume that the ratio $\bal/\tal$ is of order unity, so that
\be \label{MPl}
\tal \sim \bal \sim (16 \pi G_{\rm Newton})^{-1} \sim M_{\rm Planck}^2.
\ee
On the other hand, we shall assume that the dimensionless parameters $c_5$ and $c_6$ are
very large and of order
\be \label{largec5c6}
- c_5 \sim c_6 \sim \l \frac{M_{\rm Planck}}{H_0} \r^2 \gg1 ,
\ee
so that the masses $m_2$ and $m_0$ are of order the Hubble scale $H_0$, and can
thereby modify gravity at cosmological scales.

\section{Field equations of TG} \label{sec3}

We have indicated in Eq. \eqref{lagrangian} the dependence of the first two terms on the vierbein, the connection,
and their derivatives. Like the term $\frac{3}{2}  \widetilde{\alpha}\, F[e, A, \d A]$, all the terms
involving $F_{ij}$ or $F_{ijkl}$ depend on $e, A$ and $\d A$. Remembering that the second derivatives
$\d^2 e$ of the vierbein enter the Einsteinian term $ \frac{3}{2} \overline\alpha \, R[e, \d e, \d^2 e] $
only linearly, we easily see that the action involves (because of its, at most, quadratic nature in $F$)
only two derivatives of the fundamental fields $e$ and $A$, leading to field equations containing at
most two derivatives of $e$ and $A$.

It is convenient to write the explicit forms of the Euler-Lagrange equations following from the action Eq. \eqref{lagrangian}, as 
\be
\frac{\delta S}{\delta {e^i}_{\mu}} =- 2 |e| e^{j\mu } {\cG}_{ji} \,\, ; \,\, \frac{\delta S}{\delta A_{ij\mu}} = |e| {\cT}^{ijk} {e_k}^\mu \,.
\ee 
The explicit expression of the vierbein field equation ${\cG}_{ij}$ reads \cite{Nikiforova:2017saf}
\begin{align} \label{Geq}
{{\cal G}}_{ij} \equiv &  \frac{3}{2} \widetilde{\alpha} \left( F_{ij} - \frac{1}{2} \eta_{ij}F  \right) +
\frac{3}{2} \overline{\alpha} \left( R_{ij} - \frac{1}{2} \eta_{ij}R  \right) \nonumber  \\ 
& - \frac12 c_2 \, \eta_{ij}   + c_3 \left( F_{ki} F_{kj} + F_{kl} F_{kilj} \right)
\nonumber\\
& + c_4 \left( F_{ik}
F_{kj} + F_{lk} F_{kilj} \right) + 2 c_5 F F_{ij} \nonumber \\
& + 2 c_6 \epsilon_{klmi} F_{klmj} (\epsilon\cdot F) -
\frac{1}{2}\eta_{ij} L^{(2)}  = 0 \;,
\end{align}
where
\be 
L^{(2)}  = c_3 F_{ij}F_{ij} + c_4 F_{ij}F_{ji} + c_5 F^2 + c_6 (\epsilon \cdot F)^2 \,,
 \ee
(with $\epsilon \cdot F\equiv \epsilon^{ijkl}  F_{ijkl} $)  is the part of the Lagrangian 
that is quadratic in $F_{ijkl}$. Note that ${\cG}_{ij}$ is
not symmetric in its two indices $i j$. The gravitational equations \eqref{Geq} involve
two derivatives of the vierbein $e$ and only one derivative of the connection $A$.

To write the explicit form of the $A_{ij \mu}$ field equation ${\cT}_{ijk}$, one needs
to define the following building blocks:
\be
 H_{ijk} \equiv \frac{3 \tal}{2} \l K_{ikj} -
K_{jki} - K_{ill} \eta_{jk} + K_{jll} \eta_{ik}  \r \;,
\ee
\be
 P_{ij} \equiv c_3F_{ij} + c_4 F_{ji}\; ; \;  P\equiv \eta^{ij}P_{ij} \,,
 \ee
 \bea
 S_{ijk} &=& \frac{2}{3\tal} H_{mnk} \l \eta_{im} P_{jn}
- \eta_{jm} P_{in} -\frac{2}{3} \eta_{im}\eta_{jn} P  \right.  \nonumber \\
 &+&  \left.2c_6 \epsilon_{ijmn} (\epsilon \cdot F) \r \;.
\eea
In terms of these quantities, the connection field equation reads
\begin{align} \label{Aeq}
{{\cal T}}_{ijk} \equiv  &  \left[\eta_{ik} \left( D_m P_{jm} - \frac{2}{3}
D_j P\right) - D_i P_{jk}\right]   \nonumber\\
&  - \left[\eta_{jk} \left( D_m P_{im} -
\frac{2}{3} D_i P\right) - D_j P_{ik} \right]
\nonumber\\
& +4 c_6 \epsilon_{ijkm} D_m (\epsilon\cdot F) +  H_{ijk}  + S_{ijk} = 0 \;,
\end{align}
where the derivative $D_i$ involves the connection $A_{ij\mu}$:
\be  
D_iB_j  \equiv  e^\mu_iD_\mu B_j = e^\mu_i( \partial_\mu B_{j} + A_{ jk\mu}B_{k} ).
 \ee
The connection equations \eqref{Aeq} involve
two derivatives of the connection $A$ and only one derivative of the vierbein $e$ .

The above field equations satisfy Bianchi-type identities linked to the invariance of the action under
both diffeomorphisms and local Lorentz rotations of the vierbein. See Eqs. (17) and (20) in 
Ref. \cite{Nikiforova:2017saf}.

\section{de Sitter-like solutions of the field equations} \label{sec4}

\subsection{A torsionfull self-accelerating solution 
in absence of bare cosmological constant \cite{Nikiforova:2016ngy}}

Ref. \cite{Nikiforova:2016ngy} found that, in absence of bare cosmological constant (i.e. when
setting $c_2=0$ in the TG action), the above field equations admitted a self-accelerating
solution, i.e. a solution whose metric corresponds to an expanding de Sitter solution:
\be
{\bar g}_{\mu \nu} dx^\mu dx^\nu= - dt^2 + a^2(t) \delta_{ab} dx^a dx^b ,
\ee
where $a,b = 1,2,3$ and
\be
a(t)=e^{\lambda t}\,.
\ee
The inverse vierbein ${\bar e^{\tilde i}}_\mu$ describing this background\footnote{We use 
an overbar to denote a background solution. } solution is naturally chosen as
\be \label{baree}
{\bar e^{\tilde 0}}_0=1 \, ; \, {\bar e^{\ta}}_b = a(t) \delta^a_b.
\ee
This solution is sustained by a connection background having both ``electriclike" and
``magneticlike" frame components, namely 
\be \label{bareA}
{\bar A}_{\t0 \ta \tb}=f \, \delta_{ab}  \;, \quad
{\bar A}_{\ta\tb\tc}=g \, \varepsilon_{abc}\;,    
\ee
with time-independent connection strengths $f$ and $g$. We note that the contorsion tensor of
this connection has, as only nonzero components,
\be  \label{torsionfg}
K_{\t0 \ta \tb}= (f+\lambda) \, \delta_{a b} \,  ; \, K_{\ta\tb\tc} = g \, \varepsilon_{abc}\,.
\ee
Therefore the connection background ${\bar A}$ will contain some torsion as soon as 
either $ f \neq - \lambda$ or $ g\neq 0$. We will see in the formulas relating $f$ and $g$,
for this self-accelerating solution, to the basic parameters of TG that the self-accelerating solution
is necessarily torsionfull.

An important facet of the present work is to understand whether a torsionfull
background has different stability properties than a torsion-free background.
In order to clarify this issue, we will find convenient to contrast the properties
of the perturbations around the torsionfull, self-accelerating, de Sitter solution, Eqs. \eqref{baree}, \eqref{bareA},
above, with those around the {\it torsionless} expanding de Sitter solution (for the same value of the
Hubble expansion rate $\lambda$) considered in Ref. \cite{Nair:2008yh} .  The latter solution needs to have a non-zero value of the
bare cosmological constant $c_2$ to sustain its expansion. Here, we shall more generally show how to
construct two different one-parameter families of TG backgrounds, parametrized by a continuously varying
value of $c_2$ and  having different values of the 
connection strengths $f$ and $g$. One of these families (called ``first branch" below) 
is the torsionless de Sitter-like solutions of Ref. \cite{Nair:2008yh}, while the second family
(called ``second branch" below) is made of  torsionfull  de Sitter-like backgrounds that are $c_2$-deformed versions
of the self-accelerating solution recalled above. The crucial point is that  we shall show below that these 
two one-parameter families intersect at some point, so that the $c_2$-family of second-branch solutions
defines a way to interpolate between a torsionless de Sitter background and the torsionfull self-accelerating solution. 

\subsection{Deformation of the self-accelerating solution by a bare cosmological constant, $c_2$,
and the two intersecting branches of de Sitter-like solutions}

We generalize the self-accelerating solution found in Ref. \cite{Nikiforova:2016ngy} by allowing for
a non-zero value of $c_2$. The Lagrangian $|e| L$ given in Eq. (9) of \cite{Nikiforova:2016ngy}
must be augmented by the term
\be
\l |e| L \r^{c_2}= c_2 N a^3,
\ee
where $N$ denotes the lapse (which is set to 1 after variation).
This supplementary term only modifies the gravity equation $\delta S/\delta N$
(i.e. Eq. (10a) in \cite{Nikiforova:2016ngy}), without modifying Eqs. (10b) and (10c) there,
which correspond to $\delta S/\delta f$ and $\delta S/\delta g$. Denoting for brevity
\be \label{d2d5d6}
d_2 \equiv \frac{c_2}{9} \,; \, d_6 \equiv 16 c_6 \,; \, d_5 \equiv 4 c_5+ 32 c_6\,,
\ee
one then finds the following three independent equations that this $c_2$-modified
background has to satisfy:
\begin{subequations}
\bea \label{cosmoeqa0}
\!\!\!\!0&=& \!\!d_2\!+ \!\tal (f^2-g^2) + \!\bal \lam^2 - d_6g^2 \lam^2\!+\! 4d_6 f^2 g^2 \\
\label{cosmoeqb0} 0 &=& \tal (f+\lam) -d_5 g^2 \lam + 4d_6 g^2 f ,  \\
\label{cosmoeqc0} 0&=& g \left[ -\tal + d_5 \lam f - 2 d_6 \lam^2 + 4 d_6 f^2  \right].
\eea
\end{subequations}
A crucial point is that the last equation, Eq. \eqref{cosmoeqc0}, actually splits into two possible types of solutions:
either $g=0$, or $ -\tal + d_5 \lam f - 2 d_6 \lam^2 + 4 d_6 f^2=0$, which is an equation 
relating $f$ and $\lam$. This split defines {\it two separate branches} of solutions. 

Along the {\it first branch} (i.e. when $g=0$), we get, by inserting $g=0$ in the second equation
the result $f=-\lam$. Then, inserting these results for $g$ and $f $ in the first equation one
gets a relation determining $\lam$ in terms of $c_2$, namely
\be\label{lambranch1}
\lam^2_{\rm first \, branch}= - \frac{  d_2}{\tal+\bal}.
\ee
Note that $d_2=c_2/9$ has to be negative; which is expected as the vacuum energy density is actually $- c_2$.
In view of Eq. \eqref{torsionfg}, this first branch of de Sitter solutions (with $g=0$ and $f=-\lam$) is torsion-free. It coincides with the solutions studied in \cite{Nair:2008yh}.

To discuss the {\it second branch} of solutions, it is convenient to use some notation. Let us first define the following 
combinations of the theoretical parameters entering the TG action (using the definitions \eqref{d2d5d6})
\be \label{xid56d26}
{\xi} \equiv \frac{\bal}{\tal} \; ; \;  d_{56} \equiv \frac{d_5}{d_6}\,;\, d_{26}\equiv\frac{d_2d_6}{\tal^2} \,;\, H_6^2\equiv \frac{\tal}{d_6 }
 \; .
\ee
The first three of these quantities($\xi,d_{56},d_{26}$) are dimensionless, and will be 
all considered as being of order unity in the present work. [Note that the dimensionless quantity 
denoted $\xi$, which we shall use in this work, is the inverse of the quantity 
$\Xi \equiv \frac{\tal}{\bal} \equiv \frac1{\xi}$ used in Ref. \cite{Nikiforova:2017xww}.]
On the other hand, the quantity $H_6$
(with $H_6>0$) defined last, has the same dimension as the Hubble expansion rate (as well as 
that of $f$ and $g$) and provides a convenient fiducial Hubble expansion rate (hence its notation).

We then define other dimensionless quantities of order unity that involve the quantities $f$, $g$ and $\lam$ entering our cosmological solutions. Namely,
\be
\delta \equiv - \frac{f}{\lam} \; ; \; h \equiv  \frac{g}{\lam} \,;
\ee
and
\be
 \hlam\equiv \frac{\lam}{H_6}\,;\, \bar x \equiv \frac{g^2}{H_6^2} \equiv h^2 \hlam^2 \,.
\ee
In terms of these quantities,  Eqs. \eqref{cosmoeqa0}, \eqref{cosmoeqb0}, \eqref{cosmoeqc0} read
\begin{subequations}
\bea \label{cosmoeqa}
0&=& \!\!d_{26} - \bar x + \! (\delta^2+ \xi) \hlam^2+( 4\delta^2-1)  \bar x \hlam^2,\\
\label{cosmoeqb} 0 &=& 1-\delta -  (d_{56} + 4\delta)  \bar x,  \\
\label{cosmoeqc} 0&=& h \left[ -1 + (4 \delta^2 -2 - d_{56} \delta) \hlam^2  \right].
\eea
\end{subequations}
 Using Eq. \eqref{cosmoeqb}, we can express $\delta$ as a function of $\bar x$:
 \be \label{delx}
 \delta(\bx) =\frac{1-d_{56} \,\bx}{1+ 4 \,\bx}\,.
 \ee
 Replacing this result in Eq. \eqref{cosmoeqc} yields (for the second branch) $\hlam$ as a function of $\bx$:
 \bea
 &&\hlam^2(\bx)=\frac1{4 \delta^2 -2 - d_{56} \delta} \nonumber \\
 &=&\frac{(1+4\, \bx)^2}{2- d_{56} +(d_{56}^2-12 d_{56}-16)\bx +(8 d_{56}^2-32) \bx^2}\,. \nonumber\\
 \eea
 Substituting this relation in Eq. \eqref{cosmoeqa} then yields an equation
 relating $\bx$ to the basic TG parameters $d_{26}, d_{56}, \xi$:
 \be \label{cubiceq}
 \frac{P_3(\bx; d_{56}, \xi)}{P_2(\bx; d_{56})}= - d_{26} \,,
 \ee
 where $P_3$ and $P_2$ are two polynomials in $\bx$ that are respectively cubic and quadratic. They
 read
 \bea
 P_3(\bx; d_{56}, \xi)=1 + \xi + (1 + 8 \xi- d_{56} ) \bx  \nonumber \\
 + (8 + 16 \xi + 4 d_{56}) \bx^2   
 +( 16  - 4  d_{56}^2) \bx^3\,,
 \eea
 \bea
 P_2(\bx; d_{56}, \xi)&=&2 - d_{56}  + (d_{56}^2-12 d_{56} -16) \bx \nonumber \\
 &+& (8 d_{56}^2-32) \bx^2\,.
 \eea
 Ref. \cite{Nikiforova:2016ngy} has considered the case $d_{26}=0$ (zero bare cosmological constant 
 $c_2\equiv 9 d_2=0$),
 and proved that the corresponding cubic equation $P_3(\bx; d_{56}, \xi)=0$ had a unique,
 positive, real solution, say $\bx_0>0$ under the condition $d_5 + 2 d_6 <0$, i.e. (remembering that $d_6=16 c_6>0$)
 \be \label{ineqd56}
 -d_{56} \equiv -2 - \frac14 \frac{c_5}{c_6} >2.
 \ee
 This self-accelerating solution necessarily has $g=\pm H_6 \sqrt{\bx_0} \neq0$ as well as a 
 a corresponding value of $\delta=- f/\lam$ given by
 \be
  \delta(\bx_0) =\frac{1-d_{56} \,\bx_0}{1+ 4 \,\bx_0}\,.
 \ee
 It is easily seen that, as $-d_{56}$ varies between $2$ and $+\infty$ (with $\bx_0$ taking any
 positive value),  $\delta(\bx_0)$ will take all values in the interval $(\frac12, +\infty)$.
 As $g\neq0$ and (generically) $\delta \neq 1$ (except when $-d_{56} = 4$, in which case
 $\delta(\bx_0)=1$, independently of $\bx_0$), this self-accelerating solution is torsionfull
 (see Eq. \eqref{torsionfg}).

 We studied the one-parameter deformation of the latter self-accelerating solution defined by solving
 the cubic in $\bx$, Eq. \eqref{cubiceq}, with some non-zero (negative or positive) value of $d_{26}$
 (i.e. some nonzero value of the vacuum energy parameter $c_2$).
 When $ d_{26}$ is positive (negative bare cosmological constant)
 and increases away from zero, the unique positive solution $\bx_0$ 
 continuously evolves into a unique, larger solution  $\bx$ of  Eq. \eqref{cubiceq}. [The corresponding
 larger value of $g^2$compensates for the additional negative cosmological constant.]
 On the other hand, when $- d_{26}$ is taken to be positive and increasing away from zero
 (positive cosmological constant), $\bx_0$ evolves into a smaller solution $\bx$.
 We found that this one-parameter family of $d_{26}$-deformed
 avatars of the self-accelerating solution of Ref. \cite{Nikiforova:2016ngy} is continuously
connected\footnote{The second branch is algebraically defined by considering, for all values of  $d_{26}$, the
set of solutions of $ -\tal + d_5 \lam f - 2 d_6 \lam^2 + 4 d_6 f^2=0$, together with Eqs. \eqref{cosmoeqa}, \eqref{cosmoeqb}. In some cases, $d_{26}$ varies continuously but not monotonically along this branch.} to a vanishing value of $\bx$, i.e. a vanishing value of $g$.
This happens when 
\be
- d_{26}=\frac{1+\xi}{2-d_{56}}, 
\ee
(where we recall the definitions Eqs. \eqref{xid56d26} with Eqs. \eqref{d2d5d6})
at which point the corresponding value of $\hlam^2$ along this second branch coincides with
the value of $\hlam^2$ along the first branch, as given by Eq. \eqref{lambranch1}. In addition,
the limiting value of $ \delta = -f/\lam$ along the second branch also coincides with its
value along the first branch, i.e. $\delta(\bx \to 0)=1$. 

In other words, the $c_{2}$-deformed one-parameter family of torsionfull second-branch solutions
interpolates between the torsionfull self-accelerating solution and the torsionfree de Sitter solution
of the first branch discussed above. However, the merging of these two branches of solutions is
not smooth. It should be viewed as the {\it transversal} crossing of two curves that have a common point,
with different tangents at the common point. We will use below the existence of the second branch
of solutions as a conceptual tool to contrast the effect on the stability of cosmological
perturbations of turning on a torsionfull background (second branch), versus having an always
a torsionless one (first branch, along which $g=0$ and $f=-\lam$).

\subsection{Expressing TG parameters in terms of the dimensionless parameters $\delta, h$
characterizing the de Sitter-like solutions}

Above we discussed what are the equations that determine, in principle, how the
physical quantities, $f,g,\lam$, entering the self-accelerating solution (and its  $c_{2}$-deformation)
depend on the basic parameters entering the TG action (such as $\tal, \bal, c_2, c_5, c_6, \ldots$,
modulo the intermediate definitions \eqref{d2d5d6}).
However, in our stability analysis below, we will not directly need such relations. It will be more
useful to work with the inverse relations, i.e., how to relate  the parameters entering the TG action,
such as $c_2, c_5, c_6, \xi \equiv \bal/\tal$ to the dimensionless parameters $\delta, h$ characterizing our de Sitter-like
solutions. 
From Eqs. \eqref{cosmoeqa0},\eqref{cosmoeqb0}, \eqref{cosmoeqc0}, we respectively get
\begin{align}
& \frac{c_2}{9 \tal \lambda^2}+ \xi = \frac12 (h^2- \delta^2- \delta ) \;, \label{xi+d2}  \\
&c_6= \frac{\tal}{\lam^2} \frac{h^2 +\delta - \delta^2 }{32 \, h^2 \,   (4\delta^2-1) }\;,  \label{c6}
\\
& c_5=- \frac{\tal}{\lam^2}\frac{h^2 + (\delta -1)^2 }{4 \, h^2 \, (2\delta-1)}  \;.  \label{c5}  
\end{align}
Note that Eq. \eqref{c5}, together with the necessary inequality $c_5<0$, \eqref{ineqs}, implies that 
\cite{Nikiforova:2017xww}
\be
\delta > \frac12.
\ee
In addition, Eq. \eqref{c6}, together with the necessary inequality $c_6>0$, Eq. \eqref{ineqs}, implies that we must have
\be \label{lowerboundh}
h^2 > \delta^2- \delta \,.
\ee
We recall that the dimensionless ratio $\tal/\lam^2$ is an extremely large number (while we assume that
$\delta$ and $h$ are of order unity). We will see below that the large number $\tal/\lam^2$ can be
scaled out of the perturbation equations, so  that we shall be able to express the stability conditions
only in terms of $\delta$ and $h$ and a couple of other dimensionless parameters of order unity.

In the first relation \eqref{xi+d2} we must have $\xi>0$ (see \eqref{ineqs}), but $\frac{c_2}{9\tal \lam^2}$ can
have any sign. [This is what allows to have $h=0$ along this second branch.] On the other hand,
if we consider the self-accelerating solution (i.e. when $c_{2}=0$), we get the link
\cite{Nikiforova:2017saf,Nikiforova:2017xww}
\be\label{hdelxiSA}
h^2= \delta^2+ \delta + 2 \xi, \; \; (c_{2}=0)\,,
\ee
and the following lower bound on the square of $h=g/\lam$:
\be \label{SAlowerbound}
h^2 > \delta +\delta^2 >\frac34, \; \; (c_{2}=0)\,,
\ee
where we used, in the last inequality,  the fact that $\delta > \frac12$.

 \section{Parametrization of cosmological perturbations in TG: $SO(3)$, Fourier, and helicity decomposition}
 \label{sec5}
 
 The de Sitter-like background solutions considered above were expressed in a coordinate system
 where the spatial geometry is Euclidean. It will be convenient to first  rewrite them in a conformally flat form, i.e.
 \be
{\bar g}_{\mu \nu} dx^\mu dx^\nu= e^{2\phi(\eta)}[- d\eta^2 + \delta_{ab} dx^a dx^b] ,
\ee
where $ \eta = \int e^{-\lambda t}dt = -\frac{1}{\lambda}e^{-\lambda t}$ is the conformal time, and
\be
 e^{\phi(\eta)} \equiv a(\eta) = -\frac1{\lam \eta} \,.
\ee
In terms of the basic variables of TG, say taken in the form ${e^i}_\mu$ and $A_{i j \mu}$, 
the background values of the TG fields read
\be \label{bareec}
{\bar e^{\tilde 0}}_0=  e^{\phi(\eta)}\, ; \, {\bar e^{\ta}}_b =  e^{\phi(\eta)} \delta^a_b \,,
\ee
\be \label{bareAc}
{\bar A}_{\t0 \ta b}= e^{\phi(\eta)} f \, \delta_{ab}  \;, \quad
{\bar A}_{\ta\tb c}= e^{\phi(\eta)} g \, \varepsilon_{abc}\;.   
\ee
These expressions differ from the ones written in Eqs. \eqref{baree}, \eqref{bareA} above because,
on the one hand, the coordinate $x^0$ now refers to the conformal time $\eta$, and because
we are now working with the connection components $A_{i j \mu}$, with a spacetime index as last index.

Then we can decompose, as usual \cite{Mukhanov:1990me}, the most general perturbations of these backgrounds into 
irreducible representations of the three-dimensional rotation group $SO(3)$. As the most general 
representations of $SO(3)$ that can appear in a decomposition of TG involve spins 0, 1 or 2 
\cite{Sezgin:1979zf,Hayashi:1980qp}, we  get the most general cosmological perturbation
by allowing for all possible scalar, vector and tensor perturbations of the background fields
\eqref{bareec}, \eqref{bareAc}. We can parametrize these perturbations as follows:
\be \label{e0+e1}
{e^i}_\mu= {{\bar e}^i}_{\, \mu} +  e^{\phi(\eta)} {\epsilon^i}_\mu=e^{\phi(\eta)} \l {\delta^i}_\mu + {\epsilon^i}_\mu \r \,,
\ee
and
\be \label{A0+A1}
A_{i j \mu} = {\bar A}_{i j \mu} + a_{i j \mu}\,.
\ee
The (conformally rescaled) perturbation of the inverse vierbein, ${\epsilon^i}_\mu$ (or equivalently $\epsilon_{i \mu }\equiv \eta_{ij} {\epsilon^j}_\mu$),
and the perturbation of the connection, $ a_{i j \mu}\equiv A_{i j \mu} - {\bar A}_{i j \mu}$, will
then be decomposed into scalars, vectors and tensors. 
The presence of an overall factor $e^{\phi(\eta)}$ in front of the perturbed vierbein \eqref{e0+e1}
allows one to compute the Ricci-tensor contribution to the gravitational field equation \eqref{Geq}
by using the conformal transformation properties of the Ricci tensor. In addition, the connection
curvature $F_{ijkl}$ (which depends on the vierbein only through a factor ${e_k}^{\mu} {e_l}^{\nu}$)
has a very simple conformal variance. The only quantity having a subtle conformal variance in the
connection field equation \eqref{Aeq} is the contorsion $K_{ijk}$. Ref. \cite{Nikiforova:2017saf}
used these conformal variances to rewrite the field equations, Eqs. \eqref{Geq}, \eqref{Aeq}, in terms
of the rescaled perturbed vierbein $e^{-\phi(\eta)}{e^i}_\mu= {\delta^i}_\mu + {\epsilon^i}_\mu $.
See Eqs. (24), (25) there. Note that the latter equations are numerically equal to the original field equations ${{\cal G}}_{ij}$, Eq. \eqref{Geq}, and ${{\cal T}}_{ijk}$, Eq. \eqref{Aeq}, but expressed in terms of rescaled metric variables.
Note also that the contribution proportional to the bare cosmological constant $c_2$ will
not explicitly contribute to the perturbed field equations because it does not involve any
explicit field variable, being only multiplied by $-\frac12 \eta_{ij}$.

It is convenient to use the two gauge freedoms of TG to restrict the forms of these perturbations.
As indicated in \cite{Nikiforova:2017saf}, one can use the local Lorentz freedom to render
$\epsilon_{\tilde{i} \mu } \equiv \epsilon_{i \mu }\equiv \eta_{ij} {\epsilon^j}_\mu$ symmetric, i.e.
\be \label{symgauge}
\epsilon_{\tilde{i} \mu } = \epsilon_{\tilde{\mu} i} \,,
\ee
where the indices must be considered simply as numbers between $0$ and $3$, and where we added
a tilde on the first index to recall that it is a frame index. This completely fixes the freedom of
local Lorentz rotations. In addition, we can use the diffeomorphism freedom
to set the conformally rescaled metric perturbation $h_{\mu \nu}$,
defined by writing $g_{\mu \nu}= e^{2\phi(\eta)}( \eta_{\mu\nu} + h_{\mu \nu})$, i.e.
\be 
h_{\mu \nu} \equiv
e^{-2\phi(\eta)} e^i_\mu e_{i \nu} - \eta_{\mu \nu} = 2\epsilon_{\mu \nu} +O(\epsilon^2)\,,
\ee
into a zero-shift gauge (for the vector perturbations), and a ``longitudinal" gauge
\cite{Mukhanov:1990me} for the scalar ones, i.e. such that
\be \label{zeroshift}
h_{0a}=  2\epsilon_{0a}=0\,,
\ee
and such that the perturbations are of the form
\be
\epsilon_{\mu\nu}= \epsilon_{\mu\nu}^{\rm scalar}+\epsilon_{\mu\nu}^{\rm vector}+\epsilon_{\mu\nu}^{\rm tensor},
\ee
where the nonzero components of the scalar, vector and tensor parts are parametrized as
\begin{align} \label{scalareperts}
&\epsilon_{00}^{\rm scalar}=-\Phi \;, \\
&\epsilon_{ab}^{\rm scalar}=\Psi\delta_{ab}\;,
\end{align}
\be \label{vectoreperts}
\epsilon_{ab}^{\rm vector}=\d_{a}W_{b}+\d_{b} W_{a}\,,
\ee
\be
\epsilon_{ab}^{\rm tensor}=\pi_{ab} \,.
\ee
Here the latin indices from the beginning of the alphabet are spatial Euclidean indices ($a,b,c=1,2,3$),
$W_a$ is a transverse vector ($\d_a W_a=0$) and $\pi_{ab} $ a transverse-traceless 
(symmetric) tensor  ($\d_a\pi_{ab}=0$, $\delta_{ab} \pi_{ab}=0$). In other words, after our gauge fixing,
the gravitational perturbation $\epsilon_{\tilde{i} \mu }$ contains 6 independent components:
two scalars ($\Phi, \Psi$), the two independent components of a transverse vector ($W_a$), and the
two independent components of a transverse-traceless tensor ($\pi_{ab}$).
Using the symmetry of the background under spatial translations, these irreducible pieces are then 
decomposed into spatial Fourier integrals of the type 
\bea
\Phi(\mathbf{x})&=&\int d^3 k \, \Phi(\mathbf{k})\, e^{i \mathbf{k} \cdot\mathbf{x}},\nonumber \\
\Psi(\mathbf{x})&=&\int d^3 k \, \Psi(\mathbf{k})\, e^{i \mathbf{k} \cdot\mathbf{x}},\nonumber \\
W_a(\mathbf{x})&=&\int d^3 k \, W_a(\mathbf{k})\, e^{i \mathbf{k} \cdot\mathbf{x}},\nonumber \\
\pi_{ab}(\mathbf{x})&=&\int d^3 k \, \pi_{ab}(\mathbf{k})\, e^{i \mathbf{k} \cdot\mathbf{x}},
\eea
so that, for instance, the vector piece becomes, in Fourier space, 
\be
\epsilon_{ab}^{\rm vector}(\mathbf{k})=i k_a W_{b}(\mathbf{k})+ i k_{b} W_{a}(\mathbf{k})\,.
\ee

On the other hand, the 24 independent components of the connection perturbation $ a_{i j \mu}$ are
 correspondingly decomposed into: eight scalars ($\widetilde \xi, \chi,\sigma,\rho,\theta,Q,u,M$), six (two-component) transverse vectors ($\zeta_{a}, \nu_a, \mu_a,\kappa_a, A_a,L_a$), and two 
 (two-component) transverse-traceless tensors ($\tau_{ab}$ and $N_{ab}$)
 \be
 a_{i j \mu}=  a_{i j \mu}^{\rm scalar}+ a_{i j \mu}^{\rm vector}+ a_{i j \mu}^{\rm tensor},
 \ee
 with nonzero components (written directly in Fourier space):
 \begin{align}
a_{0a0}^{\rm scalar} & = - a_{a00}^{\rm scalar} = k_{a} \widetilde\xi \;,\nonumber 
\\
a_{0ab}^{\rm scalar} &= - a_{a0b}^{\rm scalar} = k_{a} k_{b} \chi + \delta_{ab} \sigma +
\epsilon_{abc} k_{c} \rho \;,\nonumber 
\\
a_{ab0}^{\rm scalar} &= \epsilon_{abc} k_{c} \theta \;,\nonumber 
\\
a_{abc}^{\rm scalar} &= \epsilon_{abd} k_{c} k_{d} Q + (k_{a} \epsilon_{bcd} - k_{b}
\epsilon_{acd}) k_{d} u \nonumber \\
&+ (k_{a} \delta_{bc} - k_{b} \delta_{ac} ) M  ;
\end{align}
\begin{align}
a_{0a0}^{\rm vector} & = - a_{a00}^{\rm vector} = \zeta_{a} \;,\nonumber
\\
a_{0ab}^{\rm vector} &= - a_{a0b}^{\rm vector} = k_{a} \nu_{b} + k_{b} \mu_{a} \;,\nonumber
\\
a_{ab0}^{\rm vector} &= k_{a} \kappa_{b} - k_{b} \kappa_{a} \;,\nonumber
\\
a_{abc}^{\rm vector} &= k_{a} k_{c} A_{b} - k_{b} k_{c} A_{a} + \eta_{ac} L_{b} -
\eta_{bc} L_{a} \;;
\end{align}
\begin{align}
a_{0ab}^{\rm tensor} &= - a_{a0b}^{\rm tensor} = \tau_{ab} \;,
\\
a_{abc}^{\rm tensor} &= k_{a} N_{bc} - k_{b} N_{ac} \;.
\end{align}
In addition, the various vector and tensor Fourier pieces will be decomposed into their two independent
(complex) helicity components ($ h=\pm1$ for a transverse vector and $ h=\pm2$ for a transverse-traceless tensor)
according to the general scheme
\bea \label{hdecomp}
W_a(\mathbf{k})&=& W_{(+1)} e_+^a+ W_{(-1)} e_{-}^a, \nonumber \\
\pi_{ab}(\mathbf{k})&=& \pi_{(+2)} e_+^a  e_+^b+ \pi_{(-2)} e_-^a e_-^b, 
\eea
where  
\be
e_+^a(\mathbf{k})= e_1^a(\mathbf{k}) + i \, e_2^a(\mathbf{k})\,;\, e_-^a(\mathbf{k})= e_1^a(\mathbf{k}) - i \, e_2^a(\mathbf{k})\,,
\ee
 are  two complex combinations
of two real unit vectors orthogonal to $\mathbf{k}$, so that
$( e_1^a(\mathbf{k}),e_2^a(\mathbf{k}),k^a/|\mathbf{k}|)$  form a positively-oriented 
Euclidean orthonormal  triad. Note in this respect the relations
\bea \label{kxe}
\varepsilon_{abc} k^b e_+^c &=& - i |\mathbf{k}| \, e_+^a , \nonumber \\
\varepsilon_{abc} k^b e_-^c &=& + i |\mathbf{k}| \, e_-^a \,,
\eea
that are instrumental when coding the projection of the original perturbation equations (involving
vectors or tensors) into equations for their pure-helicity components.

\section{Scaling out the large parameter $\tal/\lam^2$ from the cosmological perturbation equations}
\label{sec6}

We already mentioned that the price to obtain, within TG, a cosmologically relevant  infrared modification of GR
is to allow for a large hierarchy between the parameters entering  the general TG Lagrangian \eqref{lagrangian},
see Eqs. \eqref{MPl}, \eqref{largec5c6}. We assume that the other independent dimensionless parameter entering
the terms quadratic in the $F$ curvature in the action, namely $c_3$, is comparable to $c_5$ and $c_6$.
As for the bare vacuum-energy parameter $c_2$ (when allowed for to deform the self-accelerating
solution) it must be taken, in view of Eq. \eqref{xi+d2}, as being much smaller than the Planck scale, namely
\be
|c_2| \sim \l  M_{\rm Planck} H_0 \r^2  \sim   \l \frac{H_0}{M_{\rm Planck}} \r^2  M_{\rm Planck}^4\,.
\ee

One might a priori think that the presence of such very large and very small parameters in the action
will complicate the study of the cosmological perturbations of the de Sitter-like solutions discussed above.
Let us, however, show that a suitable rescaling of the cosmological field equations allows one to write
equations where all variables and all coefficients are of order unity. More precisely, the equations we shall
use will only involve the following dimensionless parameters of order unity:
\be \label{hatc}
\xi\equiv \frac{\bal}{\tal} \, ;\, \widehat c_2 \equiv \frac{c_2}{\tal \lam^2}\, ; \, \widehat c_n \equiv \frac{\lam^2 c_n}{\tal }\, ({\rm for} \, n=3,4,5,6)\,.
\ee
Indeed, let us consider the background variables $\eta, \delta, h$ as  being of order unity, say $O_0(1)$,
and let us also consider the perturbed variables ${\epsilon^i}_\mu$ and $a_{i j \mu}$ in Eqs.  \eqref{e0+e1},
\eqref{A0+A1} as being of order unity, say $O_1(1)$ (modulo some formally small expansion parameter, say, $\gamma$).
Then, taking into account the fact that the cosmological scale factor 
$e^{\phi(\eta)} = -1/(\lam \eta)$ involves the inverse of the Hubble expansion rate $\lam$,
the perturbed vierbein has a structure of the type
\be
{e^i}_\mu \sim \lam^{-1} \l \frac1{\eta} +  \frac{\epsilon}{\eta} \r \,,
\ee
with an inverse of the type
\be
{e_i}^\mu \sim \lam^{+1} \l \eta  + \eta \epsilon \r \,.
\ee
The perturbed connection is found to have a structure of the type
\be
A_{i j \mu} \sim \frac{\delta +h}{\eta} + a\,,
\ee
where $a$ denote the various components of $a_{ij\mu}$.

For scaling out $\lam$, what is important in the structures above is to distinguish the factors of $\lam$
from the other factors involving variables considered as being of order unity. We can denote any order-unity expression
involving the background variables $\eta, \delta, h$ as  $O_0(1)$, and any expression involving the
perturbations $\epsilon, a$ (together with coefficients involving $\eta, \delta, h$) as 
$\gamma O_1(1)$, where $\gamma$ is just a formally small book-keeping parameter.
In other words, we have the structures
\be
{e^i}_\mu \sim \lam^{-1} (O_0(1)+ \gamma O_1(1))\,;\, 
{e_i}^\mu \sim \lam^{+1} (O_0(1)+ \gamma O_1(1))\,,
\ee
\be
A_{i j \mu} \sim O_0(1)+ \gamma O_1(1)\,.
\ee
Using this notation, it is then successively found (keeping track of the presence or absence
of vierbien factors raising or lowering frame indices) that
\be
\omega_{ij\mu} \sim O_0(1)+ \gamma O_1(1)\,,
\ee
\be
\omega_{ijk}  \sim K_{ijk} \sim \lam (O_0(1)+ \gamma O_1(1))\,,
\ee
\be
R_{ijkl}  \sim R_{ij} \sim \lam^2 (O_0(1)+ \gamma O_1(1))\,,
\ee
\be
F_{i j kl} \sim F_{ij} \sim \lam^2 (O_0(1)+ \gamma O_1(1))\,.
\ee
Inserting these scalings in the field equations Eqs \eqref{Geq} , \eqref{Aeq}, it is then found that the
rescaled parameters defined in Eqs. \eqref{hatc} are such that the rescaled field equations
$\widehat{{\cal G}}_{ij} \equiv {{\cal G}}_{ij}/( \lam^2 \tal)$, 
$\widehat{{\cal T}}_{ijk}\equiv {{\cal T}}_{ijk}/(\lam \tal)$
 have the structures
\be \label{rescaledGeq}
\widehat{{\cal G}}_{ij} \equiv \frac{{{\cal G}}_{ij}}{ \lam^2 \tal} \sim \sum_{n=1}^6 \widehat c_n \l O^n_0(1)+ \gamma O^n_1(1)\r\,,
\ee
where we formally defined $\widehat c_1 \equiv \xi = \bal/\tal$, and 
\be \label{rescaledAeq}
\widehat{{\cal T}}_{ijk}\equiv \frac{{{\cal T}}_{ijk}}{\lam \tal} \sim \sum_{n=3}^6 \widehat c_n \l O^n_0(1)+ \gamma O^n_1(1)\r\,.
\ee
Note that the coefficients $\widehat c_1=\xi= \bal/\tal$ and $\widehat c_2 =c_2/(\tal \lam^2)$
are present only in the gravitational equations, 
but not in the connection ones. 

From the practical point of view, the structures Eqs. \eqref{rescaledGeq}, \eqref{rescaledAeq},
mean that we can obtain conveniently rescaled perturbation equations 
\be
\gamma \sum_n  \widehat c_n  \, O^n_1(\eta, \delta, h; \epsilon, \d \epsilon, \d^2 \epsilon, a, \d a, , \d^2 a)\,,
\ee
 simply by using the formal replacements $\lam \to 1$,
$\tal \to 1$, $\bal \to \xi$, $c_n \to \widehat c_n $ in the computation of the perturbed
cosmological equations.

In addition to this scaling out of $\lambda$ and $\tal$ there is another useful scaling property
of the perturbation equations. Indeed, as is usual in cosmological perturbation theory, the
magnitude of the (conserved) spatial wavenumber $k= |\mathbf{k}|$ can (possibly at
the price of the rescaling of some variables by some $k$ factors to give them the same dimension) be everywhere combined
with the conformal time $\eta$ so that the perturbation equations involve only the 
variable
\be \label{zdef}
z \equiv  k \eta \equiv - \frac{k_{\rm phys}}{\lam}\,.
\ee
Here, $k_{\rm phys}= k/a$ is the physical wavenumber, so that $|z|= |k \eta|$ is seen as
being equal to the ratio of the physical wavenumber to the Hubble expansion rate.
[$z$ is negative (like $\eta$), and increases towards the future.]
In other words, $1/|z|$ is the ratio of the wavelength of the considered perturbation to
the Hubble horizon radius. We will focus below on the study of the sub-horizon wavemodes,
i.e in the region where $|z| \gg 1$. These are indeed the crucial modes to consider in
a stability analysis, as the superhorizon modes ($|z| \lesssim 1$) evolve on a Hubble time scale.
We are interested here in instabilities that evolve on a scale parametrically shorter than the Hubble
time scale.

 \section{High-frequency, subhorizon dispersion laws and stability analysis of cosmological perturbations} \label{sec6bis}
 
 Before entering the details of our analysis of the stability of de Sitter-like solutions in TG,
 let us: (i) recall a few basic facts about high-frequency, subhorizon dispersion laws and their
 consequences for stability or instability of cosmological perturbations; (ii) sketch
 what will be our method for deriving dispersion laws in torsionfull backgrounds; and (iii) 
 summarize in advance our main results (whose detailed derivation will be given in the next three sections).

\subsection{High-frequency, subhorizon dispersion laws}

As explained at the end of the previous section, we are interested here in exponential instabilities in the
solutions of linearized perturbations that could evolve on a time scale parametrically shorter than the Hubble time scale.
This corresponds to focussing on the behavior of sub-horizon modes in the region where $|z| \gg 1$,
where the time-like variable $z$, which was defined in Eq. \eqref{zdef}, measures the ratio between
the physical wavenumber and the Hubble expansion rate.
We shall prove below (by a mathematical analysis of the perturbation equations) that, in the regime
$|z| \gg 1$, the general solution of the perturbation equations (for a given comoving wavenumber $\mathbf{k}$,
and for a given helicity)
behaves as a superposition of eigenmodes (featuring various values  of $\sigma$ and $\beta$)
of the form
\be \label{HFmode1}
 z^{\beta}  e^{\sigma z}e^{i \mathbf{k} \cdot\mathbf{x}} \l 1+ O\l\frac1z\r \r \,.
\ee
Here, the power-law prefactor would be important to keep if we were interested
in describing what happens when a sub-horizon mode becomes
super-horizon, and to match it to the corresponding small-$z$ solutions for super-horizon modes. 
However, such effects correspond to an evolution on the slow, Hubble scale.
Our aim here is to discuss the instabilitities that could happen on time scales much smaller than the
Hubble time. Therefore, we shall (mostly) neglect, in the following, such power-law corrections
to the mode evolution. In such an approximation, the high-frequency mode \eqref{HFmode1}
takes a simple plane-wave form, with respect to the conformal time $\eta$ and
the comoving spatial coordinates $\mathbf{x}$, say
\be \label{HFmode2}
e^{+i \omega \eta+ i \mathbf{k} \cdot\mathbf{x}}\,,
\ee
where the so-defined conformal-time frequency, $\omega$, is (remembering the definition $z\equiv k \eta$)
related to the eigenmode quantity $\sigma$ via
\be \label{sigom}
\omega= - i \sigma k  \,\; ; \;\, \sigma= i \omega/ k  \,.
\ee
As we will see,  the eigenmode quantities $\sigma$  come 
in opposite pairs  $ \sigma_{\pm \alpha}=\pm \sigma_\alpha$, where the index $\alpha$ takes $2\, j$ different values,
say $\alpha= \pm 1,\cdots, \pm j$.
We will have $j=2$ for helicities $+2$ and 0, and $j=1$ (apart from a gauge mode) for helicity $+1$.
[In addition, we will have $j=2$ for helicity $-2$, and $j=1$ (apart from a gauge mode) for helicity $-1$.]

In view of the link \eqref{sigom} each such pair implies a dispersion law of the form
\be
\omega^2 = - \sigma_\alpha^2 \mathbf{k}^2 \,,
\ee 
and the mass-shell condition determining the high-frequency propagation for the various helicity modes will then be a polynomial of the form
\be
\Pi_{\alpha=-j}^{j}\l \omega + i \sigma_\alpha k  \r = \Pi_{\alpha=1}^{j} \l \omega^2 + \sigma_\alpha^2 \mathbf{k}^2 \r \,.
\ee
[For non-zero helicities, there are two such polynomials: one for positive helicity, and another (identical) one for negative helicity.]
In flat spacetime, each free bosonic degree of freedom (d.o.f.) would have a dispersion law of the type $\omega^2 = +\mathbf{k}^2 + m^2$,
where $m$ is the mass of the field.
In the high-frequency, sub-horizon limit the mass-term is 
negligible\footnote{We are considering here a range of parameters
for which the mass terms are comparable to the Hubble rate}, and yields a simplified
dispersion law of the type $\omega^2 = +\mathbf{k}^2$ per d.o.f.. We note that such a high-frequency (large-$z$)
dispersion law corresponds to an eigenvalue $\sigma= \pm  i$ in Eq. \eqref{HFmode1} (we have set the
velocity of light to one).

A value of $\sigma$ which is pure imaginary, but different from $\pm i$, say
$\sigma = \pm i \, c_s$ would correspond to a velocity of propagation $c_s$ different from the velocity of light.
If we do not worry about the possible causality consequences of having superluminal propagation,
all the cases where the eigenvalues $\sigma$ are purely imaginary correspond to an absence
of exponential instabilities. As we shall end up finding strong exponential instabilities, we will
not worry here whether the modes that have no exponential instabilities are ghostlike or not. 
See, e.g., Ref. \cite{Rubakov:2014jja} for a review of the various instabilities in cosmology.

What we shall worry about are pairs of values of $\sigma$ that are either real or complex
(with a nonzero real part). Indeed, a real, or complex, eigenvalue $\sigma= \sigma_r + i \sigma_i$, with 
$\sigma_r \neq0$, implies a mode containing the real exponential factor
\be
 e^{\sigma_r k \eta} \,.
\ee
As the $\sigma$ eigenvalues always come in opposite pairs, this would always imply the presence of
an exponentially growing mode.

Such exponential instabilities, with a growth rate proportional to the spatial wavenumber are
called ``gradient instabilities", or ``Laplacian instabilities". For the high-frequency, subhorizon modes
we focus on, these are about the worst type
of instabilities as they imply that the smallest wavelengths grow with the fastest exponential rates.
(see, e.g., Ref. \cite{Rubakov:2014jja}). In particular, we shall find that vector perturbations
contain a pair of {\it real} values of $\sigma$, say $\sigma = \pm c_s$. This
corresponds to an imaginary propagation velocity, i.e. to the dispersion law that would exist in
an Euclidean spacetime ($ \omega^2 + c_s^2 \mathbf{k}^2 =0 \, !$) rather than in a Lorentzian one.

\subsection{Method for deriving dispersion laws}

We will discuss separately, and successively, tensor, vector and scalar perturbations around the
torsionfull second branch of de Sitter-like solutions. For each helicity $h$, with $h=\pm 2,\pm 1$ or 0,
we will start from an initial system, directly deduced from linearizing the field
equations of TG around the considered background, of $N_h$ ordinary differential equations in $z$ 
for $N_h$ unknowns. The values of $N_h$ will be
$N_{+2}=N_{-2}=3$, $N_{+1}=N_{-1}=7$, and $N_{0}=10$. This initial system of equations involves up to second
derivatives for some variables. We shall show, for each helicity $h$, how to transform this initial system 
(by eliminating some variables) into an equivalent system
of $n_h$ {\it first-order} differential equations in $n_h$ unknowns, which we will write in matrix form, i.e.
(using Einstein's summation convention)
 \be \label{matrixsystemh}
 \frac{d y^{(h)}_i(z)}{d z}= m^{(h)}_{ij}(z) y^{(h)}_j(z) \,, \, i, j=1,2,\cdots,n_h\,.
 \ee
 Here, the values of $n_h$ will be $n_{+2}=n_{-2}=4$, $n_{+1}=n_{-1}=3$, and $n_{0}=4$,
 and the variables $y^{(h)}_i(z)$ are combinations (with some, possibly $z-$dependent, coefficients) 
 of a subset of the inital variables. [The other, eliminated variables being expressed as combinations
 of the $y^{(h)}_i$'s and their first derivatives.]
 
 For each helicity $h$, we shall show that the matrix of differential coefficients $m^{(h)}_{ij}(z)$
 has a finite limit  $m^{(h), \infty}_{ij}= \lim_{z \to \infty} m^{(h)}_{ij}(z)$ as $z\to \infty$, and that the first two terms  
 of the large-$z$ expansion of $m^{(h)}_{ij}(z)$, say
\be \label{zexpmij}
 m^{(h)}_{ij}(z) =m^{(h), \infty}_{ij} + \frac{n^{(h)}_{ij}}{z} + O\l \frac1{z^2}\r\,,
 \ee
 determine the characteristics of the eigenmodes \eqref{HFmode1} describing the large-$z$ asymptotics
 of the general solution of the matrix system \eqref{matrixsystemh}. [In mathematical terms, we shall see that
 that $z=\infty$ is an irregular singular point of the differential system \eqref{matrixsystemh}.]
 In particular, the values of the exponents
 $\sigma$ entering the eigenmodes are the eigenvalues of the limiting matrix:
 \be
 m^{(h), \infty}_{ij} v^{(h)}_j = \sigma v^{(h)}_i  \,.
 \ee
 [The $O(1/z)$ matrix $n^{(h)}_{ij}$ entering the  $O(1/z)$ term in \eqref{zexpmij} then determines
 the power-law exponents  $\beta$ in the modes \eqref{HFmode1}.]
 In other words, the high-frequency dispersion law for the helicity $h$ perturbations is obtained from the
 characteristic polynomial of  $m^{(h), \infty}_{ij}$
  \be \label{cph}
P^{(h)}_{n_h}(\sigma) \equiv \det \l  m^{(h),\infty}_{ij} - \sigma \delta_{ij} \r\,,
 \ee
 simply as the following homogeneous polynomial, of degree $n_h$, in $\omega$ and $k$:
 \be \label{disph}
{\rm Disp}^{(h)}(\omega,k)= k^{n_h} P^{(h)}_{n_h}( i \frac{\omega}{k}) =0\,.
 \ee
When putting together helicities $+h$ and $-h$, the dispersion law is the product 
${\rm Disp}^{(h)}(\omega,k) {\rm Disp}^{(-h)}(\omega,k)$, which is even in $\omega$ and $k$.
[Actually, we will find that $P^{(-h)}_{n_h}(\sigma)=P^{(h)}_{n_h}(\sigma)$.]

\subsection{Summary of our results and comparison with
dispersion laws in the torsionless de Sitter-like solutions (first branch)}

Let us end this section by summarizing the end results for the  dispersion laws of the various
helicity sectors along the torsionfull de Sitter-like second-branch solutions, and by comparing them to
the dispersion laws of the torsionless first branch. We give the results for positive helicities.
The negative helicities have the same number of degrees of freedom,
and the same dispersion laws. 

The high-frequency dispersion laws along the torsionless first-branch are actually (as follows from the results of
Refs. \cite{Nair:2008yh,Nikiforova:2009qr}, and as we shall rederive below) the same as
around a flat spacetime background, and directly follow from the known helicity content of
TG excitations. Namely a massless spin-2 (having two d.o.f. of helicities $h=\pm2$), a massive spin-2 (containing five d.o.f, with helicities
$h= \pm2, \pm1,0$) and a massive pseudo-scalar (having one d.o.f with $h=0$).

Helicity $h =  2$ perturbations along the torsionfull (second-branch) solution (with $n_2=4$ describing two d.o.f.) have
the dispersion law
\be\label{disph2B2}
\omega^4 - 2 b'^{\rm tensor} \omega^2 {\mathbf k}^2 +  {\mathbf k}^4=0\,,
\ee
with
\be\label{bpdelxi0}
b'_{\rm tensor}(\delta,h,\xi)-1=\frac{(\delta - \delta^2 + h^2) ((\delta-1)^2 + h^2)^2}{2 h^2 \l  (1 + 2 \xi)\delta^2 -(1+ 3  \xi) \delta +\xi  - h^2\r}.
\ee
By contrast, the helicity $h = 2$ perturbations along the torsionless solution have the dispersion law
\be
(\omega^2- {\mathbf k}^2)^2=\omega^4 - 2 \omega^2 {\mathbf k}^2 +  {\mathbf k}^4=0\,.
\ee
Note that this dispersion law has the same structure as \eqref{disph2B2} but with a value of the coefficient 
$b'^{\rm tensor}$ equal to 1. We shall see that the absence of exponential instability requires that
$b'^{\rm tensor} \geq 1$.

Helicity $h =1$ perturbations along the torsionfull solution (with $n_1=3$) have the dispersion law
 \be \label{disph1B2}
 \omega \l  \omega^2 -   c^{\rm vector} {\mathbf k}^2 \r=0\,,
 \ee
 with
 \be \label{cvec}
 c^{\rm vector}= -\frac{( \del^2 + h^2 -1)^2}{4 h^2} \,.
 \ee
By contrast, the helicity $h = \pm1$ perturbations along the torsionless solution have the dispersion law
 \be
 \omega \l  \omega^2 - {\mathbf k}^2 \r=0 \,.
 \ee
 In both cases the factor $\omega$ describes a gauge mode.

Helicity $h =0$ perturbations along the torsionfull solution (with $n_0=4$) have the dispersion law
\be \label{disph0}
 \omega^2 \l  \omega^2 -   c^{\rm scalar} {\mathbf k}^2 \r=0 \,,
\ee
where the rather complicated expression of $c^{\rm scalar}$ in terms of $\del,h$ and $c_{35}\equiv c_3/c_5$
will be found in Eqs. \eqref{cscalar}, \eqref{Nscalar},  \eqref{Dscalar}, below.
By contrast, the helicity $h = 0$ perturbations along the torsionless solution have the dispersion law
 \be
  \l  \omega^2 - {\mathbf k}^2 \r  \l  \omega^2 - {\mathbf k}^2 \r=0\,,
 \ee
 where the two factors  $ \omega^2 - {\mathbf k}^2 $ describe the propagation of the two helicity-0
 d.o.f. (one being part of the massive spin-2 field, the other being a pseudo-scalar torsion-related field).
 We note in passing that the dispersion law \eqref{disph0} describes the same number of d.o.f.,
 though with strongly modified propagation properties (in particular the factor $ \omega^2$ 
 describes modes having zero propagation velocities).

The stability properties of the perturbations along the torsionfull solution (linked to the signs
of $b'_{\rm tensor}-1$, $c^{\rm vector}$ and   $c^{\rm scalar}$) will be discussed in detail below.
Let us only note here that the negative sign of $c^{\rm vector}$ signals the necessary presence of
gradient instabilities in the vector sector, and that the number of d.o.f. is  the same, for all helicities,
along torsionfull and torsionless backgrounds (or flat backgrounds).

 \section{Study of tensor perturbations and of their stability} \label{sec7}
 
 \subsection{Reduction to a linear system of four first-order ordinary differential equations} \label{sec7A}
 
 We start our analysis of the stability of the perturbations of the two branches  of de Sitter TG solutions
 discussed above by considering the tensor sector. We  explain in Appendix A below how we extracted 
 from the perturbed rescaled field equations, i.e. the
  $O(\gamma)$ contributions in $\widehat{{\cal G}}_{ij}$, Eq. \eqref{rescaledGeq}, and 
  $\widehat{{\cal T}}_{[ij]k}$, Eq.\eqref{rescaledAeq}, three equations describing the coupled
  propagation of the helicity $+2$ components of the tensor perturbations $\pi_{ab}$ (describing  the
 vierbein perturbation), and $\tau_{ab}$ and $N_{ab}$ (describing  the connection perturbation; see 
 Sec. \ref{sec5}). 
 
 We shall  work with the following rescaled versions of  the helicity $+2$ components of 
 $\pi_{ab}$, $\tau_{ab}$ and $N_{ab}$:
  $K_1 \equiv \pi_{(+2)}$,  $F_n\equiv k^{-1} \tau_{(+2)}$, and  $i\, N_n \equiv N_{(+2)}$.
 The factor $k^{-1} $ (with $k \equiv |\mathbf{k}|$) is introduced to render $F_n$ 
 dimensionless\footnote{Here, we use the
 fact that $a_{ij\mu}$, is a connection, and has the same dimension as the derivative of the vierbein.}, like $K_1$ and $N_n$, while the factor $i$ is introduced so as to get real propagation equations.
 We shall not explicitly deal with the helicity $-2$ components, because their propagation equations are
obtained from the ones satisfied by the helicity $+2$ components by the simple change $g\to -g$.
 
The three    helicity-$+2$ variables $K_1, F_n, N_n$ satisfy the three linear equations 
\be
\widehat E_1=0 \,, \,\, \widehat E_3=0\,,\,\, \widehat E_5=0 \,,
\ee
 where the explicit forms
of the expressions\footnote{The labelling of these three equations as 1, 3 and 5 is due to the fact that we had labelled their helicity-$-2$ counterparts as $\widehat E_2$, $\widehat E_4$ and $\widehat E_6$.} 
$\widehat E_1$, $\widehat E_3$ and $\widehat E_5$ will be found
in Appendix A below. These equations involve only the dimensionless coefficients 
 $\widehat c_5$, $\widehat c_6$ and $\xi$. Moreover, $\xi$ only enters in the last, gravity equation $\widehat E_5$. 
 The various rescalings we introduced are such that 
 $k$ and $\eta$ combine everywhere into the variable $z= k \eta$, introduced in Eq. \eqref{zdef}. 
 
Denoting henceforth a $z$-derivative by a prime, the structure of the three helicity-$+2$ equations reads
\bea \label{tensoreqs0}
\widehat E_1 &\equiv& A_{1F}(z) F_n''  + B_{1N}(z) N_n' + B_{1K}(z) K_1' \nonumber \\&+& C_{1F}(z) F_n + C_{1N}(z)N_n + C_{1K}(z)K_1\,,\nonumber \\
\widehat E_3 &\equiv&  B_{3K}(z) K_1'  +B_{3F}(z) F_n'  \nonumber \\&+& C_{3F}(z) F_n + C_{3N}(z)N_n+ C_{3K}(z) K_1\,,\nonumber \\
\widehat E_5 &\equiv& A_{5K}(z) K_1'' + B_{5K}(z) K_1'  +  B_{5F}(z) F_n' +B_{5N}(z) N_n' \nonumber \\&+& C_{5F}(z) F_n + C_{5N}(z)N_n + C_{5K}(z)K_1\,,
\eea
where the coefficients $A, B, C$ of the second, first and zeroth derivatives are all polynomials in $z$ of
degree $\leq3$. The coefficients of the latter polynomials are linear in $\widehat c_5$, $\widehat c_6$ and $\xi$, and depend polynomially on  $\delta$ and $h$. For instance, the coefficient of $N_n$ in 
$\widehat E_1$ reads
\be
(3 \widehat c_5 \delta) \, z^2 + (3 \widehat c_5 h + 48 \widehat c_6 h - 3 \widehat c_5 \delta h + 96 \widehat c_6 \delta h) \,z \,.
\ee

One can find the behavior of the general solution of the system of tensor equations \eqref{tensoreqs0}
(and, in particular, discuss the stability of its solutions) in the following way. We note that the equation
$\widehat E_3=0$ involves the variable $N_n(z)$ only {\it algebraically}, and contains only the first derivatives
of the two other variables. One can then solve the equation
$\widehat E_3=0$ for $N_n(z)$, with a result of the form
\bea \label{Nsol}
N_n(z)&=& \overline B_{NK}(z) K_1'  +\overline B_{NF}(z) F_n' \nonumber\\
 &+& \overline C_{NK}(z) K_1+ \overline C_{NF}(z) F_n ,
\eea
where now the coefficients are rational functions of $z$ (and of the parameters).

As the two other equations involve only at most the first derivative of $N_n(z)$, the replacement
of the solution \eqref{Nsol} for $N_n(z)$ leads to a linear system of two equations involving
the second derivatives of $K_1$ and $F_n$. In order to discuss, in a mathematically controlled way,
 the behavior of this 4th-order differential system, it is useful to solve this system for the highest
 derivatives, i.e., to write it in the form
 \bea \label{KFsystem}
  K_1''&=&\overline B_{KK}(z) K_1'  +\overline B_{KF}(z) F_n'  \nonumber\\
  &+& \overline C_{KK}(z) K_1+ \overline C_{KF}(z) F_n ,\nonumber\\
 F_n''&=&\overline B_{FK}(z) K_1'  +\overline B_{FF}(z) F_n'  \nonumber\\ &+& \overline C_{FK}(z) K_1+ \overline C_{FF}(z) F_n  \,.
 \eea
 Such a system is also equivalent to a linear system of four first order differential equations (in $z$)
 for the four variables $y_1=K_1, y_2=K'_1, y_3=F_n, y_4=F_n'$, i.e. a first-order $4 \times 4$ matrix system
 of the type
 \be \label{matrixsystem1}
 y_i'(z)= m_{ij}(z) y_j(z) \,,
 \ee
 where $i,j=1,2,3,4$ and where we use the summation convention.
 
 \subsection{High-frequency, sub-horizon dispersion laws}
 
 As discussed in the previous section, we are interested in controlling the solutions of 
 the matrix system \eqref{matrixsystem1} in the large-$z$ 
 limit $z \to \infty$, describing high-frequency sub-horizon modes. 
 [As we are dealing with rational functions of $z$, we can think of $z$ as being
 eventually extended to complex values, and do not need to make clear whether infinity is reached
 from positive or negative values along the real axis.]
 As thoroughly discussed in the mathematical literature (see notably Ref. \cite{Ince})
 the crucial mathematical question is whether $ z= \infty$ is a regular-singular,
 or an irregular-singular point of the differential system.
 Actually, we found that in all the cases of interest here, $ z= \infty$ is an {\it irregular-singular point} of the differential system. But it is of the least singular type (technically of rank 1 \cite{Ince}).
 More precisely, we find that the matrix $m_{ij}(z)$ has a finite limit as $z \to \infty$, and that
 this limit, say $m^\infty_{ij}$, is  a diagonalizable matrix\footnote{This will be the case for tensor and vector perturbations.
 The limiting matrix $m^\infty_{ij}$ for scalar perturbations will have a $2\times2$ Jordan block linked
 to a repeated zero eigenvalue. Anyway, in all cases the dispersion law is obtained from the
 characteristic polynomial of $m^\infty_{ij}$.}. The general theory of complex differential systems 
 (see Chapter XIX of \cite{Ince}) then proves that, modulo subleading power-law corrections,
 the general solution of the system \eqref{matrixsystem1} behaves as a linear combination
 of the eigensolutions of the  system with constant (i.e. $z$-independent) coefficients
 \be \label{matrixsystem2}
 y_i'(z)= m^\infty_{ij} \, y_j(z) \,,
 \ee
 where $m^\infty_{ij}= \lim_{z \to \infty} m_{ij}(z)$.
 
 In other words, this proves that the general solution of the system \eqref{matrixsystem1} behaves,
 to leading order as $z\to \infty$, as a linear combination of eigensolutions the type
 \be \label{asymptsol}
 y_i(z) = v_i e^{\sigma z} \,,
 \ee
 where $\sigma$ is one of the four eigenvalues of the matrix $m^\infty_{ij}$, and $v_i$ the
 corresponding eigenvector, i.e.
 \be \label{eigensystem}
 m^\infty_{ij} v_j = \sigma v_i \,.
 \ee
 The four eigenvalues $\sigma_\alpha$ (where $\alpha=1, \cdots,4$) describing the
 large-$z$ behavior of the $h=+2$ modes are the roots of the characteristic polynomial of $m^\infty_{ij}$, i.e.
 \be \label{cp0}
P^{(h=+2)}_4(\sigma) \equiv \det \l  m^\infty_{ij} - \sigma \delta_{ij} \r\,.
 \ee
 As explained in the previous section, this corresponds, via Eq. \eqref{sigom}, to a quartic dispersion law
 in $\omega$ and $k$ given by
 \be
 k^4 P^{(h=+2)}_4( i \frac{\omega}{k})=0\,.
 \ee
 Note in passing that our mathematical analysis justifies (under the condition that the matrix of coefficients, $m_{ij}(z)$, of our first-order
 differential system has a diagonalizable limit at $z=\infty$) the result of what would simply be a WKB
 search for large-$z$ solutions (large-$z$ meaning physically large frequency, $\omega \gg \lam$), with 
 modes of the approximate form
 \be
 e^{\sigma k \eta} e^{i \mathbf{k} \cdot\mathbf{x}}= e^{i \omega \eta+ i \mathbf{k} \cdot\mathbf{x}}
\ee
with a conformal-time frequency $\omega$ related (in view of $z=k \eta$) to the eigenvalue $\sigma$ 
via Eq. \eqref{sigom}.

Let us note that the use of a WKB approximation for describing 
 the large-$z$ behavior of the solutions of our systems of ordinary differential equations (ODEs) is somewhat delicate.
 One might think that one would get the correct dispersion law simply
by using the WKB ansatz, i.e. replacing the derivatives of all the variables according to the rules,
\bea \label{wkbansatz0}
 K_1' &\to& \sigma K_1, K_1'' \to \sigma^2 K_1,   F_n' \to \sigma F_n,  \nonumber \\
 F_n'' &\to& \sigma^2 F_n,  N_n' \to \sigma N_n \,,
\eea
in the original system of equations \eqref{tensoreqs0} concerning the complete set of coupled variables,
and then by computing the determinant of the resulting linear system of three algebraic equations
for the three variables $K_1, F_n, N_n$. The so-obtained WKB dispersion law would be defined,
when considering, more generally, $N$ such WKB-reduced equations in $N$ unknowns as the determinant
\be \label{wkbdet}
P_N^{\rm WKB}(\sigma) = \det \frac{\d  ({\rm WKB\!-\!\rm equations})}{\d  ({\rm variables})} \,.
\ee
However, we will see below, on explicit examples, that the naive WKB analysis of the original equations
generally gives {\it incorrect} results, namely
\be
P_N^{\rm WKB}(\sigma)  \neq C P_N(\sigma)\,,
\ee
where $C$ allows for a proportionality constant, and where $P_N(\sigma)$ is the correct dispersion
law, which is obtained  by a characteristic polynomial of an asymptotically  well behaved first-order system
as in Eq. \eqref{cp0}.

\subsection{Stability analysis of the first branch of de Sitter-like solutions}

To put in perspective the stability analysis of the second
 branch of  de Sitter solutions (which include the self-accelerated solution as a special point corresponding
 to $c_{2}=0$), let us start by discussing the tensor cosmological perturbations of the first branch
 of de Sitter solutions \cite{Nair:2008yh} (the one which needs the bare vacuum energy $c_2$ computed
 from Eq. \eqref{lambranch1} to sustain its expansion). This first branch has $f= - \lam$ and 
 $g=0$, i.e. $\delta=1$ and $h=0$. From Eq. \eqref{torsionfg}, we see that the background torsion vanishes along this first branch. The results of Refs. \cite{Nair:2008yh,Nikiforova:2009qr} concerning
 the stability of torsionfree Einstein backgrounds guarantee that there will be no exponentially growing modes
 around this first branch. It is, however, useful to directly derive the dispersion law of the
 cosmological perturbations around this first branch by using the same equations and the same
 methods that we shall use for the second branch. This indeed provides both a check of our
 equations and of our methods.
 
 The equations describing the tensor perturbations
 along the  first branch of de Sitter-like solutions of TG are obtained by restricting the general equations 
 given in Appendix A to the case $\delta=1$, $h=0$.
 In that case, the three equations \eqref{tensoreqs0} simplify a lot and read 
  \bea \label{tensoreqsB1}
 \widehat E_1 &=& - \frac32 (-K_1 + F_n \,z + 4 \,\hcf F_n \,z - 2 \,\hcf N_n \,z^2  \nonumber\\
 && + z \, K_1'  - 2 \,\hcf \,z^3 N_n' - 2 \,\hcf \,z^3  F_n''), \nonumber \\
 \widehat E_3 &=& -3 (K_1 \,z + N_n \,z + 
    4 \,\hcf N_n \,z + 2 \,\hcf F_n \,z^2 \nonumber \\
    && - 2 \,\hcf N_n \,z^3 - 
    2 \,\hcf \,z^3  F_n' ), \nonumber \\
 \widehat E_5 &=& \frac32 (-2 K_1 + F_n \,z - N_n \,z^2 + K_1 \,z^2 \xi \nonumber \\
&& -  z^2  F_n' - 
    2 \,z  \xi   K_1' + 
    z^2  \xi K_1'')\,.
\eea
Solving for $N_n$ from the second Eq. \eqref{tensoreqsB1}  ($\widehat E_3=0$) yields 
\be \label{NsolB1}
N_n= \frac{K_1 + 2 \,\hcf F_n \,z - 2 \,\hcf \,z^2 F_n'}{-1 - 4 \,\hcf + 
 2 \,\hcf \,z^2}\,.
\ee
Replacing this solution in the expressions $\widehat E_1$ and $\widehat E_5$ yields a system of two 
second-order ODEs for $K_1$ and $F_n$ whose large-$z$ expansions read
\bea \label{KFB1}
K_1''= -K_1 +\frac2z K_1' + O(\frac1{z^2}) \,,\nonumber \\
F_n''= - F_n +\frac2z F_n' + O(\frac1{z^2}) \,.
\eea
Here we have also exhibited the terms of order $1/z^1$, to give an example of their effect compared
to the leading-order terms of order $1/z^0$. At the order  $O(1/z^1)$ included, this system of
equations is decoupled, but the terms of order  $O(1/z^2)$ couple the $z$-evolutions of $K_1$ and $F_n$.
If we start by considering only the large-$z$ limit of the above differential system, it yields
(when written in first-order form, with $y_1=K_1, y_2=K_1', y_3=F_n, y_4=F_n'$) the $4 \times 4$ limit matrix
\begin{equation}
\label{mijB1}
m_{ij}^\infty =  \begin{pmatrix} 0 &1 &0&0 \\ -1 &0 &0&0 \\ 0 & 0&0&1\\0&0&-1&0 \end{pmatrix} \,.
\end{equation}
The characteristic polynomial of this matrix (for the first branch, say B1) is
\be \label{dispB1}
P_4^{\rm B1}(\sigma)= (\sigma^2+1)^2 \,.
\ee
The four eigenvalues of this matrix are $+i, -i,+i,-i$, and the corresponding dispersion law reads
\be
 (\omega^2 - \mathbf{k}^2 )^2=0 \,.
\ee
Note that this is exactly the same dispersion law as one would get (in the high-frequency limit)
in a flat spacetime. It describes two bosonic d.o.f. propagating at the velocity of light.
These two d.o.f. describe the helicity $+2$ mode of the usual massless Einsteinian graviton,
together with the helicity $+2$ mode of the massive spin-2 field of TG. In flat spacetime, the exact version
for all frequencies, of the above dispersion law would be $ (\omega^2 - \mathbf{k}^2 )  (\omega^2 - \mathbf{k}^2 -m_2^2)=0$, where $m_2$ denotes the mass of the spin-2 field, defined (on a flat background) by Eq. \eqref{m2}.
Note that we considered here only the helicity $+2$ modes. The helicity $-2$ modes would add two
more bosonic d.o.f., with the same dispersion law.

Let us also briefly discuss the effect of the subdominant $O(1/z)$ contributions in the large-$z$
expansion of the matrix, i.e. the terms $\frac2z K_1'$ and $\frac2z F_n'$ in Eq. \eqref{KFB1},
corresponding, more generally, to the terms denoted $ \frac{n^{(h)}_{ij}}{z}$ in Eq. \eqref{zexpmij}.
As briefly mentioned in the previous section, such terms 
 modify (for the present, rank-1 case) the leading-order exponential factor
$e^{\sigma z}$ associated with each eigenvalue of $m^\infty_{ij}$ by a logarithmic term in the
exponent, i.e. a power-law correction to the $e^{\sigma z}$ behavior:
\be \label{sigbetaasympt}
e^{\sigma_\alpha z} \to e^{(\sigma_\alpha z + \beta_\alpha \ln z)}= e^{\sigma_\alpha z} z^{\beta_\alpha}\,.
\ee
Let us illustrate this general fact in the simple case of the differential system \eqref{KFB1}. In that case,
the exact solutions of the above system (truncated at the $O(1/z)$ level included) are easily
found to be 
\be
(z \pm i) e^{\pm iz} =  e^{\pm iz}z \l1+ O\l \frac1z \r \r \,,
\ee
where the final $O(\frac1z)$ term would be sensitive to the $O(\frac1{z^2})$ contributions in $m_{ij}(z)$,
while the correcting power-law factor $z$ is entirely determined by $\frac{n^{(h)}_{ij}}{z}$.

As already mentioned, such power-law correction factors evolve on the slow,
Hubble scale, while our aim here is to discuss the instabilitites that could happen on
small scales. Therefore, we shall focus, in the following, on the high-frequency dispersion law
derived from the limiting matrix $m^\infty_{ij}$.

Let us also use the simple case of the branch-1 tensor perturbations to give an explicit example of the
possible failures of a direct WKB analysis of the original equations. If we apply the
WKB ansatz \eqref{wkbansatz0} directly in the original system \eqref{tensoreqsB1}, we get
three equations for three unknowns ($K_1, F_n, N_n$), and a computation of the large-$z$
behavior of its WKB determinant \eqref{wkbdet} yields
\be
\frac{P_N^{\rm WKB}(\sigma)}{ - 108 \,\hcf (1+ 4 \,\hcf) z^6} \sim (\sigma^2+1)\l\sigma^2+\frac{1+ 2 \,\hcf}{1+ 4 \,\hcf}\r \,.
\ee
This result differs from the correct result \eqref{dispB1} through the second factor involving a $\hcf$ modification of the factor $\sigma^2+1$. Note that this failure occurs in spite of the fact that the
solution for $N_n$ in terms of the other variables, Eq. \eqref{NsolB1}, is such that $N_n$
has the same large-$z$ asymptotic behavior \eqref{sigbetaasympt} as $K_1, F_n$ 
(including when considering the power law subleading term).
On the other hand, if one first solves (as discussed above) for the variable $N_n$ by using
equation $\widehat E_3=0$, and replaces this solution in the two other equations,
and then performs a direct WKB analysis of the resulting system of two second-order
equations for $K_1, F_n$ (without putting it in the form of  Eq. \eqref{KFsystem}, or
Eq. \eqref{matrixsystem1}), one gets the correct dispersion law \eqref{dispB1}.
However, it was necessary to reduce the system to the first-order form \eqref{matrixsystem1}
to prove that there existed such WKB-type solutions having the asymptotic behavior \eqref{sigbetaasympt}.
As indicated above, one can easily compute the subdominant power-law behavior from the
first-order system \eqref{matrixsystem1}.

One conclusion is that it is crucial to eliminate any auxiliary field when performing a WKB analysis.
Indeed, in the present case, we have a system whose physical initial data should include 
four independent data (corresponding to a dispersion law that is quartic in $\sigma$ or $\omega$).

\subsection{Stability analysis of the second branch of de Sitter-like solutions}

Let us now consider the second branch of solutions, for which, in general $\delta \neq1$, and, most
crucially, $h\neq0$. In that case, one must use the full form of the three tensor perturbation equations
given in Appendix A. The structure of these equations was delineated in the Subsec.\ref{sec7A} above.
The method to control the structure of the solutions was already indicated in Eqs. 
\eqref{tensoreqs0}, \eqref{Nsol}, \eqref{KFsystem} above, and is conceptually the same as the one used 
 along the first branch. We eliminate $N_n$ by solving the
second equation of the system, with a result of the form $N_n=N_{N_n}/D_{N_n}$ with, now, 
\begin{align}
N_{N_n} &=h \,K_1 + 6 \,\hcf \delta h \,K_1 - 32 \,\hcs \delta h \,K_1 + 2 \,\hcf \delta^2 h \,K_1    \nonumber   \\
  &- 
  64 \,\hcs \delta^2 h \,K_1 - 2 \,\hcf h^3 \,K_1   + 32 \,\hcs  h \,z \,F_n   - 2 \,\hcf \delta  h \,z  \,F_n   \nonumber   \\
  &+ 
  64 \,\hcs \delta  h \,z \,F_n + \,K_1 \,z + 2 \,\hcf \delta  \,z \,K_1 - 2 \,\hcf \delta^2  \,z \,K_1   \nonumber   \\
  &+ 
  2 \,\hcf h^2  \,z \,K_1+ 2 \,\hcf \delta  \,z^2 \,F_n + 
  2 \,\hcf h \,z^2 \,\d_zF_n    \nonumber   \\
  &- 
  2 \,\hcf \,z^3 \,\d_zF_n + 
  32 \,\hcs h \,z \,\d_zK_1 + 
  64 \,\hcs \delta h \,z \,\d_zK_1\,,
\end{align}
and
\be
D_{N_n}=\,z (-1 - 4 \,\hcf \delta + 2 \,\hcf h^2 - 4 \,\hcf h \,z + 2 \,\hcf \,z^2)\,.
\ee
However, the final matrix $m_{ij}(z)$
of the first-order system \eqref{matrixsystem1} is much more involved than along the first branch. 
As explained above, we
shall only consider here the large-$z$ limit of $m_{ij}(z)$.  

In the case of the first branch, $m^\infty_{ij} $ happened to be {\it independent} of both the
parameters entering TG (such as $\xi, \widehat c_5, \widehat c_6, \cdots$) and the parameters
entering the background solution. However, this is no longer the case along the second branch.
In that case, $m^\infty_{ij} $ depends both on $\xi, \widehat c_5, \widehat c_6$ and
on $\delta, h$.  As a consequence, the dispersion law (i.e. the characteristic polynomial of
$m^\infty_{ij} $) depends also on these parameters. We find that the dispersion law along
the second branch of cosmological solutions is again a fourth-order, bi-quadratic polynomial in $\sigma$,
with the following structure
\be \label{dispB20}
P_4(\sigma)^{(h=+2)\rm B2}=\sigma^4 + 2 b'_{\rm tensor} \sigma^2 + 1=0\,,
\ee
with 
\be
b'_{\rm tensor}=1 + \frac{8 \,\hcf^{\,2} (\delta - \delta^2 + h^2)^2}{-1024 \,\hcs^{\,2} (h + 2 \delta h)^2 + \xi + 
 4 \,\hcf \delta \xi} \,.
\ee
We have given above, in Eqs. \eqref{c6}, \eqref{c5}, the functional links relating $\widehat c_5, \widehat c_6$ to $\delta, h$  (independently from $\xi$). We can then express $\widehat c_5, \widehat c_6$
as functions of $\delta, h$, and thereby compute the dispersion law (as well as 
$m_{ij}(z)$ and $m^\infty_{ij} $) as functions
of $\delta, h$ and $\xi$. Note that we cannot, in general, relate $\xi$ to $\delta$ and $ h$ because
of the influence of the bare vacuum energy in Eq. \eqref{xi+d2}. It is only for the special, self-accelerating solution
(which has $c_2=0$) that one can compute $\xi$ as a function of $\delta, h$.
We then find (as already announced in section \ref{sec6bis})
\be\label{bpdelxi}
b'_{\rm tensor}(\delta,h,\xi)-1=\frac{(\delta - \delta^2 + h^2) (1- 2\delta+ \delta^2 + h^2)^2}{2 h^2 \l  (1 + 2 \xi)\delta^2 -(1+ 3  \xi) \delta +\xi  - h^2\r}.
\ee

Several facets of this result should be emphasized. First, we note that all the factors in the numerator of this
expression are constrained to be positive (recalling the lower bound \eqref{lowerboundh} on $h^2$). Therefore
the sign of $b'_{\rm tensor}(\delta,h,\xi)-1$ (and therefore, as discussed below, the stability) is determined by the last factor
in the denominator, i.e.
\bea \label{signbp}
&&{\rm sign}[b'_{\rm tensor}(\delta,h,\xi)-1]= \nonumber\\
&& {\rm sign}[(1 + 2 \xi)\delta^2 -(1+ 3  \xi) \delta +\xi  - h^2 ]\,.
\eea
Requiring stability, i.e. a positive sign, then gives an upper bound on $h^2$.
Second, we note that all the factors entering $b'_{\rm tensor}(\delta,h,\xi)-1$ vanish when considering the limit of torsionfree
 backgrounds (characterized
by the double condition $\delta=1$ and $h=0$). To see this better, let us define
\be
\bd \equiv \delta-1\,.
\ee
We can then rewrite Eq. \eqref{bpdelxi} as
\be
b'_{\rm tensor}(\delta,h,\xi)-1=\frac{(h^2 - \bd^2-\bd) ( \bd^2 + h^2)^2}{2 h^2 \l  
(1 + 2 \xi)\bd^2 +(1+ \xi) \bd  - h^2\r}.
\ee
In other words,  $b'_{\rm tensor}(\delta,h,\xi)-1$ has a $\frac00$ structure near zero torsion,
which shows that the presence of torsion (even in infinitesimal amount!)  in the 
background drastically, but subtly, affects the structure of the dispersion law.

In particular, we can study the limiting behavior of $b'_{\rm tensor}(\delta,h,\xi)$ as 
the second branch approaches its crossing with the (torsionfree) first branch.
Indeed, we see from the general equation \eqref{delx} that, if we keep fixed the TG parameters
$c_5$ and $c_6$, $\delta$ must approach $1$ as $h\to0$ in the following way:
\be \label{expdel}
\delta= 1- a h^2 + O(h^4) \,,
\ee
where we denoted
\be
a\equiv 4(24 \hcs+\hcf) \,,
\ee
which should not be confused with the scale factor $a(\eta)$.
From Eq. \eqref{c5}, one then finds that, at the crossing point between the two branches
\be \label{eq8.35}
1+ 4 \hcf =0\,,
\ee
so that
\be \label{eq8.36}
a= 96 \hcs-1\,.
\ee
As $\hcs $ can take any positive value, the coefficient $a$ can
be either positive or negative (but $> -1$).

Inserting the expansion \eqref{expdel} in the above expression for $b'_{\rm tensor}(\delta,h,\xi)-1$
yields
\be \label{bpexp}
b'_{\rm tensor}(\delta,h,\xi) =1 - \frac{1 +a}{2 (1 + a + a \xi)} h^2+ O(h^4)\,.
\ee
The first important fact to notice is that, in the limit where the second branch approaches
the torsionfree first branch ($h \to0$), we get $\lim b'_{\rm tensor}=1$, so that the
tensor dispersion law along the second branch tends to 
\be
\sigma^4+ 2 \sigma^2+1=(\sigma^2+1)^2\,,
\ee
which {\it coincides} with the dispersion law along the first branch. [We shall see later that such
a continuous behavior applies neither to the case of the helicity-1 perturbations, nor 
to the  helicity-0 ones.]
However, as soon as $h \neq0$ we will have a modified dispersion law with $\lim b'_{\rm tensor}\neq1$.

As we shall have several dispersion laws of this type, let us discuss the general conditions for the stability
of perturbations satisfying a law of the form of Eq. \eqref{dispB20}, i.e., in more physical terms
\be
\omega^4 - 2 b' \omega^2 {\mathbf k}^2 +  {\mathbf k}^4=0\,.
\ee
The solutions of this bi-quadratic equation are
\be
\omega^2= C_\pm {\mathbf k}^2\,,
\ee
with
\be
C_+= b'+ \sqrt{b'^2-1} \,;\, C_-= b'- \sqrt{b'^2-1}=\frac1{C_+}\,.
\ee
In order to avoid any instability we need $C_\pm$ to be real and positive. It is easily seen
that this requires
\be
b'>1  \; ({\rm for \; stability})\,.
\ee
In this case we will have one mode propagating with a velocity greater than one (i.e. superluminal)
and another mode propagating with (an inverse) velocity, smaller than one (subluminal). 
We shall not worry here about the physical consistency of having superluminal velocities.
What is most important is to have some type of hyperbolic propagation of perturbations.

From the expansion \eqref{bpexp} of $b'_{\rm tensor}$ near its crossing with the
(stable) first branch (which had $b'\equiv 1$), we see that the second branch, depending
on the sign and magnitude of $1+a$ can, near this crossing, either be stable, or exhibit some gradient instabilities.

\subsection{Regions of parameter space where the tensor perturbations of the second branch are stable}

\begin{figure}
\includegraphics[scale=0.5]{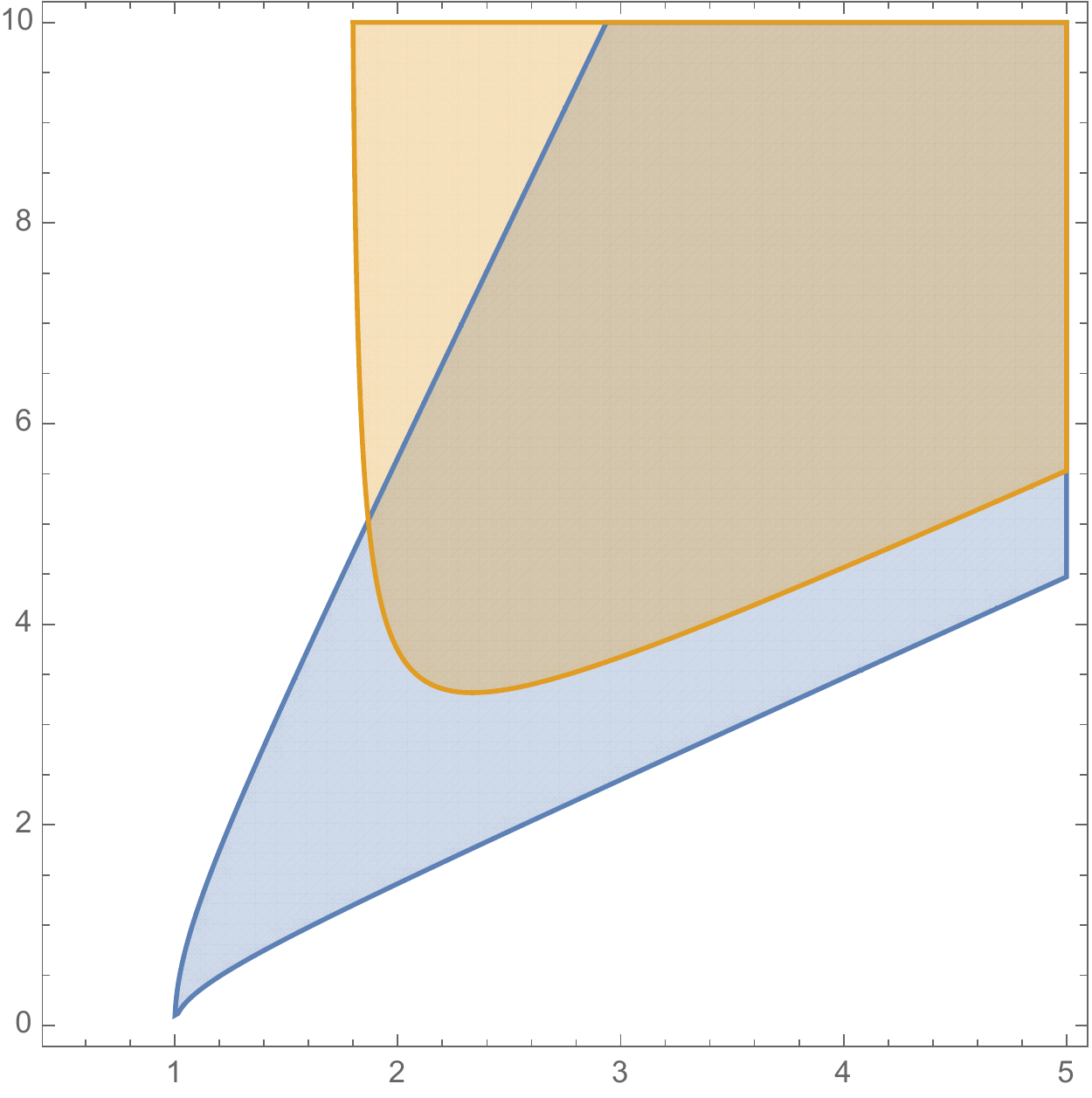}
\caption{\label{fig1}
Tensor stability regions in the $(x,y)=(\delta, h)$ plane (for $h>0$);
both for branch 2 (wedge, blue on line; for $\xi=10$), and for the self-accelerating solution
(region above the upper curve, brown on line).}
\end{figure}

Let us now consider the stability properties of the second branch all over the relevant parameter space.
For tensor perturbations around a generic point along the second branch, $b'_{\rm tensor}$
depends on the three parameters $\delta,h,\xi$. Using Eq.  \eqref{signbp} we see that a necessary
condition for stability, i.e. for $b'_{\rm tensor}-1>0$, is an upper bound  on $h^2$ 
which reads as follows  in terms of $\bd$
\be
h^2 < (1 + 2 \xi)\bd^2 +(1+ \xi) \bd \,.
\ee
Moreover, we recall that we have also the following necessary lower bound 
\eqref{lowerboundh} on $h^2$ (coming from the positivity of $c_6$)
\be
h^2 > \bd^2+\bd \,.
\ee
It is easily checked that, though the full range of variation of $\bd\equiv \delta-1$ is a priori
$\bd > -\frac12$, the above two necessary stability constraints imply that
\be
\bd >0 \, ; \, {\rm i.e.} \, \, \delta >1\,.
\ee
We conclude that, for any given value of $\xi>0$, the stability region in the $\delta, h$ plane,
along the second branch is a 
curved wedge between two hyperbolas defined by the inequalities
\be
 \bd^2+\bd < h^2 < (1 + 2 \xi)\bd^2 +(1+ \xi) \bd \;  {\rm and} \; \bd >0 \, .
\ee
This region (shown as blue online in Fig. \ref{fig1}) starts as a thin vertical line at their lower tip $\delta=1, h=0$, because the two curves defined
by the two sides of the latter inequalities have a similar parabolic shape $h^2 \propto \bd$
near their common tip $\delta=1, h=0$.

The latter stability region refers to the one-parameter family of solutions along the second branch.
The interest of this family is that it connects the torsionfree case $\delta=1, h=0$ (corresponding
to the tip of the latter stability wedge) to the self-accelerating solution (which must stay away from
the latter tip). Let us now consider the stability region of the self-accelerating solution itself. In that case
we must take into account that $\xi$ is not anymore a free parameter along the self-accelerating solution
but is related to $\delta, h$ via Eq. \eqref{hdelxiSA}. Inserting the value of $\xi$ derived from
the latter link, i.e.
\be
\xi= \frac12(h^2 - \delta^2- \delta )= \frac12(h^2 - 2-\bd^2- 3 \bd)\,,
\ee
in the stability condition $b'_{\rm tensor}-1>0$, we find as condition for stability
\be
(2 \del^2 - 3 \del -1) h^2 - 2 \del^4 + \del^3 + 4 \del^2 - 3\del>0 \,.
\ee
It is easily checked that this inequality can only be satisfied when $\del$ is larger
than the largest root of $2 \del^2 - 3 \del -1$, i.e. for
\be \label{178}
\del > \del_{\rm min}\equiv \frac14 \l 3 + \sqrt{17} \r \approx  1.78078\,.
\ee
Then, in the domain $\del>  \del_{\rm min}$, the stability region of the self-accelerating solution
is defined by the single inequality
\be \label{SAtensorstab}
h^2 > \frac{  2 \del^4 - \del^3 - 4 \del^2 + 3\del}{2 \del^2 - 3 \del -1} \,,
\ee
which is indeed found to imply the lower bound \eqref{lowerboundh}.

In Fig. \ref{fig1} we represent both an example (for $\xi=10$) of the stability region along a generic member
of the second branch (blue region on line), and the stability region of the self-accelerating solution (brown on line).
Note that the two uppermost region-bounding curves in this figure mark a limit of their
respective stability regions where the {\it denominator} of $b'_{\rm tensor}-1>0$
changes sign. This means that these upper boundaries are singular, with 
$b'_{\rm tensor}-1 \to \pm \infty$ on either side. By contrast, the lower boundary of the
(wedge-like) stability region of the generic second branch corresponds to the vanishing of the
{\it  numerator} of $b'_{\rm tensor}-1$. This would correspond to a limit where
the dispersion law is the same as in flat spacetime. However, this limit also corresponds
to a degenerate limit where $c_6 \to 0$. 

\section{Study of vector perturbations and of their stability} \label{sec8}

\subsection{Symmetry of  the gauge-fixed vector perturbation equations}

We continue our study of cosmological perturbations in de Sitter-like TG solutions by considering
the vector perturbations. While the tensor perturbations involved only three variables, the vector perturbations
now involve seven (transverse vector) variables: one ($W_a$) in the vierbien perturbation, and
six others ($\zeta_a$, $\nu_a$, $\mu_a$, $\kappa_a$, $A_a$ and $L_a$) in the connection
perturbation. 

Let us first mention that the system of vector equations has a certain symmetry that we have
used as a check on our derivation. This symmetry is a residual symmetry after the (incomplete)
gauge-fixing that we used. We have gauge-fixed the vector sector of the coordinate freedom by
imposing the zero-shift condition \eqref{zeroshift}. However, this condition will still be
satisfied if we perform a {\it time-independent} helicity-1 spatial  (infinitesimal) coordinate transformation 
$x'{}^\mu=x^\mu+ \xi^\mu$ with $\xi^0=0$ and 
$\xi^a_C = - C_a  e^{i \mathbf{k} \cdot\mathbf{x}}$  (with $k^a C_a=0$). 
Such a  diffeomorphism will act both on the vierbein and the connection as
\be
- \del_\xi e^i_{\mu}= \xi^\lam \d_\lam e^i_{\mu}+ \d_\mu\xi^\lam e^i_{\lam} \,,
\ee
\be
- \del_\xi A_{ij\mu}= \xi^\lam \d_\lam A_{ij\mu}+ \d_\mu\xi^\lam A_{ij\lam}\,.
\ee
As a consequence the spatial part of the vierbein will get out of the symmetric gauge \eqref{symgauge}.
We must therefore apply an additional compensating infinitesimal Lorentz-rotation transformation
$\omega_{ab}^C$ which is found to be (with $T_{[ab]} \equiv \frac12 (T_{ab}- T_{ba})$)
\be
\omega_{ab}^C= i k_{[a} C_{b]} e^{i \mathbf{k} \cdot\mathbf{x}}\,.
\ee
Finally, the combined coordinate-plus-Lorentz transformation $\del^{\rm tot}$
which preserves our gauge-fixing is found
to act on the vierbein and the connection as (henceforth suppressing the $e^{i \mathbf{k} \cdot\mathbf{x}}$ factor, and denoting symmetrization as $T_{(ab)} \equiv \frac12 (T_{ab} +T_{ba})$)
\be
\del^{\rm tot}e^i_{a}=  i k_{(a} C_{i)} \,,
\ee
\be
\del^{\rm tot}A_{ \t0 \ta b}=i  f e^{ \phi}  i k_{(b} C_{a)} \,,
\ee
\bea
&\del^{\rm tot}& \! \! A_{ \ta \tb c}=k_c k_{[a} C_{b]} \nonumber \\
&+& \! \! i g e^{ \phi} \l k_c \varepsilon_{abs} C^s
+  k_{[a} C_{p]}  \varepsilon_{pbc} -   k_{[b} C_{p]}  \varepsilon_{pac}\r.
\eea
Using these formulas, we can compute how the above symmetry transforms the various vector variables
parametrizing the perturbed vierbein and connection, e.g. we have: $ \del^{\rm tot} W_a = \frac12 C_a$.
We can further decompose $C_a$ into its helicity pieces: $C^a = C_{(+1)} e_+^a + C_{(-1)} e_-^a $,
and thereby derive separate symmetries of the helicity-$\pm1$ variables given by
\bea
 \del^{\rm tot} W_{(+1)} \!\!&=& \frac12 C_{(+1)} \,;\,  \del^{\rm tot} \nu_{(+1)}=\del^{\rm tot} \mu_{(+1)}=\frac{i}{2} f e^\phi C_{(+1)}\, ; \nonumber\\
\del^{\rm tot} L_{(+1)}&=&\frac{1}{2} g k e^\phi C_{(+1)}\, ;\, \del^{\rm tot} \zeta_{(+1)} =\del^{\rm tot} \kappa_{(+1)}= 0 \,;\nonumber\\
\del^{\rm tot} A_{(+1)}&=& \l\frac{1}{2}- \frac{g}{k} e^\phi \r C_{(+1)} \, .
 \eea
 We have checked that the vector perturbation equations we derived are invariant under  these correlated shift symmetries of the vector variables.

\subsection{Reducing the vector perturbations to a linear system of three first-order ordinary differential equations} 

The obtention of  the dispersion law for vector perturbations is more involved than the case
discussed above of tensor perturbations. The first reason is that we have to deal with more variables:
seven instead of three. As in the tensor case, the perturbations equations for helicity $+1$ decouple
from those for helicity $-1$. 

In all, one can derive ten vectorial equations from the gravitational and connection equations.
However, we found that the helicity$\pm1$ projections of the Bianchi identities of TG (explicitly
worked out in \cite{Nikiforova:2017saf}) imply that there are (for each helicity) three identities
among these equations. It is therefore sufficient to use only seven (independent) equations
among the ten vector equations.
It is also sufficient to deal only with the helicity $+1$ sector.

The seven, helicity $+1$, equations we worked with are given in Appendix B.
As in the case of tensor perturbations, we use the rescaled field equations
$\widehat{{\cal G}}_{ij}$, Eq. \eqref{rescaledGeq}, and 
  $\widehat{{\cal T}}_{[ij]k}$, Eq.\eqref{rescaledAeq}. In addition, we scale out $k$ and
use as independent variable $z= k \eta$.
The notation in Appendix B is the following: there are four vectorial connection equations, which are denoted 
$V_1$, $V_2$, $V_3$, $V_4$, and  three gravitational equations denoted $V_5$, $V_6$ and $V_7$.
The vector helicity $+1$ variables entering these equations are respectively denoted (keeping close to the notation used
for the corresponding vector variables $W_a, \zeta_a$, etc.):
$W_1$  ( $\equiv$ the helicity $+1$ component $W_{(1)}$ of $W_a$), 
$\zeta_1$, $\mu_1$, $\nu_1$, $\kappa_1$, $A_1$, and $L_1$.

First, we simplified these equations by replacing, respectively, $A_1$ and $\mu_1$ by
the new variables $A_3$ and $\mu_3$ defined so that
\be
A_1 \equiv A_3 - 2 L_1 \, ;\,\mu_1 \equiv \mu_3 + \nu_1 \,.
\ee

In terms of these new variables, we find that, among the seven equations $V_i$, $i=1,\cdots 7$,
four of them are algebraic in the four variables $\zeta_1$,  $\nu_1$, $\kappa_1$ and $L_1$.
More precisely $V_2$, $V_5,$ $V_6$ and $V_7$ depend only on the variables
\begin{align}
&\zeta_1(z),  \nu_1(z), \kappa_1(z), L_1(z) \,;\\
& W_1(z) , A_3(z), \mu_3(z) \, ;\, W'_1(z), A'_3(z), \mu'_3(z) \,.
 \end{align}
 When $h =g/\lam\neq0$ one finds that one can solve the set of four equations $\{V_2=0, V_5=0, V_6=0,V_7=0\}$
 in the four variables $\{\zeta_1(z),  \nu_1(z), \kappa_1(z), L_1(z) \}$, so as to get
 \begin{align} \label{solznkL}
 \zeta_1&= R_\zeta(z\,;\, W_1 , A_3, \mu_3 \, ;\, W'_1, A'_3, \mu'_3) ,\\
 \nu_1&=R_\nu(z\,;\, W_1 , A_3, \mu_3 \, ;\, W'_1, A'_3, \mu'_3),\\
 \kappa_1&=R_\kappa(z\,;\, W_1 , A_3, \mu_3 \, ;\, W'_1, A'_3, \mu'_3),\\
 L_1&=R_L(z\,;\, W_1 , A_3, \mu_3 \, ;\, W'_1, A'_3, \mu'_3)\,.
 \end{align}
 Here, $R_\zeta$ etc. are linear functions of $W_1 , A_3, \mu_3 \, ;\, W'_1, A'_3, \mu'_3$
 that are rational in $z$ (and the TG parameters).
 
 In addition, the remaining three equations $V_1, V_3, V_4$ originally depended on the following
 set of variables
  \begin{align}
  V_1 :  \; &\zeta_1,  \nu_1, \kappa_1, L_1 \,;   \nu'_1 \\
& W_1 , A_3, \mu_3 \, ;\, W'_1, A'_3, \mu'_3 \,;
  \end{align}
   \begin{align}
  V_3 :  \; &\zeta_1,  \nu_1, \kappa_1, L_1 \,;   \zeta'_1, \kappa'_1, L'_1  \\
& W_1 , A_3, \mu_3 \, ;\, A'_3 \,;\, \mu''_3 \,;
  \end{align}
   \begin{align}
  V_4 :  \; &\zeta_1,  \nu_1, \kappa_1, L_1 \,;   \nu'_1 \\
& W_1 , A_3, \mu_3 \, ;\, A'_3, \mu'_3 \,;\, A''_3 \,.
  \end{align}
  When inserting the solutions \eqref{solznkL} into the above equations $V_1, V_3, V_4$
  (we denote the results as $\overline V_1, \overline V_3, \overline V_4$),
  one is a priori generating  second derivatives of $W_1 , A_3$, and $\mu_3$.
  However, one finds that the coefficients of  $A''_3$ and  $\mu''_3$ actually {\it vanish}.
  
  At this stage, we have three equations for three unknowns ($W_1 , A_3,\mu_3$), depending on the following variables
  \begin{align}
 \{ \overline V_1, \overline V_3, \overline V_4\} :  \;  W_1 , A_3, \mu_3 \, ;\, W'_1, A'_3, \mu'_3\,;\,W''_1\,.
  \end{align} 
  
By algebraically combining these three equations, we can eliminate   $W''_1$ in two of these equations.
Actually, the third equation so obtained, namely a combination 
\be
\overline V_4^{\rm new} \equiv \overline V_4- C_{43}(z) \overline V_3\,,
\ee
which eliminates $W''_1$,  is found to depend only on $W_1 , A_3, \mu_3 \, ;\, W'_1$,
without involving the derivatives of $A_3, \mu_3$. As a consequence we can combine $\overline V_1$
with the {\it derivative} of $\overline V_4^{\rm new}$ to eliminate $W''_1$ from our system of three
equations. After these operations, we get a system of three equations involving only the variables
\be
 W_1 , A_3, \mu_3 \, ;\, W'_1, A'_3, \mu'_3\,.
\ee
This system is not quite our final system because one finds that it does not behave fully
properly in the large-$z$ limit. However, if we replace the variable $\mu_3$ by the variable
\be \label{mu4}
\mu_4(z) \equiv \frac{\mu_3(z)}{z}\,,
\ee
one ends up with a system of three first-order equations in 
\be
\{ y_1, y_2, y_3  \} \equiv \{ W_1 , A_3, \mu_4   \}\,,
\ee
which, when solved for first derivatives, yields a matrix system of the form
\be \label{matrixsystem1vec}
 y_i'(z)= m_{ij}(z) y_j(z) \, ; \, i, j=1,2,3\,,
 \ee
 where the matrix $m_{ij}(z)$ has the same good property as the matrix obtained in the
 tensor case discussed above. Namely,  the matrix $m_{ij}(z)$ has a finite limit, $m^\infty_{ij}$, as $z \to \infty$, and
 this limit yields a diagonalizable matrix. [This would not have been the case when keeping $\mu_3$.]
 
 \subsection{Dispersion law for vector perturbations along the second branch of de Sitter-like solutions:
 necessary presence of gradient instabilities}
 
 We can then apply the same mathematical results \cite{Ince} used in the tensor case above.
 The limiting system (with constant coefficients)
 \be \label{matrixsystem2vec}
 y_i'(z)= m^\infty_{ij} \, y_j(z)\,,
 \ee
 where $m^\infty_{ij}= \lim_{z \to \infty} m_{ij}(z)$, will describe the large-$z$ asymptotics of our
 solutions (modulo power-law corrections). We therefore conclude that our solutions behave,
 for large$-z$, as a linear combination of eigensolutions of the type
 \be \label{asymptsol}
 y_i(z) = v_i e^{\sigma z} \,,
 \ee
 where $\sigma$ is one of the three eigenvalues of the $3\times3$ matrix $m^\infty_{ij}$, and $v_i$ the
 corresponding eigenvector.
 
 The problem of the stability of vector perturbations is thereby reduced to the
 purely algebraic question of computing the characteristic polynomial of the $3 \times 3$ matrix
 $m^\infty_{ij}= \lim_{z \to \infty} m_{ij}(z)$. And the dispersion law for high-frequency vectorial modes
 is simply given by equating the latter characteristic polynomial  to zero
 \be
 P_3^{(h=+1)}(\sigma) = \det \l  m^\infty_{ij} - \sigma \delta_{ij} \r\,.
 \ee
 The computation of the latter characteristic polynomial yields a cubic dispersion law of the form
 \be \label{vecdispB2}
 \sigma ( \sigma^2 + c^{\rm vector}) =0 \,,
 \ee
 whose physical form (in terms of $\omega$ and $k$) was written in Eq. \eqref{disph1B2} above.
 As already announced, we found that the constant $c^{\rm vector}$ is given by
 \be \label{cvecB2}
 c^{\rm vector}= -\frac{( \del^2 + h^2 -1)^2}{4 h^2} \,.
 \ee
This dispersion law applies all along the second branch. Note that it depends neither on
the parameter $\xi$ (which is independent from $\del$ and $h$ along the second branch)
nor on the parameter $c_{35}$ (which enters the vectorial perturbation equations).

The cubic dispersion law \eqref{vecdispB2} has three roots:
$\sigma=0$ and $\sigma= \pm \sqrt{- c^{\rm vector}}$.
The vanishing root is a gauge mode which is already present in the flat space case
(see below), and which corresponds to the shift symmetry by the constant vector $C^a$ discussed above.
To have stability we would need to have only pure imaginary roots for  $\sigma= i \omega/k$.
This would require $c^{\rm vector} \geq 0$. However, we see that $-c^{\rm vector}$ is a square,
so that we have the two real roots
\be
\sigma=\pm \frac{( \del^2 + h^2 -1)}{2 h} \,.
\ee
These real roots correspond to {\it gradient instabilities} (in the helicity $+1$ sector). The same roots
are also present in the helicity $-1$ sector  (together with the gauge mode $\sigma=0$).

The only way to avoid these (strong) gradient instabilities would be to tune the parameters of TG
so that
\be \label{forvecstab}
 \del^2 + h^2=1 \, ; \, {\rm i.e.} \, f^2+g^2= \lam^2 \, ({\rm needed \, for\, stability})\,.
\ee
It is possible to tune $c_6$ and $c_5$ so as to satisfy the constraint \eqref{forvecstab}. Indeed, if
we impose
\be
c_5= - 16 \, c_6 \,,
\ee
Eqs. \eqref{c6}, \eqref{c5}, will imply the condition \eqref{forvecstab}. One can see that there
is a one-parameter family of such solutions, with $\del$ varying between $\frac12$ and $1$
(and $h^2$ correlatively varying between $\frac34$ and $0$). However, the problem is that
$\del$ being always $\leq 1$ along this family of tuned solutions, the tensor dispersion law
will necessarily be in the unstable region. Indeed, Fig. 1 (and the text around) showed that a necessary
condition for tensor stability along the second branch is $\delta > 1$. Note, in particular, that the self-accelerating solution itself
was found to require $\del > 1.78078$ (see Eq. \eqref{178}) and cannot even be tuned to reach the values
$\del \leq 1$ needed for vector stability. This shows the {\it necessary instability of the self-accelerating solution}.

 \subsection{Dispersion law for vector perturbations along the first branch of de Sitter-like solutions,
 and near the crossing between the two branches}

It is finally interesting to consider the limit where the second branch of  solutions approaches (and
crosses) the first (torsionless) branch (along which $\del=1$ and $h=0$). First, we note that the quantity
$c^{\rm vector}$ entering the dispersion law \eqref{vecdispB2} has a singular $\frac00$ structure
at its crossing with the first branch. If, however, we use the local expansion \eqref{expdel} 
for $\del$ around this crossing, we find that the roots in $\sigma$ behave as
\be
\sigma= \pm \frac{1-2a}{2}h + O(h^3) \,.
\ee
The limit at $h\to 0$ does exist and corresponds to the  dispersion law
\be
\sigma^3=0  \, ; \,{\rm as }\, \, h\to 0 \, {\rm ( second \, branch)}\,.
\ee
Two remarks are in order here. On the one hand, the marginally stable dispersion law
$\sigma^3 =0$ (i.e. $\omega^3=0$) turns into a gradient instability  as soon
as $h \neq0$, and, on the other hand, this differs from the dispersion law which holds
all along the first branch, i.e. when $h=0$ and $\del=1$.

We have computed the dispersion law along the first branch by using the same method as used
along the second branch. This requires a separate computation
because the presence of denominators $h^2$ in the second-branch dispersion law corresponds
to the fact that the elimination of the algebraic variables $\{\zeta_1(z),  \nu_1(z), \kappa_1(z), L_1(z) \}$
discussed above cannot be done in the same way. Indeed, one finds that the determinant
entering the solution for these four variables vanishes when $h\to 0$. In the case (first branch)
where $h=0$ from the start, one has to proceed slightly differently. One can, however, first eliminate
the three variables  $\{\zeta_1(z),  \nu_1(z), \kappa_1(z) \}$. This yields four equations for the
four unknowns $L_1, W_1 , A_3,\mu_3$. One then finds that, among the correspondingly reduced four
remaining equations, there is an equation which is algebraic in $L_1$. One can then eliminate $L_1$
as a second step. This yields three equations for $W_1 , A_3,\mu_3$. As before the latter system
can be written as a first-order system of the same form as \eqref{matrixsystem2vec}. Let us only cite here the
resulting dispersion law along the first branch. It is found to be
\be
 \sigma ( \sigma^2+1 )=0 \,.
 \ee
 Note that the latter dispersion law is: (i) stable (purely imaginary roots, apart from the gauge mode $\sigma=0$); 
 and (ii) coincides
 with the (high-frequency limit of the) flat spacetime one, i.e.
 \be \label{disph1B1}
 \omega (\omega^2- {\mathbf k}^2)=0\,.
 \ee
 The latter dispersion law is indeed the high-frequency limit of
  \be
 \omega (\omega^2- {\mathbf k}^2 -m_2^2)=0\,,
 \ee
(where $m_2$ denotes as above the mass of the massive spin-2 TG field) that describes, apart from the gauge
mode $\omega=0$, the helicity-$+1$ projection of the massive spin-2 excitation, 
which is (together with its helicity-$-1$ counterpart) the only physical, propagating vector mode. [There are no
 physical vectorial degrees of freedom in the pure helicity-$\pm2$ massless graviton.]
 The extra solution $\omega=0$ corresponds to the gauge-mode solution  parametrized
 by $C^a$, discussed above.
 
Let us emphasize that  the first-branch dispersion law \eqref{disph1B1} {\it does not coincide } with the $h \to 0$ limit of the dispersion law along the second branch.
 Indeed, $c^{\rm vector}_{B1} = -1$ along the first branch, while $\lim_{h\to 0} c^{\rm vector}_{B2} = 0$,
 as a limit along the second branch.

 This shows again that a torsionfull background (even an infinitesimal one)
 is a highly non trivial modification of the flat space
 dispersion laws in TG, which is prone to introducing instabilities that do not occur
 in torsionfree backgrounds.
 
\section{Study of scalar perturbations and of their stability} \label{sec9}

Finally, we come to the study of scalar perturbations and of their stability. We will be briefer than
for the other perturbations, both because their treatment is similar to what we explained above
for the tensor and vector perturbations, and because a detailed discussion of scalar perturbations 
has been recently given by one of us \cite{Nikiforova:2017saf,Nikiforova:2017xww}. We shall mainly comment
on the differences between the present treatment and the one given in the latter references.

\subsection{Deriving a linear system of four first-order ordinary differential equations from the ten scalar perturbations} 

As explained in Section \ref{sec5} above, there are ten scalar variables: two, $\Phi, \Psi$, parametrize
scalar perturbations of the vierbein, and eight, $\widetilde \xi, \chi,\sigma,\rho,\theta,Q,u,M$, parametrize scalar
perturbations of the connection. A difference with the treatment in Refs. \cite{Nikiforova:2017saf,Nikiforova:2017xww} 
is that the variables used there to parametrize perturbations were defined so as to parametrize the variations in the contorsion $K_{ij\mu}$ ($K$-parametrization), rather than in the connection $A_{ij\mu}$
($A$-parametrization).  This amounts
to a conceptually unimportant redefinition of variables, involving some mixing between vierbein and
connection variables. [We have checked that one gets the same final results using either 
the $K$-parametrization or the $A$-parametrization.]
The precise connection between the variables denoted by the same letters as here in 
Refs. \cite{Nikiforova:2017saf,Nikiforova:2017xww} and the variables used here is
\be \label{KvsA}
\sigma^A=\sigma^K\! - \d_\eta \Psi \,;\,  M^A=M^K\! - i\Psi \;;\, \widetilde{\xi}^A=\widetilde{\xi}^K\! - i\Phi \,.
\ee
As discussed in Refs. \cite{Nikiforova:2017saf,Nikiforova:2017xww}, the scalar projection of the 
perturbed field equations yields fourteen equations for the ten scalar variables. However, there are
four Bianchi-like identities between these fourteen equations. This leaves one with ten independent equations for the ten scalar unknowns. We give in Appendix C  the ten independent scalar equations we have used,
expressed in terms of our current $A$-parametrized variables, and in terms of rescaled parameters (and unknowns)
and of the variable $z=k \eta$. [We do not put a superscript $A$ on our variables.] 

Actually, following Refs. \cite{Nikiforova:2017saf,Nikiforova:2017xww}, we shall work with suitable
combinations of the equations in Appendix C and of their derivatives.  Indeed,  two (and only two) among the equations given in Appendix C contain second
derivatives of scalar variables (namely the eighth equation, which involves $\chi''$, and the ninth,
which involves $\rho''$). However, by suitable combinations of the equations and their derivatives
one can replace the latter two equations by two other equations involving (as the eight other ones)
only first derivatives of the scalar variables. At this stage, we have therefore a system of
ten equations which involve at most the first derivatives of the scalar variables.
In a second stage,  we can combine equations of the latter system so as to define two combinations that are
purely algebraic, i.e. that involve no derivatives. [The explicit expressions of these two algebraic
equations have been given in \cite{Nikiforova:2017saf}. By using the transformations  \eqref{KvsA} above, one
can reexpress them in terms of our $A$-parametrization variables.]
At this second stage, we have therefore eight equations,
 say $E_1, E_2, \cdots, E_8$, involving first derivatives and two algebraic equations, 
say $AE_1, AE_2$, involving no derivatives. We then found convenient to deal with this system of equations
in the following way (which differs both from the methods used above, and 
from the one used in Refs. \cite{Nikiforova:2017saf,Nikiforova:2017xww}). 

First, we redefine our scalar variables as follows, and separate them in two different groups,
denoted $y_I$ with $I=1,2,\cdots,6$ and  $y_A$, with $A=7, 8, 9, 10$:
\bea
\{ y_I \}_{I=1,\cdots,6}: y_1&=& \Psi \,; \, y_2= z^{-1} \chi \,; \, y_3= Q \,; \nonumber \\
\, y_4&=&u\,; \, y_5=M\,;  y_6 =\rho\,; 
\eea
where the $z$-dependent rescaling of $\chi$ is necessary to end up with a differential-coefficient
matrix $m_{ij}(z)$ having a finite limit when $z\to \infty$, and
\be
\{ y_A \}_{A=7,\cdots,10}: y_7=\sigma\,; \, y_8=\xi\,; \, y_9= \Phi\,; \, y_{10}=\theta \, .
\ee
This separation in two groups is linked to the following facts \cite{Nikiforova:2017saf}:
the variables of the second group enter equations $E_1, E_2, \cdots, E_8$ only algebraically,
and {\it do not enter} the two algebraic equations  $AE_1, AE_2$, which only involve the variables of the first group.

As a consequence of the latter facts, we can usefully consider the following system of {\it twelve equations}
\bea
E_1(y_A, y_I, y'_I) &=&0 \, ;\cdots ; \,E_8(y_A, y_I, y'_I)=0\, ; \nonumber \\
AE_1(y_I)&=&0 \, ;\, AE_2(y_I)=0 \, ;\, \nonumber \\
 \l\frac{d}{dz} AE_1\r (y_I, y'_I) &=&0 \, ;\,  \l \frac{d}{dz}AE_2\r(y_I, y'_I)=0 \, . \nonumber\\
\eea
These twelve equations define a system of  equations for the following twelve unknowns:
\be \label{12unk}
\{y'_1, y'_2, y'_3, y'_4, y'_5, y'_6, y_5, y_6, y_7, y_8, y_9, y_{10} \} \,.
\ee
We find that this linear system of twelve equations for twelve unknowns is uniquely solvable,
and thereby allows one to express the twelve quantities \eqref{12unk} in terms of the remaining
variables occurring in the system, namely $\{y_1, y_2, y_3, y_4\}$. In other words,
we have thereby not only succeeded in algebraically expressing $\{y_5, y_6, y_7, y_8, y_9, y_{10} \}$
in terms of $\{y_1, y_2, y_3, y_4\}$  (thereby showing that six scalar variables can be algebraically
eliminated), but, most importantly, we have also expressed the
derivatives $\{y'_1, y'_2, y'_3, y'_4 \}$ in terms of $\{y_1, y_2, y_3, y_4\}$.
In other words, we have obtained the following autonomous system of four first-order equations
for the four unknowns $\{y_1, y_2, y_3, y_4\}$:
\be \label{scalarmatrixsystem}
y'_i=m_{ij}(z) y_j \, ; \, {\rm where} \, i, j=1,2,3,4\,.
\ee

\subsection{Dispersion law for scalar perturbations}

The system \eqref{scalarmatrixsystem} is the analog of the tensor and vector systems  \eqref{matrixsystem1}, \eqref{matrixsystem1vec}.
Thanks to our redefinition of the variable $y_2$, we find that the $4\times4$ coefficient matrix $m_{ij}(z) $
entering this system has a finite limit at $z\to \infty$.
As before, we conclude that the dispersion law for scalar perturbations is given by the characteristic polynomial
of the $4\times4$ coefficient matrix $m^\infty_{ij} = \lim_{z\to \infty} m_{ij}(z)$, say
\be \label{cpscalar}
P_4^{\rm tensor}(\sigma) \equiv \det \l  m^\infty_{ij} - \sigma \delta_{ij} \r ;\,  i,j=1,2,3,4 \,.
 \ee
 The explicit computation of this scalar dispersion law is found to be
 \be \label{scalardispB2}
 P_4^{\rm scalar}(\sigma) = \sigma^2 \l\sigma^2 + c^{\rm scalar}(\del, h,c_{35})  \r\,,
 \ee
 i.e., in terms of $\omega$ and $k$,
 \be \label{scalardisplaw}
 \omega^2 \l  \omega^2 -   c^{\rm scalar} {\mathbf k}^2 \r=0 \,.
\ee
 The quantity $c^{\rm scalar}$ is found to be given by the complicated expression
 \be \label{cscalar}
 c^{\rm scalar}(\del, h, c_{35}) = \frac{N^{\rm scalar} }{D^{\rm scalar}}\,,
 \ee
 with
 \begin{align} \label{Nscalar}
 &N^{\rm scalar} = - \Big\{ 9 (1 - 2 \del)^2 h^2 (-5 + 4 \del + 4 \del^2 + 4 h^2) \nonumber \\
 &+ 
    3 c_{35} (-1 + 2 \del) \big[-3 \del^5 + 3 \del^6 + 17 h^2 + 2 h^4 - 3 h^6  \nonumber\\
    &+ 
       3 \del^4 (-2 + h^2) + \del^3 (6 + 26 h^2) \nonumber\\
       &+ 
       \del^2 (3 + 4 h^2 - 3 h^4) + \del (-3 - 38 h^2 + 29 h^4)\big] \nonumber\\
       &+ 
    2 c_{35}^2 \big[-13 \del^6 + 6 \del^7 + 21 \del^4 h^2 + \del^5 (7 + 6 h^2) \nonumber\\
    &+ 
       \del^3 (4 - 34 h^2 - 6 h^4) - h^2 (7 + 8 h^2 + h^4) \nonumber\\
       &+ 
       \del^2 (-7 - 8 h^2 + 33 h^4) + 
       \del (3 + 28 h^2 + 7 h^4 - 6 h^6)\big]\Big\} \,,
 \end{align}
 and
 \begin{align} \label{Dscalar}
 &D^{\rm scalar} =(-3 + 2 \del^2 + 2 h^2) \Big\{9 (1 - 2 \del)^2 h^2 \nonumber \\
 &+ 
     12 c_{35} (1 - 2 \del)^2 h^2 + 
     4 c_{35}^2 \big[-2 \del^3  \nonumber \\
     &+ \del^4 + h^2 - 2 \del h^2 + h^4 + 
        \del^2 (1 + 2 h^2)\big]\Big\} \,.
  \end{align}
  [As a numerical check on the above expressions, note that $c^{\rm scalar}(1,2,3)= -13/25$, corresponding to
   $N^{\rm scalar}=-3276$, and $D^{\rm scalar}=6300$.]
   
 Several remarks are in order concerning this scalar dispersion law.
 Let us first emphasize again that Eq. \eqref{scalardispB2} gives the scalar dispersion law all along
 the second branch of solutions (i.e. without assuming that $\xi$ is related to $\del$ and $h$).
 Therefore, it could a priori depend not only on the background-solution parameters $\del$ and $h$ but also
 on $\xi$ (as had happened for the tensor dispersion law, Eq. \eqref{bpdelxi}). However, it happens not to depend on $\xi$. [More precisely, one finds that, when expanding the
 characteristic polynomial of $m_{ij}(z) $ in inverse powers of $z$, the value of $\xi$ starts affecting the
 evolution of the scalar perturbations only at order $O(1/z^2)$.] 
 On the other hand, contrary to the previous (tensor and vector)
 dispersion laws, it depends on a TG parameter that did not enter the previous dispersion laws, namely
 $c_{35} \equiv c_3/c_5$. Let us remark that the scalar dispersion law derived here, Eq. \eqref{scalardispB2},
 has the same general structure as the result obtained in Ref. \cite{Nikiforova:2017xww}, which had studied,
 like here,  the case where the torsion background is comparable to the Hubble scale, i.e. the case where both
 $\del$ and $h$ are of order unity. [One cannot directly compare with the previous result of Ref. \cite{Nikiforova:2017saf}
 which had considered a parametrically different case.] However, the specific value of the constant $c^{\rm scalar}$
 derived here differs from the (simpler) value, namely $c^{\rm scalar}_{N}= 2 \del+1$, given there (see the
 $\Lambda \to 0$ limit of Eq. (33) in Ref. \cite{Nikiforova:2017saf}). A reexamination of the derivation in 
 Ref. \cite{Nikiforova:2017saf} has allowed us to locate a coding misprint. After correcting it, it was found that the
 method used in   Ref. \cite{Nikiforova:2017saf} leads to the value of $c^{\rm scalar}$ given in Eq. \eqref{cscalar} above. Note that this qualifies the conclusion of 
 Ref. \cite{Nikiforova:2017saf} that the scalar perturbations have no exponential instabilities. Actually,
 as we shall show next the quantity $c^{\rm scalar}(\del, h,c_{35})$ is positive in part of the
 parameter space (corresponding to stability), but negative in other regions (where there are
 gradient instabilities).
 
 Let us also emphasize a significant difference between the dispersion law  Eq. \eqref{scalardisplaw}
 and the dispersion law for scalar perturbations along torsionless backgrounds. As already mentioned
 in Sec. \ref{sec6bis}, the latter dispersion law is
 \be
  \l  \omega^2 - {\mathbf k}^2 \r  \l  \omega^2 - {\mathbf k}^2 \r=0\,,
 \ee
 where the two factors  $ \omega^2 - {\mathbf k}^2 $ describe the propagation of the two helicity-0
 d.o.f. (one being part of the massive spin-2 field, the other being a pseudo-scalar torsion-related field).
 By comparing with  Eq. \eqref{scalardisplaw}, we see that while two modes that propagated at the
 speed of light around a torsionless background now propagate at the modified velocity 
 $\sqrt{c^{\rm scalar}}$ (when it is real), two other modes that propagated at the velocity of light
 now propagate with zero velocity (factor  $\omega^2$ instead of  $\omega^2 - {\mathbf k}^2$).
 This shows again the drastic effect of having a torsionfull background.
 
 Let us emphasize that the dispersion constant $c^{\rm scalar}(\del, h,c_{35})$ 
 has the same $\frac00$ structure near
 torsionfree backgrounds that we found above for tensor and vector dispersion laws. Indeed, it is easily checked
 that, when $\del\to 1$ and $h \to 0$ both the numerator $ N^{\rm scalar} $ and the denominator
  $D^{\rm scalar} $ tend to zero. However, if we consider the specific limit \eqref{expdel} corresponding to
  approaching the torsionfree case along the second branch, we find the following limiting behavior
  \be \label{expcscalar}
  c^{\rm scalar}(\del, h,c_{35}) = 3 +  2  \frac{15 - 36 a + 16 c_{35}  -18 a c_{35}}{3 + 2 c_{35}} h^2+O(h^4)
  \ee
We therefore have tensor stability ($c^{\rm scalar}>0$, see next subsection) 
 in the vicinity of torsionfree backgrounds. However, we do not have a continuous behavior of the 
 helicity-0 dispersion law in the vicinity of torsionfree backgrounds. Indeed, the limit of the torsionfull
 dispersion law \eqref{scalardisplaw} as $\del\to 1$ and $h \to 0$ is 
 $\omega^2 \l  \omega^2 -   3 {\mathbf k}^2 \r=0 $, instead of 
 $ \l  \omega^2 - {\mathbf k}^2 \r  \l  \omega^2 - {\mathbf k}^2 \r=0$.

 \subsection{Regions of parameter space where scalar perturbations (along the second branch) are stable}
 
 In view of the dispersion law \eqref{scalardisplaw}, the condition for the stability of scalar perturbations is
 \be
 c^{\rm scalar}(\del, h,c_{35}) >0 \,.
 \ee

\begin{figure}
\includegraphics[scale=0.5]{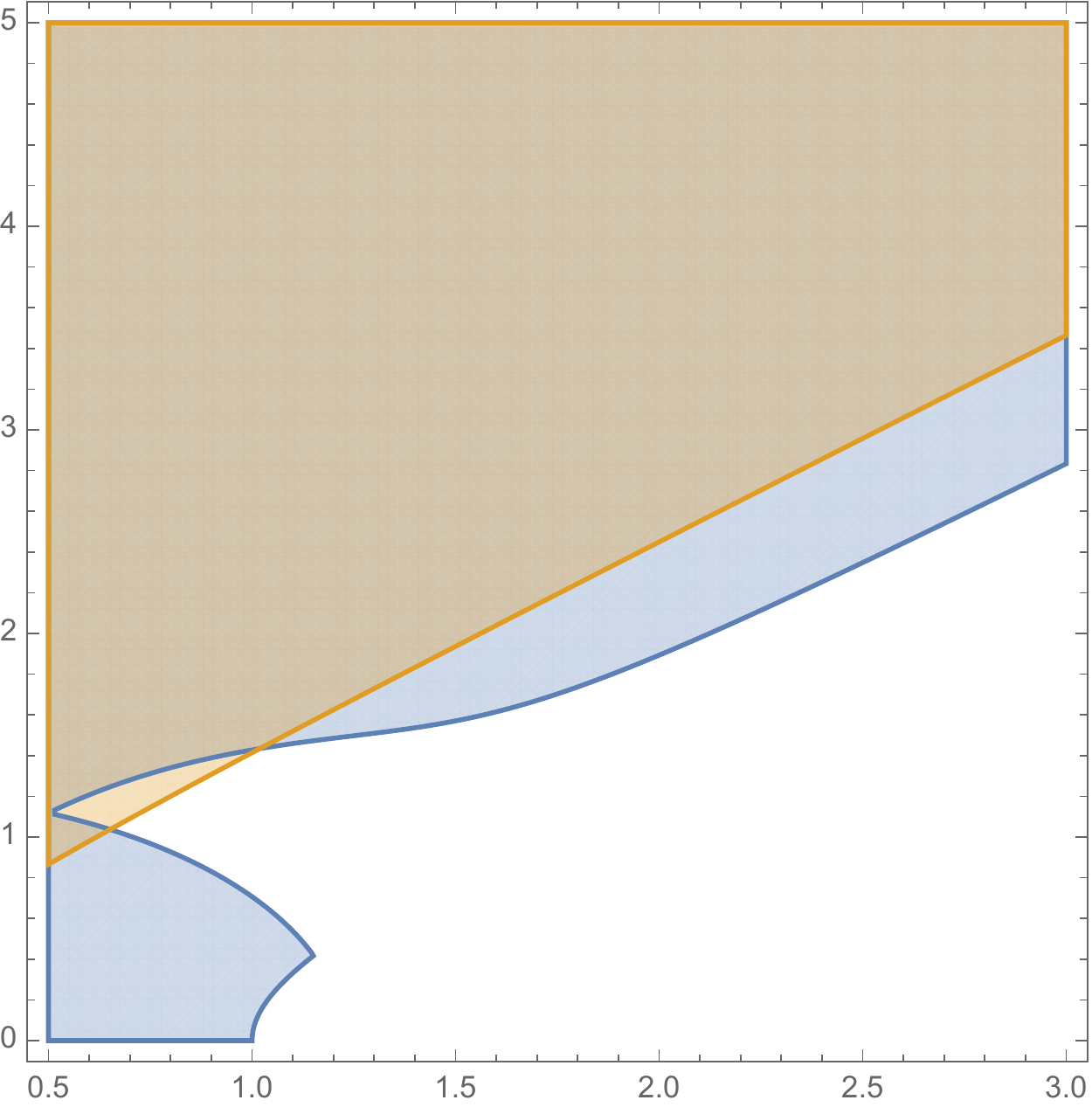}
\caption{\label{fig2}
Scalar stability regions (for $c_{35}=-2$) in the $(x,y)=(\delta, h)$ plane (for $h>0$); 
both for branch 2 (blue region on line),  and for the self-accelerating solution (part
of the previous region above the line $h^2=\del^2+\del$, shown as brown 
on line). }
\end{figure}

As we found above that there were necessary instabilities  in the vector sector, it is not worth discussing in full detail
 which regions of parameter space lead to stability in the scalar sector. Let us only say that, for each
 given  value of $c_{35}$, there is an infinite region of the $\del, h$ plane where there are no exponential
 scalar instabilities. The shape of the stable region remains qualitatively similar, though
 it undergoes very significant quantitative changes,  as $c_{35}$ varies. Let us only give one specific example of stability region
 in the $\del, h$ plane, namely the one corresponding to the specific case $c_{35} = - 2$.
The corresponding scalar stability region in the $\del,h$ plane (for the second branch)
  is shown (blue on line)  in Fig. \ref{fig2}. When restricting our attention to the self-accelerating
 solution, we must further restrict the parameters by the inequality $h^2 > \del^2 +\del$, Eq. \eqref{SAlowerbound}.
Restricting to the region above the hyperbola $h^2 - \del^2 -\del=0$ (displayed as brown on line) cuts off part of the
 previous stability region, as shown in Fig. \ref{fig2}.

\section{Conclusions} \label{sec10}

We have studied linearized perturbations of de Sitter-like solutions in a class of geometric theories
(called Torsion Gravity, or TG) that generalize General Relativity by including a propagating torsion in
addition to the usual Einsteinian metric. The class of TG theories we considered contains six parameters,
including a vacuum-energy parameter $c_2$. When considered either on a flat background (when taking
$c_2=0$), or on a torsionfree de Sitter background (when $c_2\neq 0$), these theories were found 
in previous work to be ghost-free and tachyon-free, and to feature, as propagating fields, both
massive spin-2 and massive spin-0 excitations, in addition to the Einsteinian massless spin-2 field.

We considered the two different branches of de Sitter-like solutions admitted by this class of TG theories.
The exponential expansion of the first branch of solutions (which had already been studied in \cite{Nair:2008yh}) 
is sustained by the vacuum energy $c_2$, and these solutions have zero torsion. On the other hand,
the second branch of solutions has a non-zero torsion, and its exponential expansion is jointly sustained by
its torsion background and by the vacuum-energy parameter $c_2$. This second branch can interpolate
between a torsion-free background, and a self-accelerating torsionfull solution \cite{Nikiforova:2016ngy} whose expansion
is entirely sustained by the torsion background (with $c_2=0$). We contrasted the properties of the cosmological
perturbations in these two branches of solutions as a way to understand the influence of a torsion
background on the stability of the de Sitter-like solutions.

The main new finding of the present paper is that the presence of a torsion background (and
even an infinitesimal one) generically creates 
gradient instabilities in the vector sector of the cosmological perturbations. See Eqs.  \eqref{vecdispB2},
\eqref{cvecB2}. No tuning of the TG parameters can kill these instabilities without creating instabilities
in the other sectors. We have also studied the cosmological perturbations in the tensor and scalar sectors.
The tensor sector has no exponential instabilities if the quantity $b'_{\rm tensor}-1$, Eq. \eqref{bpdelxi}
is positive. We found that this will be the case in a large region of the TG parameter space. Along a general solution
of the second branch the tensor-stability region in the plane of $\del=- f/\lam$ and $h=g/\lam$ depends
also on the theory parameter $\xi=\bal/\tal$. It is illustrated in Fig. \ref{fig1} for the value $\xi=10$. The 
tensor-stability region for the self-accelerating solution (with $c_2=0$) covers also a large
part of the $\del, h$ plane. See Eq. \eqref{SAtensorstab} and  Fig. \ref{fig1}. We finally discussed the stability
of scalar perturbations. Like tensor perturbations, scalar perturbations were found to have no exponential
instabilities in a large part of parameter space, defined by $c^{\rm scalar} >0$ with $c^{\rm scalar}$ defined
in Eqs. \eqref{cscalar}, \eqref{Nscalar}, \eqref{Dscalar}. The scalar-stability region depends on three parameters:
$\del, h$ and $c_{35}$. It is illustrated (both for the general second branch and for
the self-accelerated solution) in the specific case $c_{35}=- 2$ in Fig. \ref{fig2}.

In the course of our study we compared and contrasted the (linearized) (in)stability of torsionfull de Sitter-like solutions
to the (linearized) stability \cite{Nair:2008yh,Nikiforova:2009qr} of torsionfree solutions. This comparison showed the drastic effect on stability of having a nonzero 
torsion background. While torsionfree backgrounds exhibited (in the high-frequency limit) the same dispersion
laws as flat spacetime, the presence of even an infinitesimal torsion background drastically affected the dispersion
law by introducing, in the dispersion laws, quantities having a $\frac00$ structure in the vanishing torsion limit.
This structure is connected with some discontinuities (as the torsion goes to zero) in the dispersion laws. 
Only the tensor dispersion law has no
discontinuity between a torsionfree background (first branch) and a torsionfull one with infinitesimal torsion.
The common limit of the (high-frequency) dispersion law is $(\omega^2 - {\mathbf k}^2)^2=0$.
By contrast, while the torsionfree dispersion law for vector perturbations is $\omega (\omega^2- {\mathbf k}^2)=0$,
the corresponding vector dispersion law around an infinitesimal torsion background becomes $\omega^3=0$.
Similarly, while the torsionfree dispersion law for scalar perturbations is $(\omega^2- {\mathbf k}^2)^2=0$,
the corresponding scalar dispersion law around an infinitesimal torsion background becomes 
$\omega^2 (\omega^2- 3 {\mathbf k}^2)=0$.
These discontinuous differences with the dispersion laws  for linearized perturbations around a torsion-free background
are linked with the presence of instability-related $\frac00$ structures in the perturbations around {\it both} torsionfull and
torsionless backgrounds. Indeed, at the crossing between the two branches, the
quantity $1+ 4  \widehat c_5$ vanishes, see Eq. \eqref{eq8.35}, and this
corresponds to the vanishing of the quantity $c_5 \kappa$ that is
crucially related to the stability of the torsionless branch (see notably Eqs. (9), (11), (22), and (30) 
in Ref. \cite{Nikiforova:2009qr}).

The overall conclusion of the present work is that a torsionfull background  is a highly non
trivial modification of the propagation properties of perturbations in TG. Further work is needed to
see if generic torsionfull backgrounds are prone to introducing instabilities, or if the instabilities we found
are mainly due to the (generalized) self-accelerating nature of the special de Sitter backgrounds
we considered. There is, however, the good news
that in all cases, and in spite of significant modifications in the algebraic structure of the equations 
for perturbations, the presence of a torsion background left intact the number of propagating degrees of freedom.
In other words, TG seems to be insensitive to the Boulware-Deser-type instability phenomenon \cite{Boulware:1973my}
that generally affects theories containing a massive spin-2 excitation.

\section*{Acknowledgments} 
The authors are grateful to Valery Rubakov for many fruitful suggestions,  to Cedric Deffayet for insightful conversations,
and to Donato Bini for several helpful inputs.
T. D. thanks Alain Connes, Lavinia Heisenberg and Maxim Kontsevich for informative discussions.
V. N. thanks Guillaume Faye, Chris Kavanagh and Barry Wardell for their responsiveness in answering
her questions about the xAct package (which was instrumental  for the computations performed
in the present work).   
V. N. thanks the Institut des Hautes Etudes Scientifiques (IHES) for a visit that  initiated the 
collaboration on the present project.
The work of V. N. has been supported by the Russian Science Foundation Grant No. 14-22-00161.

\appendix

\section{Tensor perturbation equations}

As explained in Sec. \ref{sec5} above, there are three tensor perturbations: $\pi_{ab}$ in the
 vierbein perturbation, and $\tau_{ab}$ and $N_{ab}$ in the connection perturbation.   
  To get equations for the coupled propagation of the three tensor excitations $\pi_{ab}$, $\tau_{ab}$ and $N_{ab}$,
  we must consider which equations among the perturbations of the full (rescaled) field equations (i.e. the
  $O(\gamma)$ contributions in $\widehat{{\cal G}}_{ij}$, Eq. \eqref{rescaledGeq}, and 
  $\widehat{{\cal T}}_{[ij]k}$, Eq.\eqref{rescaledAeq}) contain such
  tensor contributions. Denoting spatial indices as $a,b,c,d=1,2,3$, one easily sees that there are only three ``two-index" 
  rescaled equations, namely  $ \widehat{{\cal G}}_{ab} $, $\widehat{{\cal T}}_{0ab}$ and the dualized version of
$\widehat{{\cal T}}_{abc}$, i.e. $ \widehat{{\cal T}}^*_{dc} \equiv \varepsilon_{abd}\widehat{{\cal T}}_{abc}$, which
can provide  three tensor equations for our three 
tensor unknowns $\pi_{ab}$, $\tau_{ab}$ and $N_{ab}$. 

The next step is to decompose these tensor equations into their helicity $\pm 2$ components.
When doing so, we indeed found that the latter three equations do contain helicity $\pm 2$ irreducible pieces constraining
the propagation of the three tensor excitations. 
We recall that the helicity $\pm 2$ components of any  two-index tensor are defined so that, for instance, 
 $\pi_{ab}= \pi_{(+2)} e_+^{ab} + \pi_{(-2)} e_-^{ab}$ (where we defined $e_+^{ab}\equiv e_+^{a} e_+^{b}$), or 
 $\widehat{{\cal G}}_{ab} = \widehat{{\cal G}}_{(+2)} e_+^{ab} + \widehat{{\cal G}}_{(-2)} e_-^{ab} + {\rm lower \, helicities}$. [A non-symmetric two-index tensor, such as $\widehat{{\cal G}}_{ab}$, does not even need to
 be explicitly symmetrized over $ab$ to be decomposed into helicity-$\pm2$ components because its
 antisymmetric part, being dual to a vector, will not contain any helicity-$\pm2$ piece.]
 
 We extracted the helicity $\pm 2$ components of the expressions $ \widehat{{\cal G}}_{(ab)} $, $\widehat{{\cal T}}_{0(ab)}$ and $ \widehat{{\cal T}}^*_{(ab)}$ by inserting the helicity decompositions
 of $\pi_{ab}$, $\tau_{ab}$ and $N_{ab}$ in these expressions. Then, the use of coding rules to express the pure-helicity 
 character of $e_\pm^{ab}$ allows one to extract the coefficients of $e_\pm^{ab}$ in the equations. An alternative (equivalent) way
would have been to contract the live tensor indices of the above mentioned
 two-index tensor equations by $\frac14 e_-^{ab}$ (to get the equations for the 
helicity-$+2$ components), or by $\frac14 e_+^{ab}$ (to get the equations for the 
helicity-$+2$ components). We use here the fact that $e_+^a$ and $e_-^a$ are null vectors
($e_+^a e_+^a=0$, $e_-^a e_-^a=0$) having as only nonzero scalar product 
among $e_+^a$, $ e_-^a$ and $k^a$ their mutual product: $e_+^a e_-^a=2$.  

After this helicity decomposition, we get, on the one hand, three coupled equations for the helicity $+2$ variables,
  $\pi_{(+2)}$, $\tau_{(+2)}$ and $N_{(+2)}$, and, on the other hand, three coupled equations for the helicity $-2$ variables,
  $\pi_{(-2)}$, $\tau_{(-2)}$ and $N_{(-2)}$. The two types of helicity do not mix between themselves,
  and the helicity $-2$ system is seen to be obtained from the helicity $+2$ one simply by changing
  $g \to -g$. The two latter facts are simple consequences of the structure of the field
  equations. In particular, as $g$ enters the background in the combination $g \varepsilon_{abc}$,
  the sign flip  $g \to -g$ does correspond to a different choice for the orientation of space, which indeed
  reflects as an exchange between the two helicities, see Eqs. \eqref{kxe}.

Denoting (as in our code)  $\pi_{(+2)}$ by $K_1$,  $\tau_{(+2)}$ by $F_1$, and  $N_{(+2)}$ by $N_1$,
and introducing the rescaled variables, $F_n\equiv k^{-1} \tau_{(+2)}$, and  $i\, N_n \equiv N_{(+2)}$,
we found that the first-order ($O(\gamma)$) perturbations of, respectively,  
$\widehat{{\cal T}}_{0ab}$, $ \widehat{{\cal T}}^*_{dc}$ and $ \widehat{{\cal G}}_{ab} $ yield,
using the above-defined extraction procedure, three (rescaled) equations, respectively denoted $\widehat E_1=0$,
$\widehat E_3=0$ and $\widehat E_5=0$, for $K_1$, $F_n$ and $N_n$, with

\begin{widetext}
\begin{align}
& \widehat E_1\equiv (-\frac{3\, z}{2} - 3\, \hcf \delta \,z - 3\, \hcf \delta^2 z)\, F_n    \nonumber  \\
    &+ 
  (\frac{3\, \delta}{2} - 6\, \hcf \delta + 3\, \hcf \delta^2 + 3\, \hcf \delta^3 + 6\, \hcf h^2 + 
    48\, \hcs h^2 - 3\, \hcf \delta h^2 + 96\, \hcs \delta h^2 + 48\, \hcs h\,z + 
    96\, \hcs \delta h\,z)\,K_1     \nonumber  \\
    &+ 
  (3\, \hcf h\,z + 48\, \hcs h\,z - 3\, \hcf \delta h\,z + 96\, \hcs \delta h\,z + 
    3\, \hcf \delta \,z^2)\,N_n     \nonumber  \\
    & + (- \frac{3\, z }{2} + 
 3\, \hcf \delta \,z  - 
 3\, \hcf \delta^2 z  + 
 3\, \hcf h^2 z  )\,\d_zK_1 +( - 
 3\, \hcf h\,z^2  + 
 3\, \hcf z^3)\,\d_zN_n + 
 3\, \hcf z^3\, \d^2_zF_n\;, 
    \end{align}
    \begin{align}
    &\widehat E_3\equiv (-3\, h - 18\, \hcf \delta h + 96\, \hcs \delta h - 6\, \hcf \delta^2 h + 
    192\, \hcs \delta^2 h + 6\, \hcf h^3 - 3\, z - 6\, \hcf \delta \,z + 6\, \hcf \delta^2 z - 
    6\, \hcf h^2 z)\,K_1    \nonumber  \\
    & + 
  (-96\, \hcs h\,z + 6\, \hcf \delta h\,z - 192\, \hcs \delta h\,z - 6\, \hcf \delta \,z^2)\,F_n + 
  (-3\, z - 12\, \hcf \delta \,z + 6\, \hcf h^2 z - 12\, \hcf h\,z^2 + 6\, \hcf z^3)\,N_n     \nonumber  \\
    &+(- 
 6\, \hcf h\,z^2  + 
 6\, \hcf z^3 )\,\d_zF_n - ( 
 96\, \hcs h\,z  + 
 192\, \hcs \delta h\,z )\,\d_zK_1\;,
    \end{align}
    \begin{align} 
 &\widehat E_5 \equiv  (\frac{3\, \delta \,z}{2} + 3\, \hcf \delta^2 z - 3\, \hcf \delta^3 z + 48\, \hcs h^2 z + 
    3\, \hcf \delta h^2 z + 96\, \hcs \delta h^2 z - 48\, \hcs h\,z^2 - 
    96\, \hcs \delta h\,z^2)\,F_n     \nonumber  \\
    &+ 
  (\frac{3\, h\,z}{2} + 3\, \hcf \delta h\,z - 48\, \hcs \delta h\,z - 3\, \hcf \delta^2 h\,z - 
    96\, \hcs \delta^2 h\,z + 3\, \hcf h^3 z - \frac{3\, z^2}{2} - 3\, \hcf \delta \,z^2 + 
    3\, \hcf \delta^2 z^2 - 3\, \hcf h^2 z^2)\,N_n     \nonumber  \\
    &+ 
  (-\frac{3\, \delta}{2} - \frac{3\, \delta^2}{2} + 3\, \hcf \delta^2 - 6\, \hcf \delta^3 + 
    3\, \hcf \delta^4 + \frac{3\, h^2}{2} - 48\, \hcs h^2 + 6\, \hcf \delta h^2 - 
    192 \hcs \delta h^2 - 6\, \hcf \delta^2 h^2    \nonumber  \\
    & - 192\, \hcs \delta^2 h^2 + 3\, \hcf h^4 + \frac{
    3\, z^2 \xi}{2})\,K_1 +( - \frac{3}{2} z^2  + 
 3\, \hcf \delta \,z^2  - 
 3\, \hcf \delta^2 z^2  + 
 3\, \hcf h^2 z^2)\, \d_zF_n     \nonumber  \\
    &- 
 3\, z \xi \d_zK_1 + 
 (48\, \hcs h\,z^2  + 
 96\, \hcs \delta h\,z^2 )\,\d_zN_n + 
 \frac{3}{2} z^2 \xi \,\d^2_zK_1   \;.
\end{align}
\end{widetext}


\section{Vector perturbation equations}

We extracted seven vector perturbations equations for the various helicity-$+1$ components 
$W_1$  ( $\equiv$ the helicity $+1$ component $W_{(1)}$ of $W_a$), 
$\zeta_1$, $\mu_1$, $\nu_1$, $\kappa_1$, $A_1$, and $L_1$ of the vector perturbations
in the following way. As in the tensor case, we start from the $O(\gamma)$ pieces in the rescaled field
equations $\widehat{{\cal T}}_{[ij]k}$, Eq.\eqref{rescaledAeq}, and $\widehat{{\cal G}}_{ij}$, Eq. \eqref{rescaledGeq}.
We use the same extraction procedure as above, except that we decompose now any vectorial equation
into, e.g. $ \widehat{{\cal G}}_{a0} = \widehat{{\cal G}}_{(+1)} e_+^a + \widehat{{\cal G}}_{(-1)} e_-^a$.
Equations containing more than one index are conveniently dualized into vectorial equations.
We so derived four helicity-$+1$ equations (denoted $V_1$ to $V_4$) from the connection field equations,
and three other equations (denoted $V_5$ to $V_7$) from the gravity field equations.
In these equations, we have everywhere eliminated the TG basic coefficient $c_4$ by using 
the constraint $c_4= -c_3 - 3c_5$. In addition, several vector variables 
have been rescaled as follows: $W_1^{\rm new}= k W_1; A_1^{\rm new}== k  A_1;
\zeta_1^{\rm new}= k^{-1} \zeta_1; L_1^{\rm new}= k^{-1} L_1$.  We henceforth skip the superscripts ``new".

The explicit forms of the left-hand sides of the seven helicity-$+1$ equations $V_1=0$ to $V_7=0$ 
are listed below (with the field equation from which they were extracted)
  
\begin{widetext}
Equation $\widehat{{\cal T}}_{[a0]0} \to V_1$:
\begin{align}
&(\frac{3 z}{2} + 2  \hcthree h^2 z - 3  \hcf h^2 z -  \hcthree h z^2 - 
 2  \hcthree z\,\delta^2)A_1 + 
  (3 z + 4 \hcthree h^2 z - 6 \hcf h^2 z - \hcthree z^3 - 4 \hcthree z\,\delta^2) L_1    \nonumber  \\
 &+ 
  (-\frac{3 z}{2} + 3 \hcf h^2 z + 3 \hcf z\,\delta - 
    3 \hcf z\,\delta^2)W_1 + (4 \hcthree h^2 z + 6 \hcf h^2 z - 4 \hcthree h z^2 - 
    6 \hcf h z^2 + \hcthree z^3 - 
    4 \hcthree z\,\delta^2) \zeta_1     \nonumber  \\
 & + (8 i\,  \hcthree h z\,\delta + 
    6 i\,  \hcf h z\,\delta - 4 i\, \hcthree z^2 \delta - 
    3 i\, \hcf z^2 \delta) \kappa_1 + (48 i\, \hcs h z + 
    4 i\, \hcthree h z\,\delta + 9 i\, \hcf h z\,\delta + 96 i\, \hcs h z\,\delta -
     i\, \hcthree z^2 \delta) \mu_1      \nonumber  \\
 &+ (-48 i\, \hcs h z - 
    4 i\, \hcthree h z\,\delta - 9 i\, \hcf h z\,\delta - 96 i\, \hcs h z\,\delta -
     i\, \hcthree z^2 \delta - 3 i\, \hcf z^2 \delta) \nu_1 + 
 2 \hcthree z^2 \delta \,\d_zA_1      \nonumber  \\
 &+ 
 4 \hcthree z^2 \delta \,\d_z
  L_1 + (-2 i\, \hcthree h z^2 - 3 i\, \hcf h z^2 + 
    i\, \hcthree z^3) \,\d_z\mu_1 + (2 i\, \hcthree h z^2 + 3 i\, \hcf h z^2 - i\, \hcthree z^3 - 
    3 i\, \hcf z^3) \,\d_z\nu_1 \;.
\end{align}
    
Equation $\varepsilon^{abc}\widehat{{\cal T}}_{[ab]0} \to V_2$:
\begin{align}
&z (96 \hcs h + 192 \hcs h \delta)W_1  + 
  z (-192 \hcs h - 16 \hcthree h \delta - 36 \hcf h \delta - 
    384 \hcs h \delta + 6 \hcf z\,\delta)L_1\nonumber \\
 & + 
 z (-96 \hcs h - 8 \hcthree h \delta - 18 \hcf h \delta - 
    192 \hcs h \delta + 2 \hcthree z\,\delta + 6 \hcf z\,\delta)A_1 + 
 z (-16 \hcthree h \delta - 12 \hcf h \delta + 8 \hcthree z\,\delta + 
    6 \hcf z\,\delta) \zeta_1 \nonumber \\
 &+ 
 z (8 i\, \hcthree h^2 - 8 i\, \hcthree h z + 2 i\, \hcthree z^2 - 8 i\, \hcthree \delta^2 - 
    12 i\, \hcf \delta^2) \kappa_1 + 
 z (3 i\, + 4 i\, \hcthree h^2 - 2 i\, \hcthree h z - 6 i\, \hcf h z - 4 i\, \hcthree \delta^2 + 
    6 i\, \hcf \delta^2) \mu_1\nonumber \\
 & + 
 z (-3 i\, - 4 i\, \hcthree h^2 - 2 i\, \hcthree h z - 6 i\, \hcf h z + 2 i\, \hcthree z^2 + 
    6 i\, \hcf z^2 + 4 i\, \hcthree \delta^2 - 6 i\, \hcf \delta^2) \nu_1 - 
 2 \hcthree z^2 (-2 h + z) \,\d_zA_1 \nonumber \\
 &- 
 4 \hcthree z^2 (-2 h + z) \,\d_zL_1 + 
 z (4 i\, \hcthree z\,\delta + 6 i\, \hcf z\,\delta) \,\d_z\mu_1 + 
 z (-4 i\, \hcthree z\,\delta - 6 i\, \hcf z\,\delta) \,\d_z\nu_1  \;.
\end{align}
  
Equation $\widehat{\cal T}^{*a} \equiv \widehat{\cal T}_{0[ab]}\varepsilon^{abc} \to V_3$:
\begin{align}
&(-2 \hcthree h z - 3 \hcf h z - 48 \hcs h z + 4 \hcthree h z\,\delta + 
 3 \hcf h z\,\delta - 96 \hcs h z\,\delta +  \hcthree z^2 \delta  )A_1    \nonumber  \\
 &+ 
  (48 \hcs h z + 96 \hcs h z\,\delta)W_1+ 
  (-4 \hcthree h z - 6 \hcf h z - 96 \hcs h z + 8 \hcthree h z\,\delta + 
    6 \hcf h z\,\delta - 192 \hcs h z\,\delta + 4 \hcthree z^2 \delta + 
    3 \hcf z^2 \delta)L_1      \nonumber  \\
 &+ (-4 \hcthree h z - 6 \hcf h z + 96 \hcs h z + 
    8 \hcthree h z\,\delta + 18 \hcf h z\,\delta + 192 \hcs h z\,\delta + 
    3 \hcf z^2 \delta) \zeta_1     \nonumber  \\
 & + (-3 i\, z - 4 i\, \hcthree h^2 z + i\, \hcthree z^3 + 
    3 i\, \hcf z^3 - 4 i\, \hcthree z\,\delta - 6 i\, \hcf z\,\delta + 
    4 i\, \hcthree z\,\delta^2 - 6 i\, \hcf z\,\delta^2) \kappa_1      \nonumber  \\
 &+( \frac{3 i z}{
    2} - 2 i\, \hcthree h^2 z - i\, \hcthree h z^2 - 2 i\, \hcthree z\,\delta + 
    3 i\, \hcf z\,\delta + 2 i\, \hcthree z\,\delta^2 + 
    3 i\, \hcf z\,\delta^2) \mu_1      \nonumber  \\
 &+ (-\frac{3 i\, z}{2} + 2 i\, \hcthree h^2 z + 
    3 i\, \hcthree h z^2 + i\, \hcthree z^3 + 2 i\, \hcthree z\,\delta - 3 i\, \hcf z\,\delta - 
    2 i\, \hcthree z\,\delta^2 - 3 i\, \hcf z\,\delta^2) \nu_1      \nonumber  \\
 &+ (-3 \hcf h z^2 -
     \hcthree z^3 - 3 \hcf z^3) \,\d_z
  A_1 + (-6 \hcf h z^2 - 3 \hcf z^3) \,\d_z
  L_1 + (4 \hcthree h z^2 + 6 \hcf h z^2 - 2 \hcthree z^3 - 
    3 \hcf z^3) \,\d_z\zeta_1      \nonumber  \\
 &+ (4 i\, \hcthree z^2 \delta + 
    6 i\, \hcf z^2 \delta) \,\d_z\kappa_1 + (-2 i\, \hcthree z^3 - 3 i\, \hcf z^3) \,\d^2_z\mu_1 + (2 i\, \hcthree z^3 + 3 i\, \hcf z^3) \,\d^2_z\nu_1  \;.
\end{align}
 
Equation $\widehat{{\cal T}}^{**}_e\equiv \varepsilon^{cde}\varepsilon^{abc}\widehat{{\cal T}}_{[ab]d}  \to V_4$:
\begin{align}
& z (6 - 8 \hcthree h^2 - 12 \hcf h^2 - 8 \hcthree h z - 12 \hcf h z - 2 \hcthree z^2 - 
    8 \hcthree \delta + 8 \hcthree \delta^2)L_1 \nonumber \\
 &+ 
 z (3  - 4 \hcthree h^2 - 6 \hcf h^2 - 2 \hcthree h z - 
    4 \hcthree \delta + 4 \hcthree \delta^2)A_1 + 
  z (-3 - 6 \hcf h^2 - 6 \hcf \delta + 6 \hcf \delta^2)W_1 \nonumber \\
 &+ 
 z (-6 - 8 \hcthree h^2 + 12 \hcf h^2 + 2 \hcthree z^2 - 8 \hcthree \delta + 
    8 \hcthree \delta^2) \zeta_1 \nonumber \\
 &+ 
 z (8 i\, \hcthree h - 192 i\, \hcs h - 16 i\, \hcthree h \delta - 36 i\, \hcf h \delta - 
    384 i\, \hcs h \delta + 6 i\, \hcf z\,\delta) \kappa_1 \nonumber \\
 &+ 
 z (4 i\, \hcthree h + 12 i\, \hcf h + 96 i\, \hcs h - 8 i\, \hcthree h \delta - 
    6 i\, \hcf h \delta + 192 i\, \hcs h \delta - 
    2 i\, \hcthree z\,\delta) \mu_1 \nonumber \\
 &+ 
 z (-4 i\, \hcthree h - 12 i\, \hcf h - 96 i\, \hcs h + 8 i\, \hcthree h \delta + 
    6 i\, \hcf h \delta - 192 i\, \hcs h \delta + 6 i\, \hcthree z\,\delta + 
    6 i\, \hcf z\,\delta) \nu_1 + 
 8 \hcthree z^2 \delta \,\d_z\zeta_1 \nonumber \\
 &+ 
 4 i\, \hcthree z^2 (-2 h + z) \,\d_z\kappa_1 + 
 z (-6 i\, \hcf h z + 2 i\, \hcthree z^2) \,\d_z\mu_1 + 
 z (6 i\, \hcf h z + 2 i\, \hcthree z^2 + 6 i\, \hcf z^2) \,\d_z\nu_1 \nonumber \\
 &- 4 \hcthree z^3 \,\d^2_zA_1 - 
 8 \hcthree z^3 \,\d^2_zL_1   \;.
\end{align}
 
Equation $\widehat{{\cal G}}_{a0} \to V_5$:
\begin{align}
&(2 \hcthree h^2 z + 3 \hcf h^2 z - 2 \hcthree z\,\delta^2)A_1 + 
  (4 \hcthree h^2 z + 6 \hcf h^2 z + 2 \hcthree h z^2 + 3 \hcf h z^2 - 
    4 \hcthree z\,\delta^2)L_1      \nonumber  \\
 &+ (4 \hcthree h^2 z + 6 \hcf h^2 z + 96 \hcs h^2 z - 
    2 \hcthree h z^2 - 3 \hcf h z^2 - 48 \hcs h z^2 + 3 z\,\delta + 
    192 \hcs h^2 z\,\delta - 96 \hcs h z^2 \delta - 
    4 \hcthree z\,\delta^2) \zeta_1     \nonumber  \\
 & + (-3 i\, h z + \frac{3 i\, z^2}{2} + 
    8 i\, \hcthree h z\,\delta + 6 i\, \hcf h z\,\delta + 
    96 i\, \hcs h z\,\delta - 2 i\, \hcthree z^2 \delta + 
    192 i\, \hcs h z\,\delta^2) \kappa_1      \nonumber  \\
 &+ (4 i\, \hcthree h z\,\delta + 
    3 i\, \hcf h z\,\delta) \mu_1 + (-4 i\, \hcthree h z\,\delta - 
    3 i\, \hcf h z\,\delta - 
    2 i\, \hcthree z^2 \delta) \nu_1      \nonumber  \\
 &+ (-\frac{3 z^2}{2} + 
    2 \hcthree z^2 \delta) \,\d_z
  A_1 + (-3 z^2 + 4 \hcthree z^2 \delta) \,\d_zL_1 - 
 \frac{3}{2} z^2 \xi\, \,\d_z
  W_1      \nonumber  \\
 &+ (-2 i\, \hcthree h z^2 - 3 i\, \hcf h z^2 - 48 i\, \hcs h z^2 - 
    96 i\, \hcs h z^2 \delta) \,\d_z\mu_1      \nonumber  \\
 &+ (2 i\, \hcthree h z^2 + 3 i\, \hcf h z^2 + 48 i\, \hcs h z^2 + 
    96 i\, \hcs h z^2 \delta) \,\d_z\nu_1   \;.
\end{align}

Equation $\widehat{{\cal G}}_{0a} \to V_6$:
\begin{align}
&(-48 \hcs h^2 z - \frac{3 z\,\delta}{2} - 6 \hcthree h^2 z\,\delta - 
 6 \hcf h^2 z\,\delta - 96 \hcs h^2 z\,\delta + 
 2 \hcthree z\,\delta^3 )A_1     \nonumber  \\
 &+ 
  (-96 \hcs h^2 z - 48 \hcs h z^2 - 3 z\,\delta - 
    12 \hcthree h^2 z\,\delta - 12 \hcf h^2 z\,\delta - 
    192 \hcs h^2 z\,\delta - 4 \hcthree h z^2 \delta - 
    6 \hcf h z^2 \delta - 96 \hcs h z^2 \delta + 
    4 \hcthree z\,\delta^3)L_1      \nonumber  \\
 &+ (-12 \hcthree h^2 z\,\delta - 
    12 \hcf h^2 z\,\delta + 4 \hcthree h z^2 \delta + 
    6 \hcf h z^2 \delta + 
    4 \hcthree z\,\delta^3) \zeta_1      \nonumber  \\
 &+ (4 i\, \hcthree h^3 z - 2 i\, \hcthree h^2 z^2 - 
    12 i\, \hcthree h z\,\delta^2 - 12 i\, \hcf h z\,\delta^2 + 
    2 i\, \hcthree z^2 \delta^2) \kappa_1      \nonumber  \\
 &+ (\frac{3 i\, h z}{2} + 2 i\, \hcthree h^3 z - 
    48 i\, \hcs h z\,\delta - 6 i\, \hcthree h z\,\delta^2 - 
    6 i\, \hcf h z\,\delta^2 - 
    96 i\, \hcs h z\,\delta^2) \mu_1      \nonumber  \\
 &+ (-\frac{3}{2} i\, h z - 2 i\, \hcthree h^3 z - \frac{
    3 i\, z^2}{2} - 2 i\, \hcthree h^2 z^2 + 48 i\, \hcs h z\,\delta + 
    6 i\, \hcthree h z\,\delta^2 + 6 i\, \hcf h z\,\delta^2 + 
    96 i\, \hcs h z\,\delta^2 + 
    2 i\, \hcthree z^2 \delta^2) \nu_1     \nonumber  \\
 & + (2 \hcthree h^2 z^2 - 
    2 \hcthree z^2 \delta^2) \,\d_z
  A_1 + (4 \hcthree h^2 z^2 - 4 \hcthree z^2 \delta^2) \,\d_z
  L_1      \nonumber  \\
 &- \frac{3}{2} z^2 \xi\, \,\d_z
  W_1+ (4 i\, \hcthree h z^2 \delta + 
    6 i\, \hcf h z^2 \delta) \,\d_z\mu_1 + (-4 i\, \hcthree h z^2 \delta - 
    6 i\, \hcf h z^2 \delta) \,\d_z\nu_1   \;.
\end{align}

Equation $\widehat{{\cal G}}^*_{c}= \varepsilon^{abc}\widehat{{\cal G}}_{[ab]} \to V_7$:
\begin{align}
&(\frac{3 h z}{2} + 2 \hcthree h^3 z + 3 \hcf h^3 z - 4 \hcthree h z\,\delta - 
 3 \hcf h z\,\delta - 48 \hcs h z\,\delta - 
 6 \hcthree h z\,\delta^2 - 3 \hcf h z\,\delta^2 - 
 96 \hcs h z\,\delta^2  )A_1    \nonumber  \\
 &+ 
  (3 h z + 4 \hcthree h^3 z + 6 \hcf h^3 z + \frac{3 z^2}{2} + 2 \hcthree h^2 z^2 + 
    3 \hcf h^2 z^2 - 8 \hcthree h z\,\delta - 6 \hcf h z\,\delta - 
    96 \hcs h z\,\delta - 2 \hcthree z^2 \delta + 3 \hcf z^2 \delta      \nonumber  \\
 &- 
    12 \hcthree h z\,\delta^2 - 6 \hcf h z\,\delta^2 - 
    192 \hcs h z\,\delta^2 - 2 \hcthree z^2 \delta^2 - 
    3 \hcf z^2 \delta^2)L_1  + (3 h z + 4 \hcthree h^3 z - 6 \hcf h^3 z - \frac{3 z^2}{
    2}  - 2 \hcthree h^2 z^2 + 3 \hcf h^2 z^2      \nonumber  \\
 &- 8 \hcthree h z\,\delta - 
    6 \hcf h z\,\delta - 96 \hcs h z\,\delta + 2 \hcthree z^2 \delta + 
    3 \hcf z^2 \delta - 12 \hcthree h z\,\delta^2 - 18 \hcf h z\,\delta^2 - 
    192 \hcs h z\,\delta^2 + 2 \hcthree z^2 \delta^2 - 
    3 \hcf z^2 \delta^2) \zeta_1      \nonumber  \\
 &+ (4 i\, \hcthree h^2 z + 96 i\, \hcs h^2 z - 
    2 i\, \hcthree h z^2 - 48 i\, \hcs h z^2 + 3 i\, z\,\delta + 
    12 i\, \hcthree h^2 z\,\delta + 18 i\, \hcf h^2 z\,\delta + 
    192 i\, \hcs h^2 z\,\delta - 4 i\, \hcthree h z^2 \delta      \nonumber  \\
 &- 
    12 i\, \hcf h z^2 \delta - 96 i\, \hcs h z^2 \delta - 
    4 i\, \hcthree z\,\delta^2 - 6 i\, \hcf z\,\delta^2 - 4 i\, \hcthree z\,\delta^3 + 
    6 i\, \hcf z\,\delta^3) \kappa_1    + (2 i\, \hcthree h^2 z + 6 i\, \hcf h^2 z + 
    48 i\, \hcs h^2 z + \frac{3 i\, z\,\delta}{2}      \nonumber  \\
 &+ 6 i\, \hcthree h^2 z\,\delta + 
    3 i\, \hcf h^2 z\,\delta + 96 i\, \hcs h^2 z\,\delta - 
    2 i\, \hcthree z\,\delta^2 + 3 i\, \hcf z\,\delta^2 - 2 i\, \hcthree z\,\delta^3 - 
    3 i\, \hcf z\,\delta^3) \mu_1      \nonumber  \\
 &+ (-2 i\, \hcthree h^2 z - 6 i\, \hcf h^2 z - 
    48 i\, \hcs h^2 z - 2 i\, \hcthree h z^2 - 6 i\, \hcf h z^2 - 48 i\, \hcs h z^2 - \frac{
    3 i\, z\,\delta}{2} - 6 i\, \hcthree h^2 z\,\delta      \nonumber  \\
 &- 
    3 i\, \hcf h^2 z\,\delta - 96 i\, \hcs h^2 z\,\delta - 
    4 i\, \hcthree h z^2 \delta - 96 i\, \hcs h z^2 \delta + 
    2 i\, \hcthree z\,\delta^2 - 3 i\, \hcf z\,\delta^2 + 2 i\, \hcthree z\,\delta^3 + 
    3 i\, \hcf z\,\delta^3) \nu_1      \nonumber  \\
 &+ (2 \hcthree h z^2 + 48 \hcs h z^2 + 
    4 \hcthree h z^2 \delta + 12 \hcf h z^2 \delta + 
    96 \hcs h z^2 \delta) \,\d_z
  A_1      \nonumber  \\
 &+ (4 \hcthree h z^2 + 96 \hcs h z^2 + 8 \hcthree h z^2 \delta + 
    24 \hcf h z^2 \delta + 192 \hcs h z^2 \delta) \,\d_zL_1      \nonumber  \\
 &+ (-\frac{3 i\, z^2}{2} - 2 i\, \hcthree h^2 z^2 + 3 i\, \hcf h^2 z^2 + 
    2 i\, \hcthree z^2 \delta + 3 i\, \hcf z^2 \delta + 
    2 i\, \hcthree z^2 \delta^2 - 
    3 i\, \hcf z^2 \delta^2) \,\d_z\mu_1     \nonumber  \\
 & + (\frac{3 i\, z^2}{2} + 2 i\, \hcthree h^2 z^2 - 3 i\, \hcf h^2 z^2 - 
    2 i\, \hcthree z^2 \delta - 3 i\, \hcf z^2 \delta - 
    2 i\, \hcthree z^2 \delta^2 + 
    3 i\, \hcf z^2 \delta^2) \,\d_z\nu_1  \;.
\end{align}
\end{widetext}


\section{Scalar perturbation equations}

The ten scalar equations written below have been obtained by projecting the $O(\gamma)$ pieces of
$\widehat{{\cal G}}_{ij}$ , and $\widehat{\cal T}_{ijk}$ (expressed, as explained in the
text, in our $A$-parametrization) onto ten scalars denoted $\widehat{{\cal G}}_{00}$,
$\widehat{{\cal G}}_{{\hat k}0}$, $\widehat{{\cal G}}_{0{\hat k}}$,  $\widehat{{\cal G}}^{(k\otimes k)}$,  $\widehat{{\cal G}}^{(\epsilon k)}$, $\widehat{{\cal T}}_{{\hat k}00}$, $\widehat{{\cal T}^*}_{{\hat k}0}$,  $\widehat{{\cal T}}^{(\delta)}$, $\widehat{{\cal T}}^{(\epsilon k)}$, and $\widehat{{\cal T}}^{(\epsilon k \otimes k)}$.
Here, $\widehat{{\cal G}}_{00}$ is the $00$ component of $\widehat{{\cal G}}_{ij}$, and 
(denoting ${\hat k}^a\equiv k^a/k$)
$\widehat{{\cal G}}_{{\hat k}0} \equiv {\hat k}^a \widehat{{\cal G}}_{a0}$, 
$\widehat{{\cal G}}_{0{\hat k}} \equiv {\hat k}^a \widehat{{\cal G}}_{0a}$,
$\widehat{{\cal T}}_{{\hat k}00} \equiv {\hat k}^a \widehat{{\cal T}}_{a00} $,
$\widehat{{\cal T}^*}_{{\hat k}0} \equiv \epsilon^{abc}{\hat k}_c{\cal T}_{ab0} $,
while $\widehat{\cal G}^{(\delta)}$, $\widehat{{\cal G}}^{(k\otimes k)}$  ,  $\widehat{{\cal G}}^{(\epsilon k)}$,  $\widehat{{\cal T}}^{(\delta)}$ , $\widehat{{\cal T}}^{(\epsilon k)}$, and $\widehat{{\cal T}}^{(\epsilon k \otimes k)}$ are defined by the following decompositions:
\be
\widehat{{\cal G}}_{ab}=\delta_{ab}\widehat{\cal G}^{(\delta)} + {\hat k}_a{\hat k}_b\widehat{\cal G}^{(k\otimes k)}\\ +  \epsilon_{abc}{\hat k}_c \widehat{\cal G}^{(\epsilon k)}\,, \nonumber
\ee
\be
\widehat{{\cal T}}_{0ab}=\delta_{ab}\widehat{\cal T}^{(\delta)} + {\hat k}_a{\hat k}_b\widehat{\cal T}^{(k\otimes k)} \\+ \epsilon_{abc}{\hat k}_c\widehat{\cal T}^{(\epsilon k)}\,, \nonumber
\ee

\be
\widehat{{\cal T}}_{abc}=\epsilon_{abc}\widehat{\cal T}^{(\epsilon)} + \epsilon_{abd}{\hat k}_c{\hat k}_d\widehat{\cal T}^{(\epsilon k \otimes k)} \\+  (\delta_{ac}{\hat k}_b - \delta_{bc}{\hat k}_a)\widehat{\cal T}^{(\delta \otimes k)}\,.\nonumber
\ee
We did not use the equations $\widehat{\cal G}^{(\delta)}$ , $\widehat{\cal T}^{(k\otimes k)}$ , $\widehat{\cal T}^{(\epsilon)}$ and $\widehat{\cal T}^{(\delta \otimes k)}$, which are redundant because of
the TG  Bianchi identities \cite{Nikiforova:2017saf}.
In addition, among the scalar variables $\widetilde \xi, \chi,\sigma,\rho,\theta,Q,u,M$ parametrizing the
perturbed connection one needs to rescale some of them by suitable powers of $k$ to make them dimensionless, as are
the scalar vierbein variables $\Phi, \Psi$, namely  $\chi^{\rm new}= k \chi$, $\sigma^{\rm new}= k^{-1}  \sigma$,
$Q^{\rm new}=k Q$ and $u^{\rm new}= k u$. We henceforth skip the superscripts ``new".

\begin{widetext}
Equation $\widehat{{\cal G}}_{00}$:
\begin{align}
& (6 h z - 16 \hcthree h^3 z - 16 \hcfour h^3 z - 48 \hcf h^3 z + 
    16 \hcthree h z\,\delta^2 + 16 \hcfour h z\,\delta^2 + 
    48 \hcf h z\,\delta^2 - 384 \hcs h z\,\delta^2)u  \nonumber \\ 
    &+ 
  (3 h z - 8 \hcthree h^3 z - 8 \hcfour h^3 z - 24 \hcf h^3 z + 
    8 \hcthree h z\,\delta^2 + 8 \hcfour h z\,\delta^2 + 
    24 \hcf h z\,\delta^2 - 192 \hcs h z\,\delta^2)Q  \nonumber \\ 
    &+
  (3 i\,z^2 - 8 i\,\hcthree h^2 z^2 - 8 i\,\hcfour h^2 z^2 - 24 i\,\hcf h^2 z^2 + 
    8 i\,\hcthree z^2 \delta^2 + 8 i\,\hcfour z^2 \delta^2 + 
    24 i\,\hcf z^2 \delta^2)M + 
 48 i\,\hcs h z^2 \theta  \nonumber \\ 
    & + (-4 i\,\hcthree z^2 \delta - 
    4 i\,\hcfour z^2 \delta - 12 i\,\hcf z^2 \delta) \widetilde{\xi} - 
 192 i\,\hcs h z^2 \delta \rho + (9 z\,\delta - 
    24 \hcthree h^2 z\,\delta - 24 \hcfour h^2 z\,\delta  \nonumber \\ 
    & - 
    72 \hcf h^2 z\,\delta + 576 \hcs h^2 z\,\delta + 
    24 \hcthree z\,\delta^3 + 24 \hcfour z\,\delta^3 + 
    72 \hcf z\,\delta^3) \sigma   + (144 \hcs h^2 + 12 \hcthree \delta^2 + 
    12 \hcfour \delta^2 + 36 \hcf \delta^2 - 
    9 \xi\,) \Phi\,   \nonumber \\ 
    &+ (3 z\,\delta - 8 \hcthree h^2 z\,\delta - 
    8 \hcfour h^2 z\,\delta - 24 \hcf h^2 z\,\delta + 
    192 \hcs h^2 z\,\delta + 8 \hcthree z\,\delta^3 + 8 \hcfour z\,\delta^3 + 
    24 \hcf z\,\delta^3) \chi  \nonumber \\ 
    & + (9 h^2 + 144 \hcs h^2 - 24 \hcthree h^4 - 
    24 \hcfour h^4 - 72 \hcf h^4 - 9 \delta^2 + 12 \hcthree \delta^2 + 
    12 \hcfour \delta^2 + 36 \hcf \delta^2 + 48 \hcthree h^2 \delta^2   \nonumber \\ 
    &+ 
    48 \hcfour h^2 \delta^2 + 144 \hcf h^2 \delta^2 - 
    1152 \hcs h^2 \delta^2 - 24 \hcthree \delta^4 - 24 \hcfour \delta^4 - 
    72 \hcf \delta^4 + 3 z^2 \xi) \Psi   - 
 48 \hcs h z^2 \,\d_zQ   \nonumber \\ 
    &- 
 96 \hcs h z^2 \,\d_z
  u + (12 \hcthree z^2 \delta + 12 \hcfour z^2 \delta + 
    36 \hcf z^2 \delta) \,\d_z\sigma + (4 \hcthree z^2 \delta + 4 \hcfour z^2 \delta + 
    12 \hcf z^2 \delta) \,\d_z\chi - 
 9 z \xi \,\d_z\Psi  \;.
\end{align}

Equation $\widehat{{\cal G}}_{{\hat k}0}$:
\begin{align}
&z (2 i\,\hcthree h z - 2 i\,\hcfour h z)u  + 
  z (-2 \hcthree h^2 + 2 \hcfour h^2 + 4 \hcthree \delta^2)M +   \nonumber \\ 
    &
 z (3 h - 8 \hcthree h^3 - 8 \hcfour h^3 - 24 \hcf h^3 + 2 \hcthree h \delta + 
    10 \hcfour h \delta + 24 \hcf h \delta - 96 \hcs h \delta + 
    8 \hcthree h \delta^2 + 8 \hcfour h \delta^2 + 24 \hcf h \delta^2 - 
    192 \hcs h \delta^2) \theta   \nonumber \\ 
    &+ 
 z (2 \hcthree h^2 - 2 \hcfour h^2 + 96 \hcs h^2 + 3 \delta - 
    8 \hcthree h^2 \delta - 8 \hcfour h^2 \delta - 24 \hcf h^2 \delta + 
    192 \hcs h^2 \delta   \nonumber \\ 
    &+ 4 \hcthree \delta^2 + 8 \hcfour \delta^2 + 
    24 \hcf \delta^2 + 8 \hcthree \delta^3 + 8 \hcfour \delta^3 + 
    24 \hcf \delta^3) \widetilde{\xi\,}   \nonumber \\ 
    &+ 
 z (6 \hcthree h \delta - 2 \hcfour h \delta) \rho + 
 4 i\,\hcthree z^2 \delta \sigma - 3 i\,z \xi\, \Phi\,   \nonumber \\ 
    &+(3 z^2 - 8 \hcthree h^2 z^2 - 8 \hcfour h^2 z^2 - 24 \hcf h^2 z^2 + 
    4 \hcthree z^2 \delta + 8 \hcfour z^2\delta + 24 \hcf z^2 \delta + 
    8 \hcthree z^2 \delta^2 + 8 \hcfour z^2\delta^2 + 
    24 \hcf z^2 \delta^2) \,\d_zM   \nonumber \\ 
    &- 
 2 h z^2 [\hcthree - \hcfour + 48 (\hcs + 2 \hcs \delta)] \,\d_z\rho - 3 i\,z^2 \xi\, \,\d_z\Psi  \;.
 \end{align}

Equation $\widehat{{\cal G}}_{0{\hat k}}$:
\begin{align}
& z (-96 i\,\hcs h z - 4 i\,\hcthree h z\,\delta + 4 i\,\hcfour h z\,\delta - 
    192 i\,\hcs h z\,\delta)u   \nonumber \\ 
    &+ 
  z (96 \hcs h^2 + 3 \delta - 12 \hcfour h^2 \delta - 
    24 \hcf h^2 \delta + 192 \hcs h^2 \delta + 8 \hcthree \delta^2 + 
    8 \hcfour \delta^2 + 24 \hcf \delta^2 + 4 \hcthree \delta^3 + 
    8 \hcfour \delta^3 + 24 \hcf \delta^3)M   \nonumber \\ 
    &+ 
 z (-4 \hcthree h^3 + 8 \hcthree h \delta^2 - 4 \hcfour h \delta^2) \theta + 
 z (-8 \hcthree h^2 \delta + 4 \hcfour h^2 \delta + 
    4 \hcthree \delta^3) \widetilde{\xi}   \nonumber \\ 
    &+ 
 z (3 h - 4 \hcthree h^3 - 8 \hcfour h^3 - 24 \hcf h^3 + 8 \hcthree h \delta + 
    8 \hcfour h \delta + 24 \hcf h \delta - 96 \hcs h \delta + 
    12 \hcfour h \delta^2 + 24 \hcf h \delta^2 - 
    192 \hcs h \delta^2) \rho   \nonumber \\ 
    &+ 
 z (3 i\,z - 4 i\,\hcthree h^2 z - 8 i\,\hcfour h^2 z - 24 i\,\hcf h^2 z + 
    8 i\,\hcthree z\,\delta + 8 i\,\hcfour z\,\delta + 24 i\,\hcf z\,\delta + 
    4 i\,\hcthree z\,\delta^2 + 8 i\,\hcfour z\,\delta^2 + 
    24 i\,\hcf z\,\delta^2) \sigma   \nonumber \\ 
    &- 3 i\,z \xi \Phi   + 
 4 \hcthree z^2 (-h^2 + \delta^2) \,\d_zM + 
 4 (\hcthree - \hcfour) h z^2 \delta \,\d_z\rho - 
 3 i\,z^2 \xi \,\d_z\Psi  \;.
 \end{align}

Equation $\widehat{{\cal G}}^{(k\otimes k)}$:
\begin{align}
& (\frac{3 h z}{2} - 5 \hcthree h^3 z - 5 \hcfour h^3 z - 12 \hcf h^3 z + 
    3 \hcthree h z\,\delta + 3 \hcfour h z\,\delta + 12 \hcf h z\,\delta - 
    48 \hcs h z\,\delta    \nonumber \\
    &+ 5 \hcthree h z\,\delta^2 + 5 \hcfour h z\,\delta^2 + 
    12 \hcf h z\,\delta^2 - 96 \hcs h z\,\delta^2)Q + 
  (-\frac{3 h z}{2} + 5 \hcthree h^3 z + 5 \hcfour h^3 z + 12 \hcf h^3 z - 
    3 \hcthree h z\,\delta - 3 \hcfour h z\,\delta    \nonumber \\
    &- 12 \hcf h z\,\delta + 
    48 \hcs h z\,\delta - 5 \hcthree h z\,\delta^2 - 5 \hcfour h z\,\delta^2 - 
    12 \hcf h z\,\delta^2 + 96 \hcs h z\,\delta^2)u    \nonumber \\
    &+ 
  (\frac{3 i\,z^2}{2} - 5 i\,\hcthree h^2 z^2 - 5 i\,\hcfour h^2 z^2 - 12 i\,\hcf h^2 z^2 + 
    3 i\,\hcthree z^2 \delta + 3 i\,\hcfour z^2 \delta + 
    12 i\,\hcf z^2 \delta + 5 i\,\hcthree z^2 \delta^2    \nonumber \\
    &+ 
    5 i\,\hcfour z^2 \delta^2 + 
    12 i\,\hcf z^2 \delta^2)M    + (-48 i\,\hcs h z^2 - 
    96 i\,\hcs h z^2 \delta) \theta    \nonumber \\
    &+ (\frac{3 i\,z^2}{2} - 
    3 i\,\hcthree h^2 z^2 - 3 i\,\hcfour h^2 z^2 - 12 i\,\hcf h^2 z^2 + 
    5 i\,\hcthree z^2 \delta + 5 i\,\hcfour z^2 \delta + 
    12 i\,\hcf z^2 \delta + 3 i\,\hcthree z^2 \delta^2 + 
    3 i\,\hcfour z^2 \delta^2 + 
    12 i\,\hcf z^2 \delta^2) \widetilde{\xi}    \nonumber \\
    &+ (-48 i\,\hcs h z^2 - 
    96 i\,\hcs h z^2 \delta) \rho + 
 \frac{3}{2} z^2 \xi \Phi  + (48 \hcs h^2 z + \frac{3 z\,\delta}{2} - 
    5 \hcthree h^2 z\,\delta - 5 \hcfour h^2 z\,\delta - 
    12 \hcf h^2 z\,\delta    \nonumber \\
    &+ 96 \hcs h^2 z\,\delta + 3 \hcthree z\,\delta^2 + 
    3 \hcfour z\,\delta^2 + 12 \hcf z\,\delta^2 + 5 \hcthree z\,\delta^3 + 
    5 \hcfour z\,\delta^3 + 12 \hcf z\,\delta^3) \chi    \nonumber \\
    &+ 
 \frac{3}{2} z^2 \xi \psi + (48 \hcs h z^2 + 
    96 \hcs h z^2 \delta) \,\d_z
  Q + (-48 \hcs h z^2 - 96 \hcs h z^2 \delta) \,\d_z
  u    \nonumber \\
    &+ (-\frac{3 z^2}{2} + 3 \hcthree h^2 z^2 + 3 \hcfour h^2 z^2 + 12 \hcf h^2 z^2 - 
    5 \hcthree z^2 \delta - 5 \hcfour z^2 \delta - 12 \hcf z^2 \delta - 
    3 \hcthree z^2 \delta^2 - 3 \hcfour z^2 \delta^2 - 
    12 \hcf z^2 \delta^2) \,\d_z\chi   \;.
    \end{align}

Equation $\widehat{{\cal G}}^{(\epsilon k)}$:
\begin{align}
& (-\frac{3 h z}{2} + 3 \hcthree h^3 z + 5 \hcfour h^3 z + 12 \hcf h^3 z - 
    \hcthree h z\,\delta - 5 \hcfour h z\,\delta - 12 \hcf h z\,\delta + 
    48 \hcs h z\,\delta    \nonumber \\
    &+ \hcthree h z\,\delta^2 - 5 \hcfour h z\,\delta^2 - 
    12 \hcf h z\,\delta^2 + 96 \hcs h z\,\delta^2)M  + 
  (\frac{3 i\,z^2}{2} - 3 i\,\hcthree h^2 z^2 - 5 i\,\hcfour h^2 z^2 - 12 i\,\hcf h^2 z^2 + 
    i\,\hcthree z^2 \delta + 3 i\,\hcfour z^2 \delta    \nonumber \\
    &+ 12 i\,\hcf z^2 \delta + 
    3 i\,\hcthree z^2 \delta^2 + 5 i\,\hcfour z^2 \delta^2 + 
    12 i\,\hcf z^2 \delta^2)u + (-2 \hcthree h^2 z - 48 \hcs h^2 z - \frac{
    3 z\,\delta}{2} + \hcthree h^2 z\,\delta + 7 \hcfour h^2 z\,\delta + 
    12 \hcf h^2 z\,\delta    \nonumber \\
    &- 96 \hcs h^2 z\,\delta - 3 \hcthree z\,\delta^2 - 
    5 \hcfour z\,\delta^2 - 12 \hcf z\,\delta^2 - \hcthree z\,\delta^3 - 
    3 \hcfour z\,\delta^3 - 12 \hcf z\,\delta^3) \theta   + (\frac{3 h z}{2} - 
    \hcthree h^3 z - 3 \hcfour h^3 z - 12 \hcf h^3 z    \nonumber \\
    &+ \hcthree h z\,\delta + 
    5 \hcfour h z\,\delta  + 12 \hcf h z\,\delta - 48 \hcs h z\,\delta + 
    \hcthree h z\,\delta^2 + 7 \hcfour h z\,\delta^2 + 12 \hcf h z\,\delta^2 - 
    96 \hcs h z\,\delta^2) \widetilde{\xi}    \nonumber \\
    &+ (-2 \hcfour h^2 z + 
    48 \hcs h^2 z + \frac{3 z\,\delta}{2} + \hcthree h^2 z\,\delta - 
    5 \hcfour h^2 z\,\delta - 12 \hcf h^2 z\,\delta + 
    96 \hcs h^2 z\,\delta + \hcthree z\,\delta^2 + 3 \hcfour z\,\delta^2 + 
    12 \hcf z\,\delta^2    \nonumber \\
    &+ 3 \hcthree z\,\delta^3 + 5 \hcfour z\,\delta^3 + 
    12 \hcf z\,\delta^3) \rho + (-2 i\,\hcfour h z^2 + 48 i\,\hcs h z^2 + 
    4 i\,\hcthree h z^2 \delta + 
    96 i\,\hcs h z^2 \delta) \sigma    \nonumber \\
    &+ (-2 \hcthree h z^2 - 48 \hcs h z^2 + 
    4 \hcfour h z^2 \delta - 96 \hcs h z^2 \delta) \,\d_zM    \nonumber \\
    &+ (-\frac{3 z^2}{2} + \hcthree h^2 z^2 + 3 \hcfour h^2 z^2 + 12 \hcf h^2 z^2 - 
    3 \hcthree z^2 \delta - 5 \hcfour z^2 \delta - 12 \hcf z^2 \delta - 
    \hcthree z^2 \delta^2 - 3 \hcfour z^2 \delta^2 - 
    12 \hcf z^2 \delta^2) \,\d_z\rho  \;.
\end{align}

Equation $\widehat{{\cal T}}_{{\hat k}00}$:
\begin{align}
&\frac{1}{3}  z (2 i\,\hcthree h z + 2 i\,\hcfour h z)Q + 
 \frac{1}{3}  z (16 i\,\hcthree h z + 4 i\,\hcfour h z)u + 
 \frac{1}{3}  z (-9 - 18 \hcthree h^2 - 6 \hcfour h^2 - 2 \hcthree z^2 - 2 \hcfour z^2 + 
    12 \hcthree \delta^2)M   \nonumber \\ 
    &+ 
 \frac{1}{3} z (-18 \hcthree h \delta + 6 \hcfour h \delta) \theta + 
 \frac{1}{3} z (6 \hcthree h^2 - 6 \hcfour h^2 + 2 \hcthree z^2 + 2 \hcfour z^2 - 
    12 \hcthree \delta^2) \widetilde{\xi}   \nonumber \\ 
    &+ 
 \frac{1}{3} z (288 \hcs h + 6 \hcthree h \delta - 18 \hcfour h \delta + 
    576 \hcs h \delta) \rho + 4 i\,\hcthree z^2 \delta \sigma + 
 \frac{1}{3} z (2 i\,\hcthree z\,\delta + 2 i\,\hcfour z\,\delta) \chi   \nonumber \\ 
    &- 
 3 i\,z \Psi - 4 \hcthree z^2 \delta \,\d_zM + 
 2 (-\hcthree + \hcfour) h z^2 \,\d_z\rho + 
 \frac{1}{3} z (2 i\,\hcthree z^2 + 2 i\,\hcfour z^2) \,\d_z\chi  \;.
 \end{align}

Equation $\widehat{{\cal T}^*}_{{\hat k}0}$:
\begin{align}
&-32 i\,\hcs  z^2 \delta Q + 
  z (96 \hcs h + 2 \hcthree h \delta - 6 \hcfour h \delta + 
    192 \hcs h \delta)M + 
  z (-2 i\,\hcthree z\,\delta + 2 i\,\hcfour z\,\delta - 64 i\,\hcs z\,\delta)u   \nonumber \\ 
    &+ 
 z (-4 \hcthree h^2 + 16 \hcs z^2 + 2 \hcthree \delta^2 - 
    2 \hcfour \delta^2) \theta + 
 z (-6 \hcthree h \delta + 2 \hcfour h \delta) \widetilde{\xi}   \nonumber \\ 
    &+ 
 z (3 + 4 \hcthree h^2 + 32 \hcs z^2 - 6 \hcthree \delta^2 - 
    2 \hcfour \delta^2) \rho + 
 z (-4 i\,\hcfour h z + 96 i\,\hcs h z) \sigma - 48 i\,\hcs h z \Phi   \nonumber \\ 
    &+ 
 32 i\,\hcs h z^2 \chi + 48 i\,\hcs h z \Psi - 
 4 \hcthree h z^2 \,\d_zM + 
 16 i\,\hcs z^3 \,\d_zQ   \nonumber \\ 
    &+ 
 32 i\,\hcs z^3 \,\d_zu + 
 z (2 \hcthree z\,\delta - 2 \hcfour z\,\delta) \,\d_z\rho   \;.
 \end{align}

Equation $\widehat{{\cal T}}^{(\delta)}$:
\begin{align}
&\frac{1}{6}  (14 \hcthree h z + 14 \hcfour h z - 288 \hcs h z + 2 \hcthree h z\,\delta + 
    2 \hcfour h z\,\delta - 1344 \hcs h z\,\delta)u   \nonumber \\ 
    &+ 
 \frac{1}{6}  (10 \hcthree h z + 10 \hcfour h z - 288 \hcs h z - 2 \hcthree h z\,\delta - 
    2 \hcfour h z\,\delta - 960 \hcs h z\,\delta)Q   + 
 \frac{1}{6}  (10 i\,\hcthree z^2 \delta - 2 i\,\hcfour z^2 \delta)M   \nonumber \\ 
    &+ 
  \frac{1}{6} (12 i\,\hcfour h z^2 - 192 i\,\hcs h z^2) \theta + 
 \frac{1}{6} (-10 i\,\hcthree z^2 \delta + 2 i\,\hcfour z^2 \delta) \widetilde{\xi} + 
 \frac{1}{6} (12 i\,\hcthree h z^2 - 384 i\,\hcs h z^2) \rho   \nonumber \\ 
    &+ 
 \frac{1}{6} (18 z + 1152 \hcs h^2 z - 12 \hcthree z^3) \sigma + 
 \frac{1}{6} (18 + 24 \hcthree h^2 + 24 \hcfour h^2 - 576 \hcs h^2) \Phi   \nonumber \\ 
    &+ 
\frac{1}{6} (9 z + 384 \hcs h^2 z - 2 \hcthree z\,\delta - 2 \hcfour z\,\delta - 
    2 \hcthree z\,\delta^2 - 2 \hcfour z\,\delta^2) \chi   \nonumber \\ 
    &+ 
 \frac{1}{6} (48 \hcthree h^2 + 48 \hcfour h^2 - 1152 \hcs h^2 - 18 \delta - 
    3456 \hcs h^2 \delta) \Psi - 
 \frac{1}{3} i\,(5 \hcthree - \hcfour) z^3 \,\d_zM   \nonumber \\ 
    &- 
 \frac{1}{3} (5 \hcthree + 5 \hcfour - 96 \hcs) h z^2 \,\d_zQ + 
\frac{1}{6} (-14 \hcthree h z^2 - 14 \hcfour h z^2 + 384 \hcs h z^2) \,\d_zu + 
 \frac{1}{6} (-2 i\,\hcthree z^3 - 2 i\,\hcfour z^3) \,\d_z\widetilde{\xi}  \nonumber \\ 
    & + 3 z \,\d_z\Psi + 
 \frac{1}{6} (2 \hcthree z^3 + 2 \hcfour z^3) \,\d^2_z\chi  \;.
 \end{align}

Equation $\widehat{{\cal T}}^{(\epsilon k)}$:
\begin{align}
&-32 i\,\hcs  z^2 \delta Q + 
  (\hcthree h z - \hcfour h z + 48 \hcs h z - 3 \hcthree h z\,\delta + 
    \hcfour h z\,\delta + 96 \hcs h z\,\delta)M    \nonumber \\
    &+ 
  (i\,\hcthree z^2 \delta - i\,\hcfour z^2 \delta - 64 i\,\hcs z^2 \delta)u + ( \frac{
    3 z}{2} + 2 \hcthree h^2 z + 16 \hcs z^3 + \hcthree z\,\delta - \hcfour z\,\delta - 
    3 \hcthree z\,\delta^2 - \hcfour z\,\delta^2) \theta    \nonumber \\
    &+ (-\hcthree h z + 
    \hcfour h z + 48 \hcs h z + \hcthree h z\,\delta - 3 \hcfour h z\,\delta + 
    96 \hcs h z\,\delta) \widetilde{\xi}    \nonumber \\
    &+ ( \frac{3 z}{2} - 2 \hcthree h^2 z + 
    32 \hcs z^3 - 3 \hcthree z\,\delta - \hcfour z\,\delta + \hcthree z\,\delta^2 - 
    \hcfour z\,\delta^2) \rho   + (-2 i\,\hcthree h z^2 + 
    96 i\,\hcs h z^2) \sigma    \nonumber \\
    &+ 96 i\,\hcs h z\,\delta \phi + 
 32 i\,\hcs h z^2 \chi - 
 96 i\,\hcs h z\,\delta \psi + (-\hcthree h z^2 - 
    \hcfour h z^2) \,\d_zM + 
 16 i\,\hcs z^3 \,\d_z
  Q   \nonumber \\
    & + (i\,\hcthree z^3 - i\,\hcfour z^3 + 32 i\,\hcs z^3) \,\d_z
  u + (-\hcthree z^2 \delta + \hcfour z^2 \delta) \,\d_z\theta + (\hcthree h z^2 - \hcfour h z^2) \,\d_z\widetilde{\xi} + (-\hcthree z^3 + \hcfour z^3) \,\d^2_z\rho   \;.
 \end{align}

Equation $\widehat{{\cal T}}^{(\epsilon k \otimes k)}$:
\begin{align}
&-2 i\,\hcfour h  z^2M + 
  ( \frac{3 z}{2} + \hcthree h^2 z + \hcfour h^2 z + \hcthree z^3 - \hcfour z^3 - 
    2 \hcthree z\,\delta - 2 \hcfour z\,\delta) u  \nonumber \\
    & + 
  (- \frac{3 z}{2} - \hcthree h^2 z - \hcfour h^2 z + 2 \hcthree z\,\delta + 
    2 \hcfour z\,\delta)Q + (i\,\hcthree z^2 \delta - 
    i\,\hcfour z^2 \delta) \theta - 
 2 i\,\hcthree h z^2 \widetilde{\xi}    \nonumber \\
    &+ (-i\,\hcthree z^2 \delta + 
    i\,\hcfour z^2 \delta) \rho + (-48 \hcs h z - \hcthree h z\,\delta - 
    \hcfour h z\,\delta - 96 \hcs h z\,\delta) \chi    \nonumber \\
    &+ (i\,\hcthree z^3 - 
    i\,\hcfour z^3) \,\d_z\rho + (\hcthree h z^2 + \hcfour h z^2) \,\d_z\chi  \;.
 \end{align}
 \end{widetext}


\end{document}